\documentclass[twocolumn,times,tighten]{aastex631}

\usepackage{caption}

\makeatletter
\newcommand\footnoteref[1]{\protected@xdef\@thefnmark{\ref{#1}}\@footnotemark}
\makeatother

\usepackage{soul}
\newcommand{\del}[1]{\textcolor{magenta}{{\iffalse{#1}\fi}}}

\received{ \today}
\revised{} 
\accepted{} 

\shorttitle{The FAST GPPS survey VI: The discovery of 473 new pulsars}
\shortauthors{Han et al.}

\begin{document}


\title{The FAST Galactic Plane Pulsar Snapshot survey:
  VI. The discovery of 473 new pulsars\footnote{See the essays in the news and Views}}

\correspondingauthor{J.L. Han: hjl@bao.ac.cn}

\author[0000-0002-9274-3092]{J.~L. Han} %
\author{D.~J. Zhou} 
\author{C. Wang}
\author{W.~Q. Su}
\author{Yi Yan}
\author{W.~C. Jing} 
\author{Z.~L. Yang}
\author{P.~F. Wang}
\author{T. Wang}
\author{J. Xu}
\author{N.~N. Cai}
\author{J.~H. Sun}

\affil{National Astronomical Observatories, Chinese Academy of Sciences, Beijing 100101, China}
\affil{School of Astronomy and Space Sciences, University of Chinese Academy of Sciences, Beijing 100049, China}

\author{Q.~L. Yang}
\affil{National Astronomical Observatories, Chinese Academy of Sciences, Beijing 100101, China. Email: hjl@nao.cas.cn}

\author{R.~X. Xu}
\affil{Department of Astronomy, Peking University, Beijing 100871, China}
\affil{Kavli Institute for Astronomy and Astrophysics, Peking University, Beijing 100871, China}

\author{H.~G. Wang}         
\affil{Department of Astronomy, School of Physics and Materials Science, Guangzhou University, Guangzhou 510006, China}

\author{X.~P. You}
\affil{School of Physical Science and Technology, Southwest University, Chongqing 400715, China}

\begin{abstract}
The Five-hundred-meter Aperture Spherical radio Telescope (FAST) is the most sensitive telescope at the $L$-band (1.0–1.5 GHz) and has been used to carry out the FAST Galactic Plane Pulsar Snapshot (GPPS) survey in the last 5 yr. Up to now, the survey has covered one-fourth of the planned areas within $\pm10^{\circ}$ from the Galactic plane visible by FAST, and discovered 751 pulsars. After the first publication of the discovery of 201 pulsars and one rotating radio transient (RRAT) in 2021 and 76 RRATs in 2023, here we report the discovery of 473 new pulsars from the FAST GPPS survey, including 137 new millisecond pulsars and 30 new RRATs. We find 34 millisecond pulsars discovered by the GPPS survey which can be timed with a precision better than 3 $\mu$s by using FAST 15 minute observations and can be used for pulsar timing arrays. The GPPS survey has discovered eight pulsars with periods greater than 10 s including one with 29.77 s. The integrated profiles of pulsars and individual pulses of RRATs are presented. During the FAST GPPS survey, we also detected previously known pulsars and updated parameters for 52 pulsars. In addition, we discovered two fast radio bursts plus one probable case with high dispersion measures indicating their extragalactic origin.
\end{abstract}


\section{Introduction}

Pulsars are rotating neutron stars with radiation beams sweeping across the Earth. They are the condensed remnants of dead stars distributed widely in the Milky Way. There are two approaches for stars to evolve into neutron stars. The stars with intermediate masses can evolve into red giants, and finally, the neutron stars are born during the supernova so that some young pulsars have been detected within or beside supernova remnants \citep[e.g.][]{kmj+96, lzc+24}. Stars with a mass of less than 8 MSun can evolve into white dwarfs, and then they can accrete materials from the companion if they are in a binary. When the total mass exceeds the Chandrasekhar limit, the white dwarf can collapse to form a neutron star \citep{acrs82}. Such formed neutron stars have been recycled to a short spin period. Though the distribution of pulsars in the Milky Way is unknown, one may expect their distribution similar to stars. The distribution is strongly dispersed due to the large kick velocities of neutron stars gained from the supernova explosion \citep[e.g.][]{ll94, hllk05}.

The rotation periods of currently detected pulsars are distributed in several groups from a millisecond to tens of seconds \citep{hrs+06, chr+22}. Pulsars with a period less than 30 ms are millisecond pulsars \citep{bkh+82}, often in binary systems. Pulsars have periods ranging from 1.4 ms to 76 s. There have been 10 pulsars identified with periods greater than 10 s \citep[e.g.][]{chr+22}, including two we found previously \citep{hww+21}. The larger pulse duty cycles of millisecond pulsars are generally thought \citep[see, e.g.][]{kll+99} to reflect their wide emission beams compared to those of normal pulsars \citep{lm88}. 

\begin{figure*}[t]
\centering 
\includegraphics[width=0.99\textwidth]{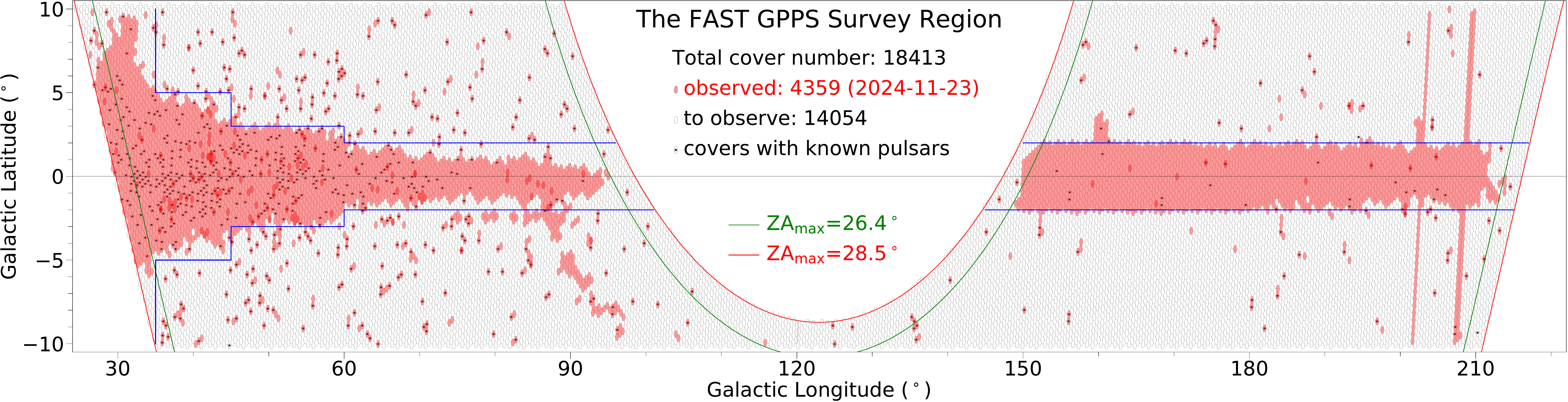}   
\caption{The current progress of the FAST GPPS survey which observes the visible sky regions within  $\pm10\degr$ from the Galactic plane. Each point is a {\it cover} that is a sky area filled by 76 beams of four adjacent pointings by a FAST ``snapshot mode” observation. The observed covers are marked with filled color. The covers marked with a star are those with one or more known pulsars. The zenith angle limit for the FAST GPPS survey was set to 26$\degr$.4  for a full telescope gain due to the feed illumination and now extends to 28$\degr$ .5 without significant degradation of the gain and system noise.
}
   \label{gppsSky}
\end{figure*}

There have been many pulsar surveys, some done around the Galactic plane for pulsars in the Galactic disk \citep{clj+92, jlm+92, mld+96, mlc+01, cfl+06, psf+22, pbs+23}, some done over the whole visible sky \citep{kjv+10, bck+13, blr+13, kbj+18, tkt22, ttk+24}, and some done in high or intermediate latitude regions for halo pulsars \citep{jbo+09, bcm+16, bsb+19, mlk+24, zbd+24}. Because of the tenuous interstellar medium in the Galactic halo, the dispersion measures (DMs) of pulsars in the halo are often small and can be easily discovered at lower frequencies \citep{bsm+23, tt23}. Pulsars in the Galactic disk suffer more radio wave dispersion, especially those behind HII regions \citep{hww+21}.  For example, \citet{mlk+24} recently surveyed an area of the Galactic plane, and found that the distant pulsars with high DM often have a large scattering which prevents the detection at low frequencies. To detect pulsars in more distant parts of the Galactic disk with larger DMs and large scattering effect, pulsar surveys must be carried out at higher frequencies \citep{clj+92, jlm+92, xwhh11}. However, pulsars get much weaker at higher frequencies because they have steep spectra \citep{hwxh16}, and pulsars can only be detected by sensitive observations. Also, any improvements in channel bandwidth and sampling rate could lead to a better sensitivity for short- period pulsars, which is very important to detect millisecond pulsars \citep[e.g.][]{bbb+11}. 

The significance of pulsar surveys is not just to find pulsars, but to find pulsars with important implications for physics. For example, the Green Bank Telescope 350 MHz drift-scan survey \citep{blr+13} found a pulsar with a 2.01 M$_\odot$ \citep{afw+13}. In the multibeam pulsar survey with the Arecibo telescope \citep{cfl+06}, a millisecond pulsar was found in a highly eccentric orbit \citep{crl+08}. When the survey data were searched by using the single pulse module, many rotating radio transient (RRAT) sources were discovered \citep{mll+06, bbj+11, zhx+23} as were fast radio bursts  \citep[FRBs,][]{lbm+07, zhj+23}.

\begin{table*}
\centering
\caption{Examples of the Second Massive Sample of Pulsars Discovered by the FAST GPPS survey.}
\label{gppstab9exam}
\setlength{\tabcolsep}{2.5pt} \renewcommand{\arraystretch}{0.8}  \tabletypesize{\scriptsize} 
\begin{tabular}{lccrccccrrrrr}
\hline\noalign{\smallskip}
 Name$^*$    &	gpps No.  &  Period    & DM      &  RA(2000)   &   DEC(2000)  &   GL  &    GB  &  $S_{\rm 1.25GHz}$  & $D_{\rm NE2001}$ & $D_{\rm YMW16}$ & FWHM & Ref.\\
             &            &   (s)   & (**) &  (hh:mm:ss) & ($\pm$dd:mm) &  ($^{\circ}$) & ($^{\circ}$) & ($\mu$Jy) & (kpc)  & (kpc)   &  ($\degr$)   &  \\ 
\hline
J1853$-$0009g  & gpps0202 & 0.72110 &  443.1 & 18:53:36.2 & $-$00:09 &  33.0354 & $-$0.5556 &  12.7 & 7.2 & 5.5 &    7.0\\
J1906$+$0352g  & gpps0203 & 0.28560 &  372.0 & 19:06:27.2 & $+$03:52 &  38.0986 & $-$1.5628 &  33.7 & 7.9 & 11.2&    8.5\\
J1837$+$0528g  & gpps0204 & 0.00626 &  120.8 & 18:37:40.5 & $+$05:28 &  36.2535 & $+$5.5477 &  18.9 & 3.8 & 5.9 &   17.9\\
J1840$+$0151g  & gpps0205 & 1.75041 &   73.6 & 18:40:58.6 & $+$01:51 &  33.3997 & $+$3.1769 &  20.6 & 2.5 & 2.7 &    5.4\\
J1854$+$0358g  & gpps0206 & 0.59084 &  486.5 & 18:54:04.0 & $+$03:58 &  36.7715 & $+$1.2287 &  18.7 & 9.2 & 11.9&    6.1\\
J1858$-$0055g  & gpps0207 & 2.84662 &  270.7 & 18:58:05.8 & $-$00:55 &  32.8614 & $-$1.9064 &  16.5 & 6.3 & 5.6 &   10.0\\
J1939$+$2452g  & gpps0208 & 2.40253 &  240.5 & 19:39:09.8 & $+$24:52 &  60.3194 & $+$1.4247 &   7.7 & 8.1 & 8.6 &    4.6\\
J1933$+$2037g  & gpps0209 & 0.79886 &  100.0 & 19:33:01.7 & $+$20:37 &  55.9258 & $+$0.5929 &  23.6 & 4.2 & 3.3 &   11.6\\
J1951$+$2528g  & gpps0210 & 0.00231 &  205.1 & 19:51:27.0 & $+$25:28 &  62.2360 & $-$0.6780 & 286.8 & 7.3 & 7.9 &   63.8\\
... & ... & ... & ... & ... & ... &... & ... &... & ... & ... & ... & ... \\
\hline 
\end{tabular}
\tablecomments{
See Table A1 for a complete table of pulsars from gpps0202 to gpps0751 and also http://zmtt.bao.ac.cn/GPPS/ for updates.\\
$^a$ "g" indicates the temporary nature, due to position uncertainty of about 1'.5.\\ 
$^b$ DM values in this and the following tables all have a unit of pc~cm$^{-3}$.\\
The last column is for references, and it happens that there are no references for these first pulsars.} 
\end{table*}

\begin{figure}
  \centering
  \includegraphics[width=0.98\columnwidth]{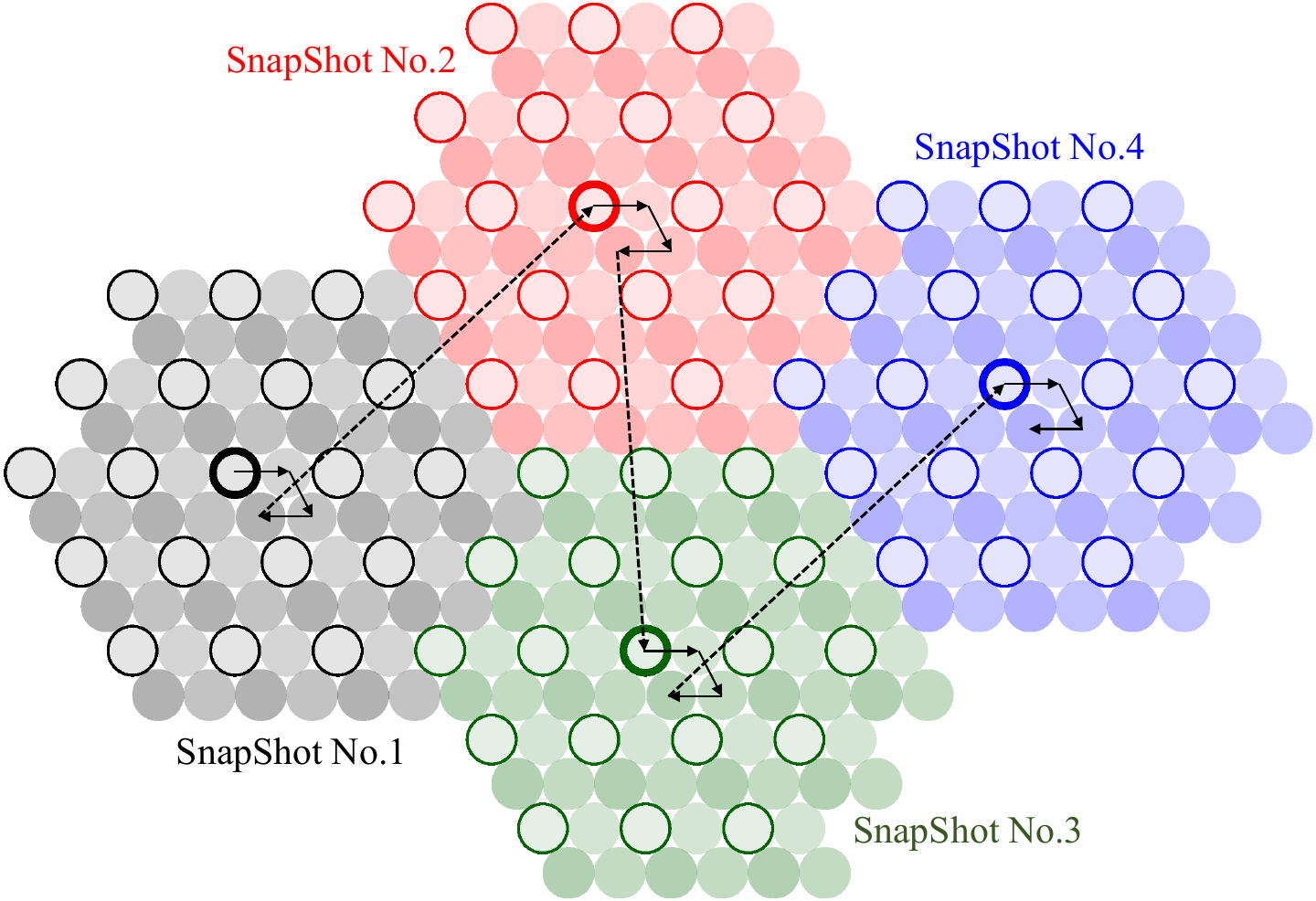}
  \caption{ The newly designed ``snapshotZ mode” for FAST to observe the four adjacent {\it covers} which saves the slewing time by 13 minutes. Each {\it cover} is observed by four adjacent paintings by FAST with the 19-beam $L$-band receiver using the ``snapshot mode” \citep{hww+21}. }
  \label{snapshotZ}
\end{figure}

\begin{figure*}[t]
\centering
\includegraphics[width=0.24\textwidth]{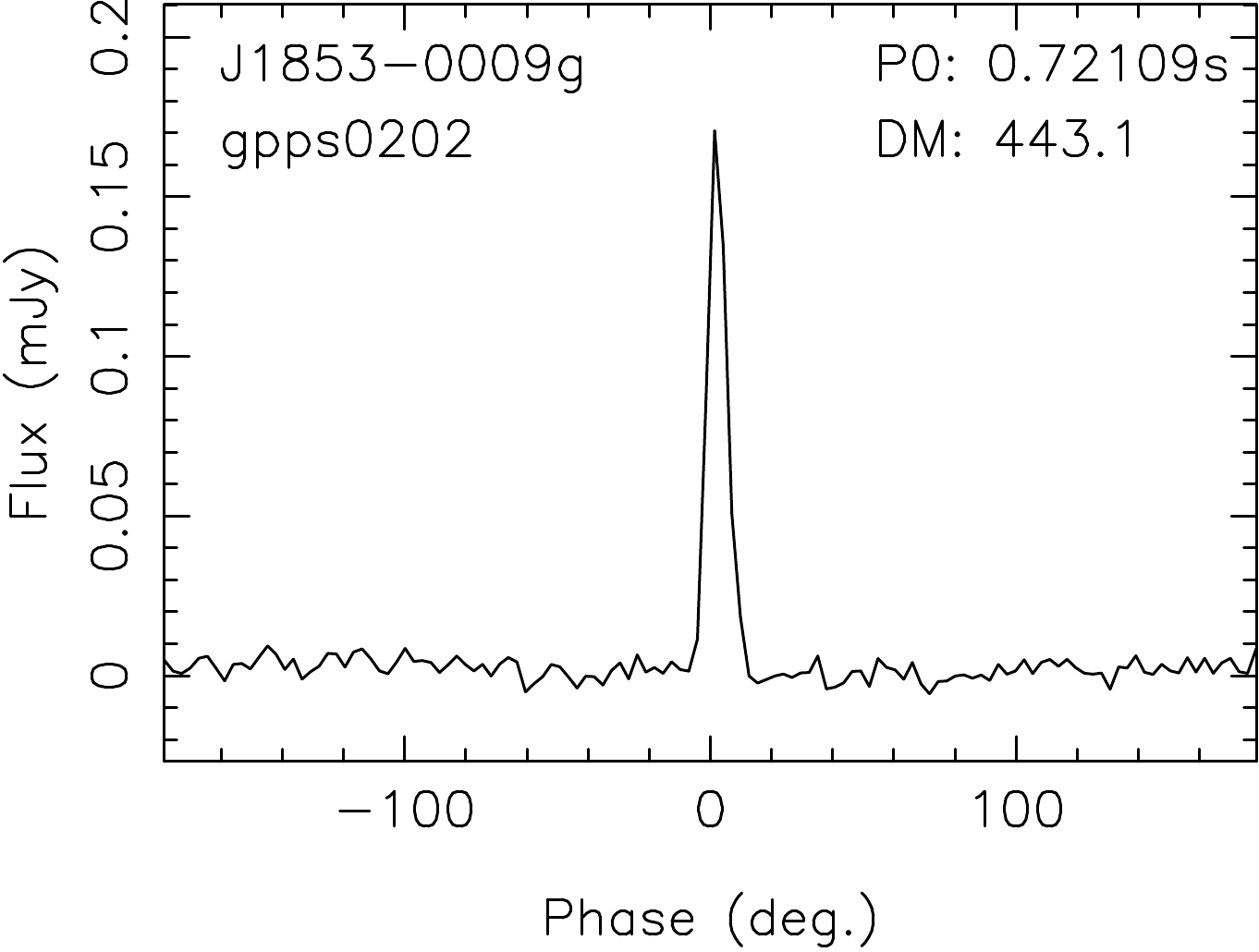}
\includegraphics[width=0.24\textwidth]{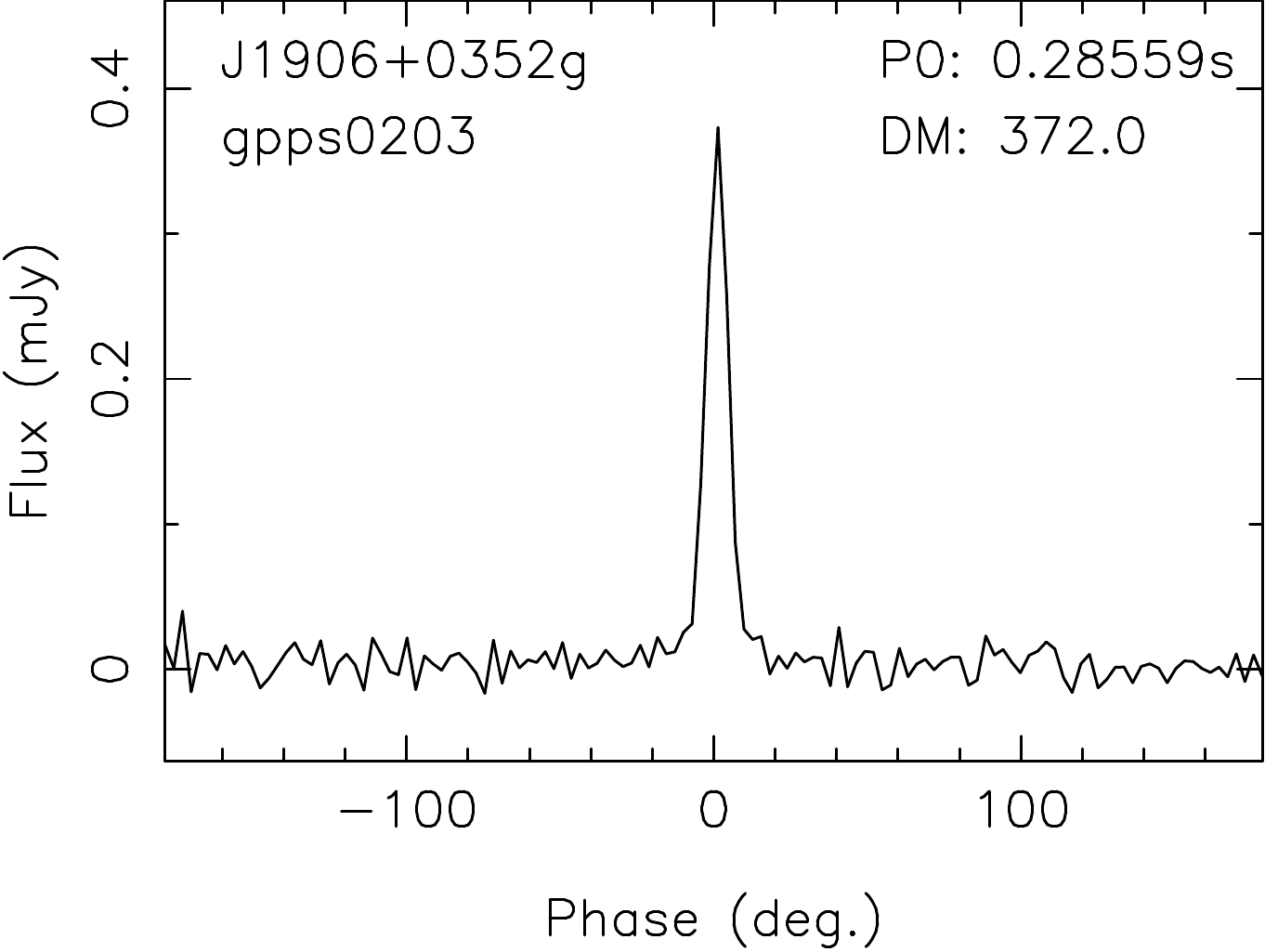}
\includegraphics[width=0.24\textwidth]{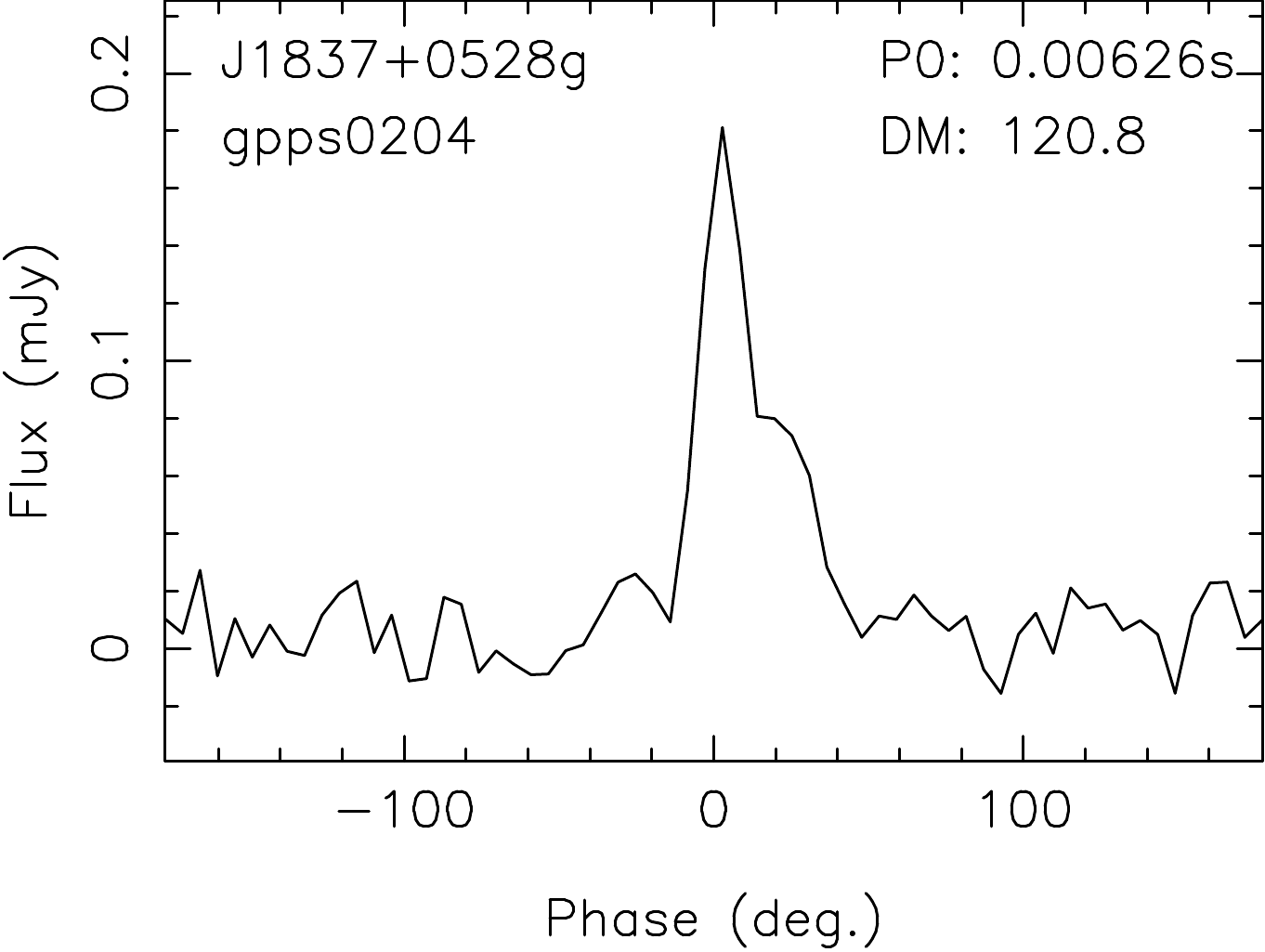}
\includegraphics[width=0.24\textwidth]{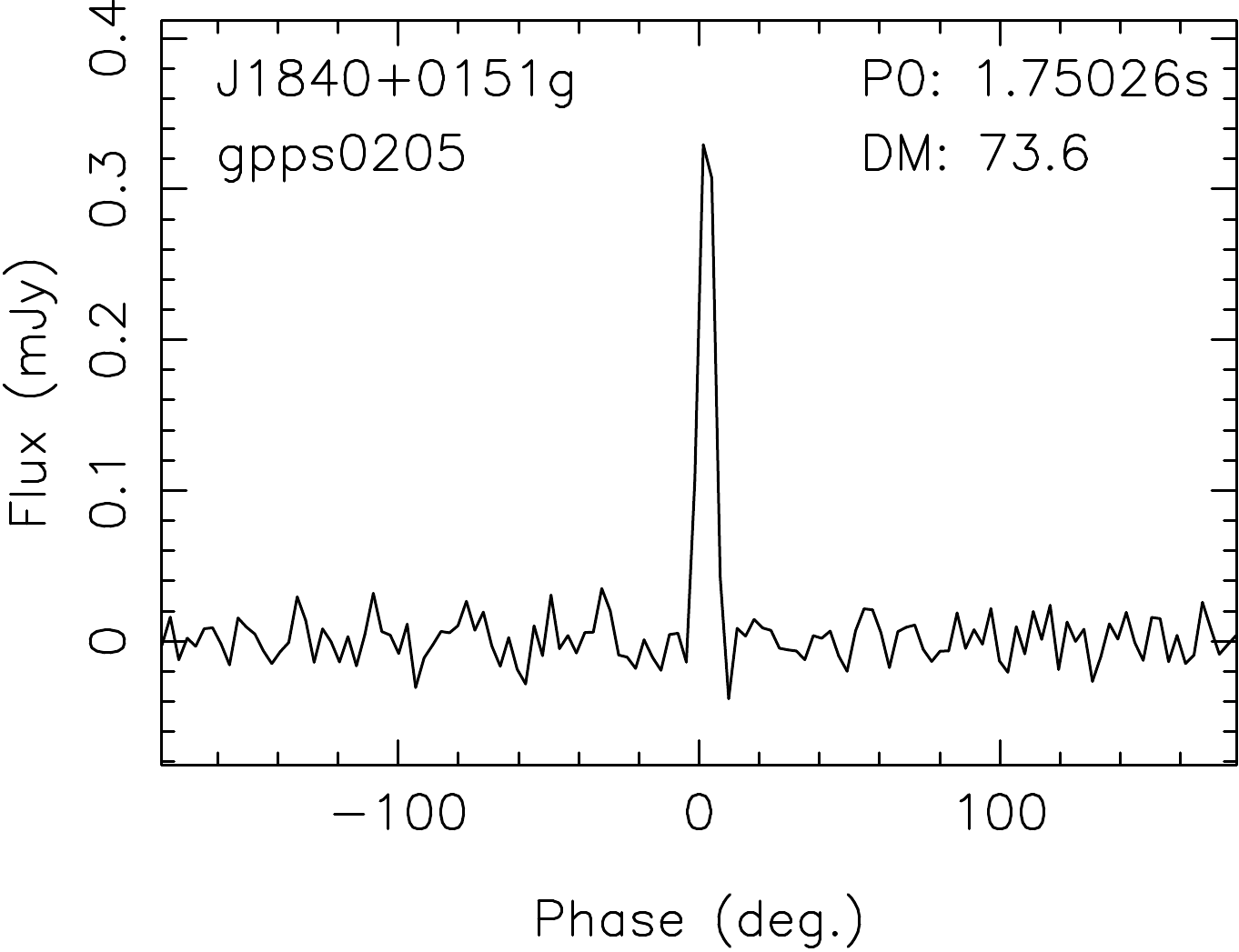}
\includegraphics[width=0.24\textwidth]{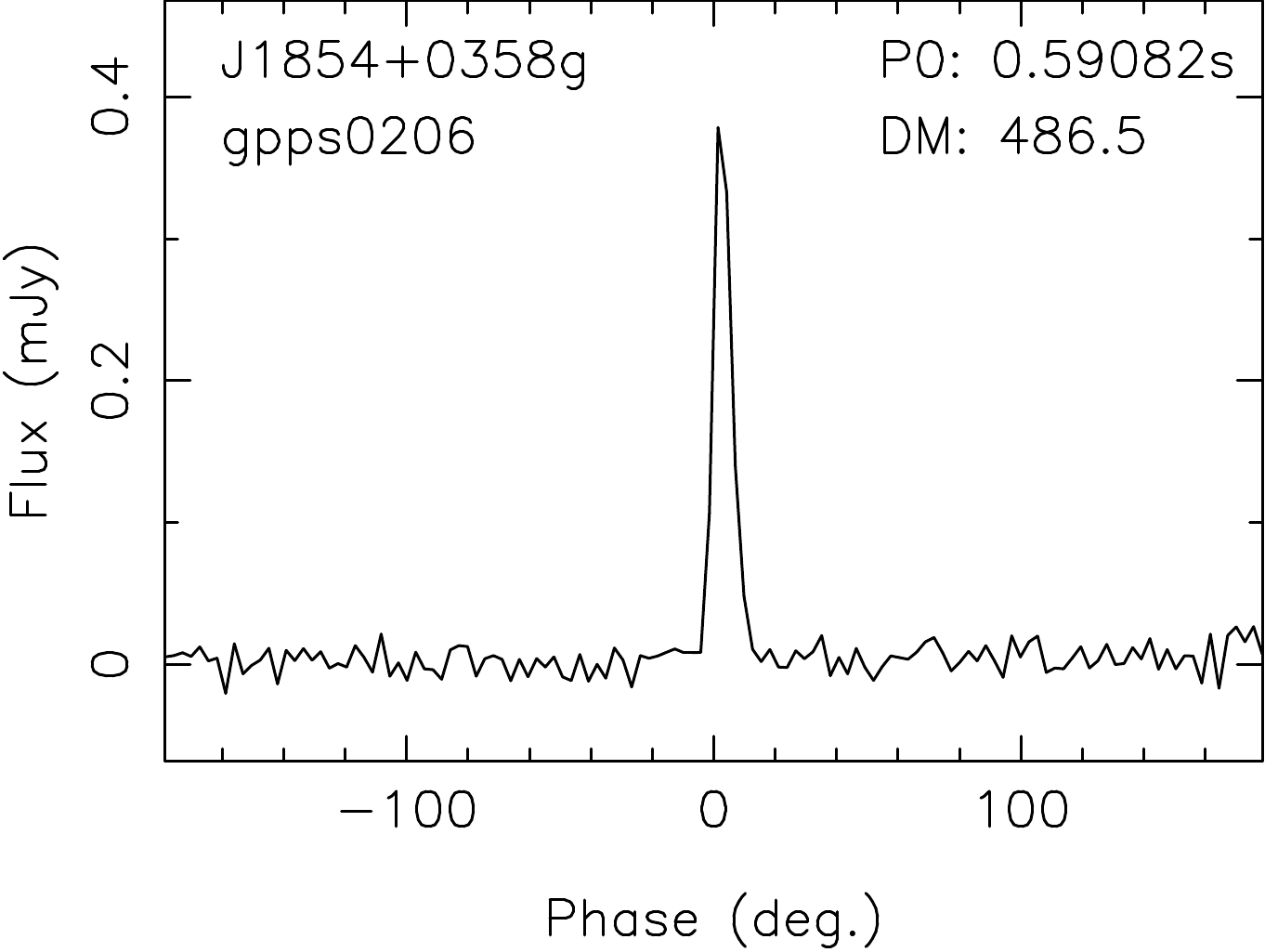}
\includegraphics[width=0.24\textwidth]{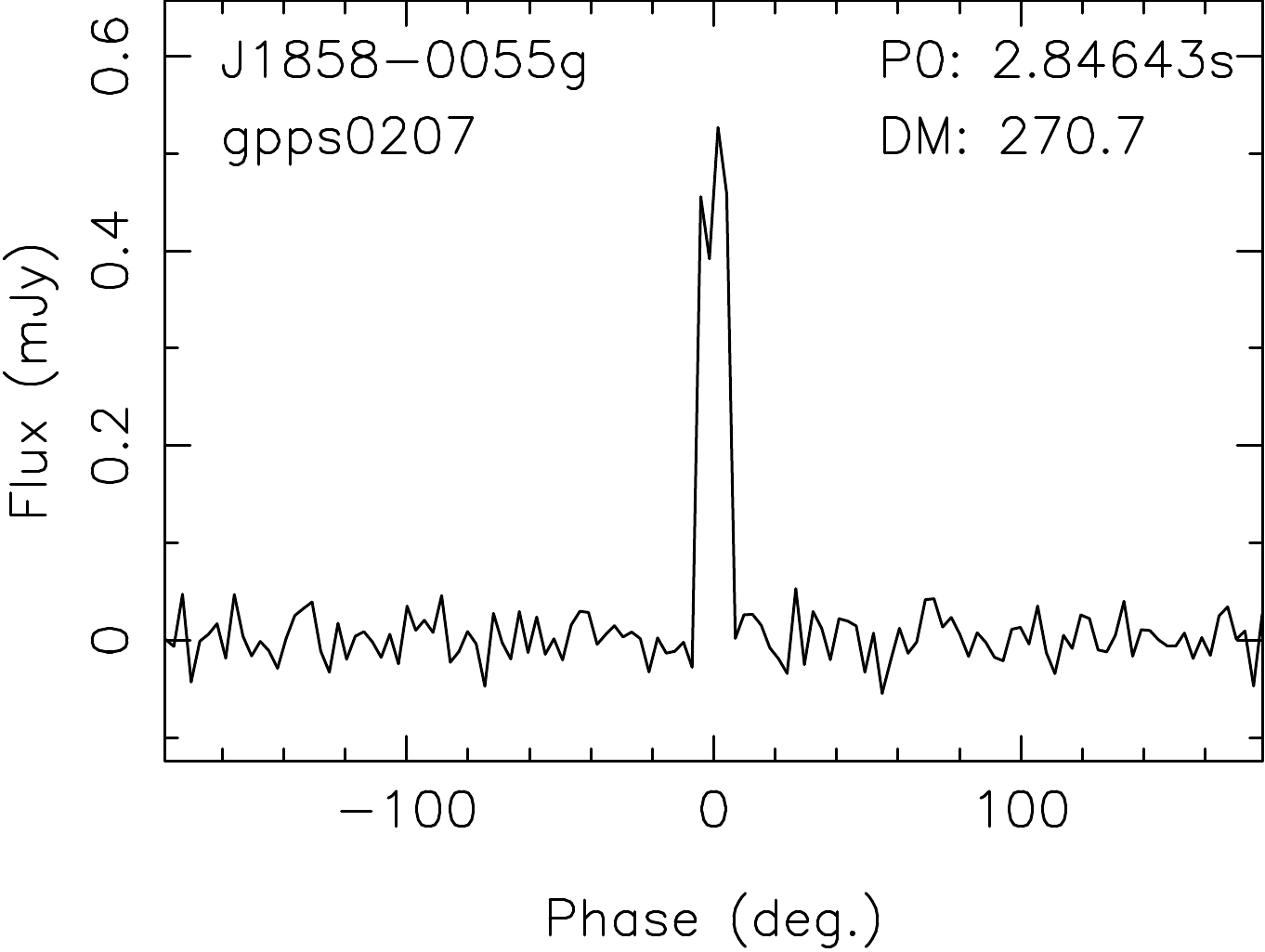}
\includegraphics[width=0.24\textwidth]{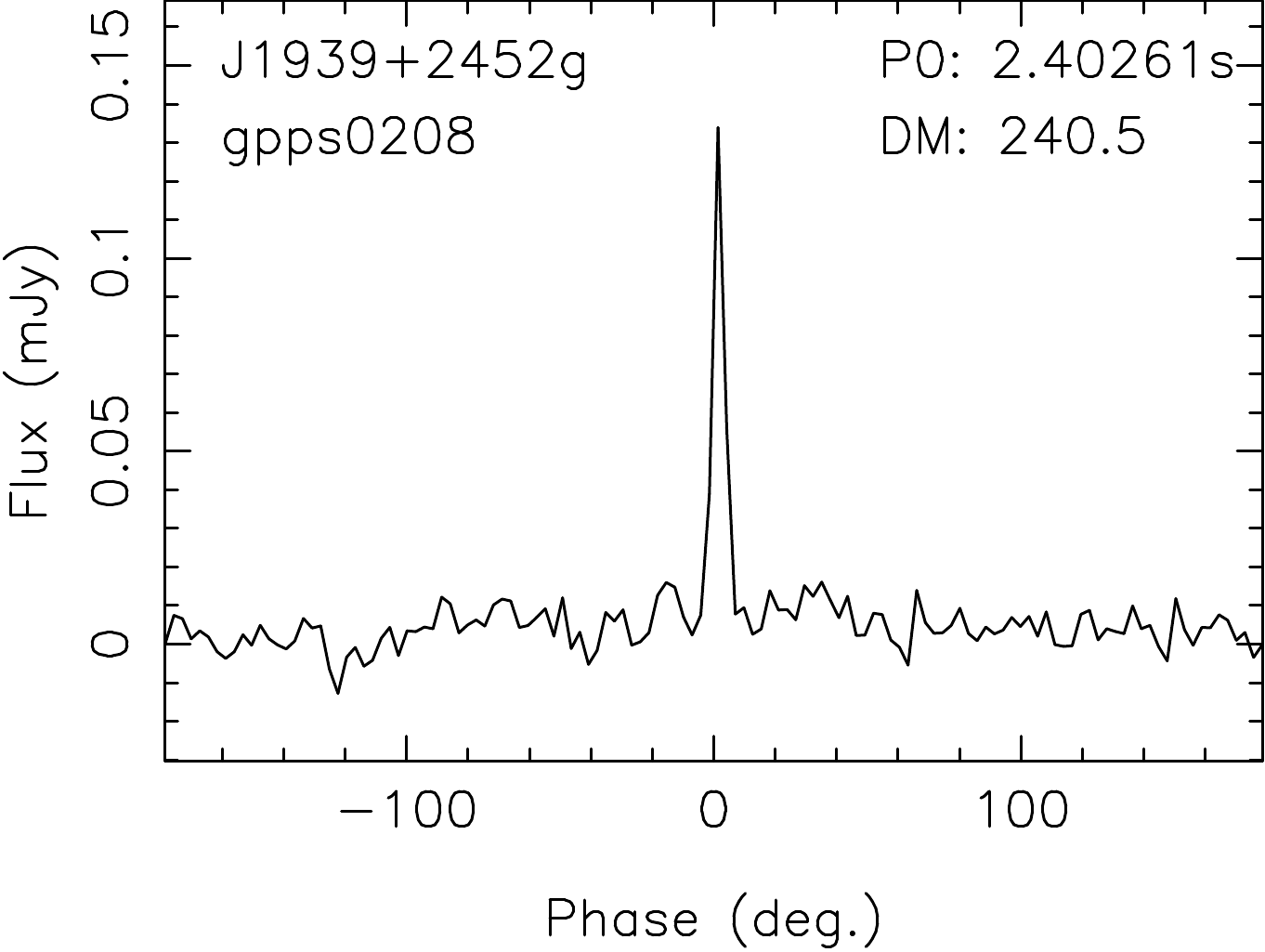}
\includegraphics[width=0.24\textwidth]{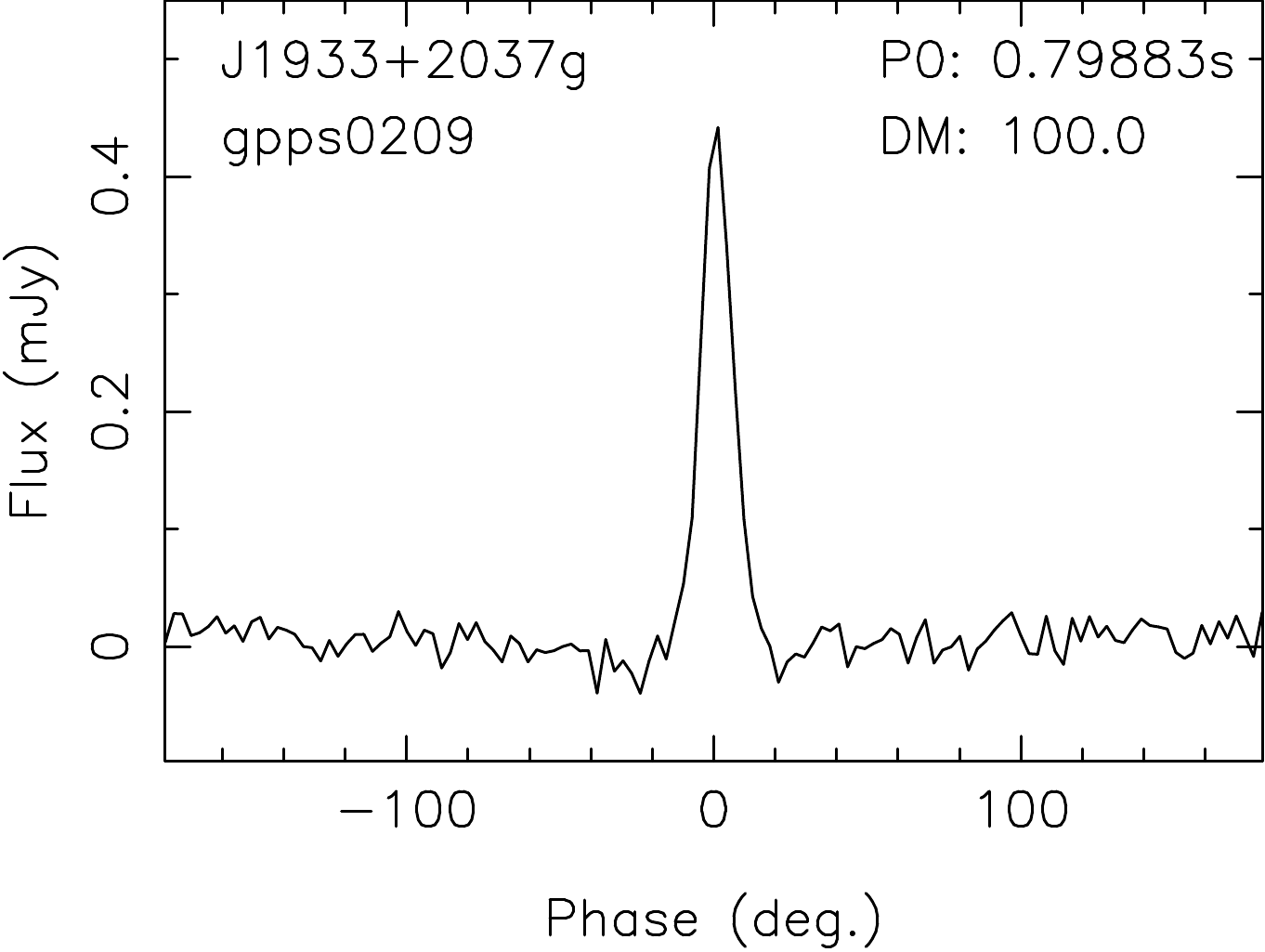}
\caption{Eight examples of integrated profiles of newly discovered pulsars. The profiles are plotted in the full rotation phase of 360$^{\circ}$ of a rotation period. The pulsar name, gpps number, period, and DM are noted in each panel. Plots for all pulsars are given in Fig.~A\ref{gppsPSRprof} in the Appendix.}
\label{prof20in_text}
\end{figure*}

The Five-hundred-meter Aperture Spherical radio Telescope \citep[FAST,][]{nan06}, mounted with the $L$-band 19-beam receiver \citep{jth+20}, is the most sensitive single-dish radio telescope in the world to search  {for} pulsars \citep{hww+21} and to detect hydrogen lines \citep{hhh+22} or recombination lines 
\citep{hhh+22a}. Since 2020 we have been carrying out the FAST Galactic Plane Pulsar Snapshot (GPPS) survey\footnote{http://zmtt.bao.ac.cn/GPPS/}, and 201 pulsars and 1 RRAT have been published in the first paper \citep{hww+21}. Subsequently, we reported 76 RRATs discovered during the survey \citep{zhx+23}. We have also published the timing solutions for 30 pulsars by \citet{shw+23}. There are many interesting discoveries. For example, PSR J1953+1844 (gpps0190) is a binary pulsar with an orbital period of 53 minutes \citep{plj+23} that is probably a descendant of an ultracompact X-Ray binary \citep{yhj+23}. PSR J1928+1815 (gpps0121) is an eclipsing millisecond pulsar in a highly compact binary with an orbital period of 3.6 hr, which is the product of the common envelope phase \citep{yhz+24}. By using the FAST GPPS data, we have also discovered five FRBs \citep{zhj+23}.

Up to now, we have discovered 751 pulsars, among which we have 107 RRATs, 177 millisecond pulsars, and 160 binary pulsars. In this paper, we publish the second massive discovery of 473 new pulsars after the first and the second GPPS discovery papers \citep{hww+21, zhx+23}. In the companion papers in this volume, we discuss the six millisecond pulsars in compact orbits with massive white dwarf companions \citep{yhw+24}, and present primary results for 116 binary systems \citep{why+24}.

\begin{figure}[ht]
  \centering
  \includegraphics[width=0.98\columnwidth]{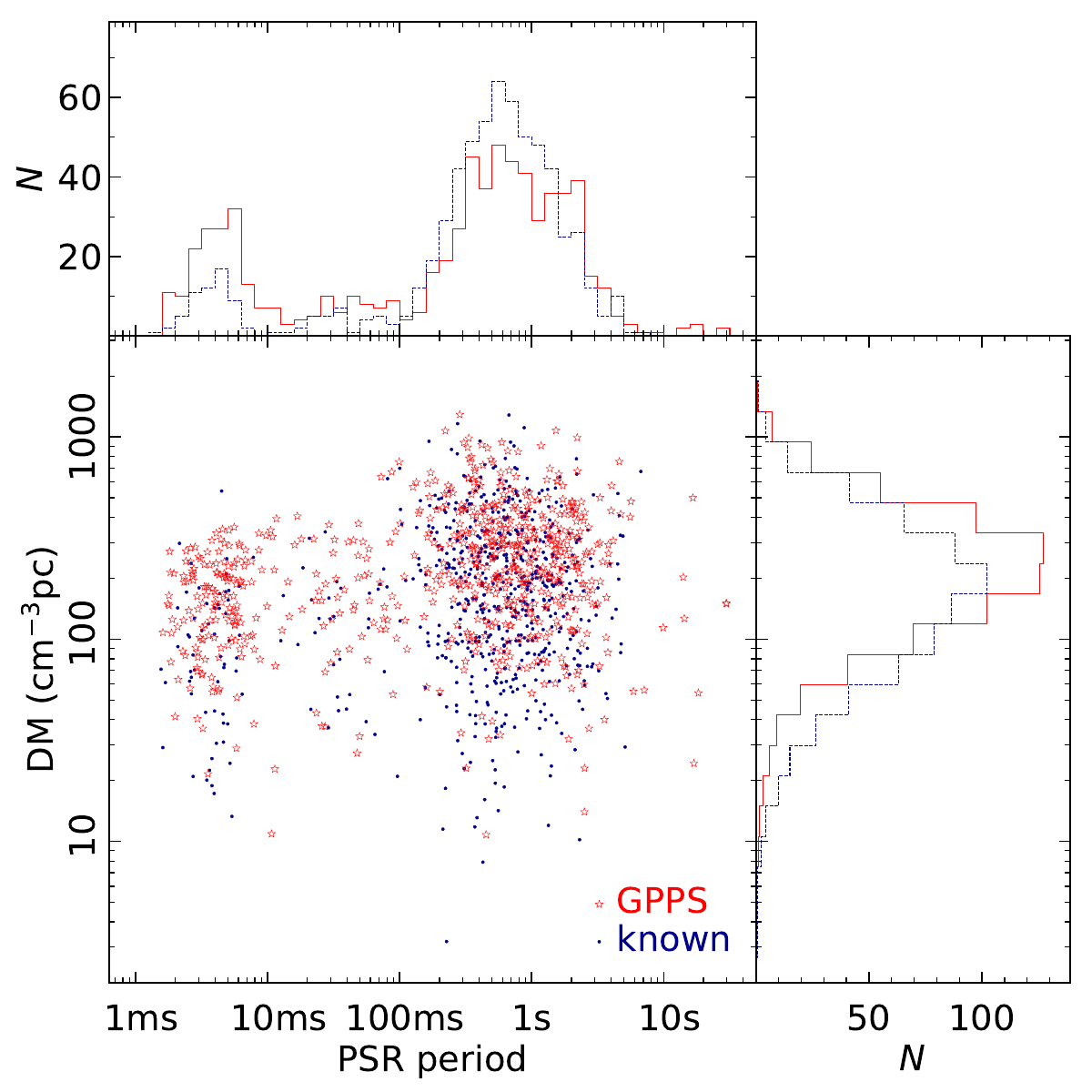}
  \caption{The distribution of DMs and periods of the newly discovered pulsars in the GPPS survey compared to those of previously known pulsars.}
  \label{pdm}
\end{figure}

\begin{figure}
  \centering
  \includegraphics[width=0.98\columnwidth]{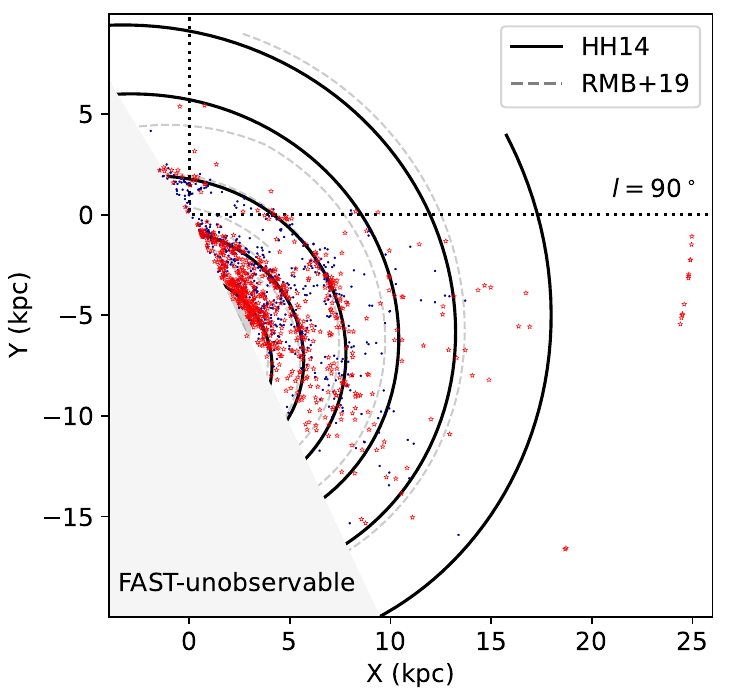}
  \includegraphics[width=0.98\columnwidth]{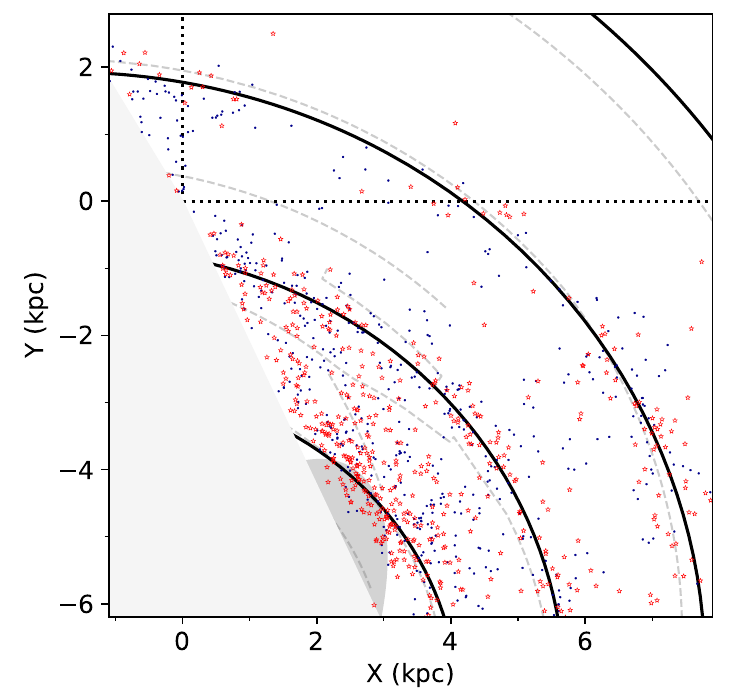}
  \caption{The pulsar distribution in the Galactic disk for the FAST-visible area. The newly discovered pulsars by the FAST-GPPS survey are marked as stars, and the previously known pulsars by crosses. The solid and dashed curves indicate the spiral arms \citep[][HH14, RMB+19]{hh14, rmb+19}  from the inner to outer: the Scutum Arm, the Sagittarius Arm (the first interior to the Sun), the Perseus arm (the first outside the Sun), the Perseus+1 arm, the Perseus+2 arms, and the Perseus+3 arm, respectively. The zoomed region is shown in the lower panel to see the high density of pulsars. We discovered many pulsars around the Scutum Arm, the Sagittarius Arm, and also some in the Perseus arm. }
  \label{dis-disk}
\end{figure}

\section{The current status of the FAST GPPS Survey and follow-up observations}

The FAST GPPS Survey plans to survey the visible sky area within $\pm10\degr$ from the Galactic plane, see details about the survey strategy in \citet{hww+21}. The main purpose is to find new pulsars, The main purpose is to find new pulsars, especially pulsars with short spin periods or short orbital periods. To do the survey, we developed the ``snapshot mode" for FAST observations. Each snapshot observation consists of four adjacent pointings with quick switches of 19 beams to cover a hexagonal sky area of 0.1575 square degrees. The sky area covered by these 76 beams is defined to be a {\it cover} \citep[see Fig. 4 in]{hww+21}. 
There are 18,413 planned {\it covers} for the survey. for the survey. After FAST observations in the last 5 yr, we have made good progress on the FAST GPPS Survey, and have observed 4072 {\it covers}, as shown in Figure~\ref{gppsSky}. 

The FAST GPPS observations have been carried out with a full gain of the FAST within the zenith angle of originally $26.4^{\circ}$ but now $28.5^{\circ}$ because we found the snapshot mode can work down to $28.5^{\circ}$ and the telescope gain does not have obvious degradation. In 2023, we tested and developed the "snapshotZ mode" (see Fig.~\ref{snapshotZ}), which can finish the {\it 4 adjacent covers} by using much less slewing time between the covers. 

Because of the interstellar medium, radio pulses are dispersed differently at different frequencies. To find pulsars, one has to de- disperse the signal and then add the pulse power from all frequency channels. General pulsar searching involves two steps. One is to consider the dispersion of pulses at different frequencies and de-disperse the power. In this case, the dispersion inside frequency channels is ignored. The other key step is to find the period of the pulses by using the fast Fourier transform (FFT). Currently, the widely used package for pulsar searching is PRESTO developed by \citet{Ransom01}. We have developed the pipeline \citep{hww+21} based on PRESTO for the standard pulsar search pipeline and also the acceleration search for binaries. 

We have already acquired more than 12PB survey data, which have been processed in at least three runs by using three clusters of computational nodes, one run for normal pulsar search, one for single pulse search \citep{zhx+23} and the other for acceleration search for binaries. The data processing frame for the FAST GPPS survey was presented in \citet{hww+21} and has not been changed. Nevertheless, almost all flow charts have been updated and optimized for many versions for the goal of fast processing of the survey data. For example, we have also used different strategies for mitigation of radio frequency interference, the normal {\sc rfifind} for normal pulsar searching, and the new outlier detection algorithm called Inter-Quartile Range Mitigation (IQRM) developed by \citet{mrs22} has been used for the radio frequency interference cleaning during acceleration search of close binaries.

\begin{table*}[th]
\centering
\footnotesize
\renewcommand\arraystretch{0.74}
\caption{Key parameters of 52 Known Pulsars are updated by the FAST observations}
\begin{tabular}{lllrllclhrcc}
\hline\noalign{\smallskip}
PSR Name & Ref. & Period & DM           & RA(2000)    &   DEC(2000)  & Updated & \hfil Period & Epoach for P & DM ($\sigma$) & RA(2000)   &   DEC(2000) \\
         &      & \hfil  (s)      &   (pc cm$^{-3}$)     & hh:mm:ss  & $\pm$dd:mm:ss & items   &  \hfil (s)   & \hfil (MJD)  & (pc cm$^{-3}$) & hh:mm:ss & $\pm$dd:mm \\ 
         (1) & (2) & \hfil (3) & (4) & \hfil (5) & \hfil (6) & (7) & \hfil (8) & (9) & (10)\hfil & (11) & (12) \\
\hline
J0555$+$3948     & [1]   & 1.14691 & 37    & 05:55       & $+$39:48      & Posi      & 1.14690  & 59358.273535 & 36.2(9)    & 05:55:07.61 & $+$39:51:38 \\
J0608$+$00       & [2]   & 1.0762  & 48    & 06:08:49    & $+$00:39      & Posi      & 1.07616  & 59636.544596 & 48.7(9)    & 06:08:47.49 & $+$00:44:18 \\
J1808$+$00       & [3]   & 0.42512 & 141   & 18:08:19    & $+$00:34      & Posi      & 0.42512  & 59383.752566 & 149.1(3)   & 18:08:26.55 & $+$00:38:25 \\
J1818$+$03       & [4]   & 0.799   & 99    & 18:18       & $+$03         & Posi      & 0.79919  & 60348.063356 & 98.7(6)    & 18:19:15.23 & $+$03:22:48 \\
J1822$+$02       & [4]   & 1.508   & 103.8 & 18:22       & $+$02         & Posi      & 1.50815  & 60601.376531 & 105.8(12)  & 18:22:43.14 & $+$02:29:14 \\
J1827$+$0022     & [5]   & 0.3753  & 96    & 18:27       & $+$00:22      & Posi      & 0.37517  & 59374.767128 & 95.8(6)    & 18:27:05.35  & $+$00:25:02 \\
J1830$-$0131     & [6]   & 0.15251 & 95.7  & 18:30:19.61 & $-$01:31:48.1 & P         & 0.45754  & 59454.544345 & 97.3(4)    & 18:30:17.41 & $-$01:31:56 \\
J1833$-$0209     & [7]   & 0.29193 & 325.4 & 18:33:05.41 & $-$02:09:16.4 & P         & 0.58386  & 59485.470234 & 325.9(5)   & 18:33:04.71 & $-$02:09:06 \\
J1835$+$00       & [8]   & 0.79008 & 136.1 & 18:35       & $+$00         & Posi      & 0.79007  & 59743.761277 & 134.6(6)   & 18:35:48.98 & $+$00:31:29 \\  
J1837$+$03       & [8]   & 0.01070 & 115.5 & 18:37       & $+$03         & Posi      & 0.01070  & 59573.183162 & 115.665(9) & 18:37:29.47 & $+$03:55:32 \\
J1838$-$01       & [9]   & 0.18330 & 320.4 & 18:38:30    & $-$01:01      & Posi      & 0.18330  & 59485.473821 & 311.9(1)   & 18:38:42.71 & $-$01:00:50 \\
J1839$+$0543     & [10]  & 0.05793 & 113.8 & 18:39:00    & $+$05:43      & Posi      & 0.05793  & 59186.239591 & 113.9(1)   & 18:39:25.73 & $+$05:42:36 \\
J1840$+$03       & [8]   & 0.00583 & 80.9  & 18:40       & $+$03         & Posi      & 0.00583  & 60358.071487 & 80.79(4)   & 18:40:11.41 & $+$03:22:58 \\
J1844$-$0256     & [11]  & 0.27296 & 822   & 18:44:53    & $-$02:56:42   & Posi      & 0.27299  & 59903.318761 & 824.8(9)   & 18:44:29.22 & $-$02:54:17 \\
J1846$-$0049     & [5]   & 0.02124 & 45.4  & 18:46       & $-$00:49      & Posi      & 0.02124  & 59511.374079 & 45.22(5)   & 18:46:14.01 & $-$00:50:08 \\  
J1848$+$12       & [12]  & 0.75473 & 139   & 18:48:30    & $+$12:50      & Posi      & 0.75471  & 59409.662255 & 127.9(6)   & 18:48:05.80  & $+$12:54:28 \\
J1854$-$0154     & [5]   & 0.68039 & 590   & 18:54       & $-$01:54      & Posi,DM   & 0.68059  & 60023.998247 & 569.4(8)   & 18:54:32.30 & $-$01:50:38 \\
J1858$-$0200     & [5]   & 0.48732 & 191   & 18:58       & $-$02:00      & Posi,P,DM & 1.46204  & 59509.405226 & 182.6(12)  & 18:57:46.04 & $-$01:54:52 \\
J1917$+$2441     & [5]   & 0.00440 & 82    & 19:17       & $+$24:41      & Posi      & 0.00440  & 59700.851367 & 82.07(1)   & 19:17:32.83 & $+$24:41:18 \\  
J1919$+$04       & [8]   & 0.00396 & 142.8 & 19:19       & $+$04         & Posi      & 0.00396  & 59550.314648 & 142.780(5) & 19:18:59.07 & $+$04:00:26 \\ 
J1925$+$19       & [13]  & 1.91635 & 328   & 19:25:26    & $+$19:04      & Posi      & 1.91638  & 59854.557819 & 333.2(15)  & 19:25:14.00 & $+$19:07:36 \\  
J1929$+$00       & [3,14]& 1.16690 & 42.8  & 19:29:28    & $+$00:26      & Posi      & 1.16691  & 59404.703745 & 43.8(9)    & 19:29:26.12 & $+$00:31:09 \\
J1931$+$30       & [3,15]& 0.58213 & 53.8  & 19:31:28    & $+$30:35      & Posi      & 0.58213  & 59395.829309 & 53.5(5)    & 19:31:10.41 & $+$30:32:04 \\
J1934$+$0906     & [5]   & 0.00466 & 72.4  & 19:34       & $+$09:06      & Posi      & 0.00466  & 59911.397634 & 72.37(4)   & 19:34:13.57 & $+$09:06:28 \\
J1936$+$13       & [8]   & 0.00434 & 168.0 & 19:36       & $+$13         & Posi      & 0.00434  & 59547.372147 & 167.97(1)  & 19:36:05.53 & $+$12:59:18 \\
J1938$+$14a      & [4]   & 1.661   & 114   & 19:38       & $+$14         & Posi      & 1.66145  & 59440.648283 & 113.4(13)  & 19:37:39.56 & $+$14:16:48 \\
J1939$+$10       & [3,15]& 2.31144 & 74.0  & 19:39:11    & $+$10:45      & Posi      & 2.30873  & 59990.033297 & 73.2(18)   & 19:39:21.16 & $+$10:49:28 \\
J1940$+$14       & [16]  & 1.279   & 70    & 19:40:15    & $+$14:37      & Posi      & 1.27436  & 60513.735978 & 69.5(10)   & 19:40:06.14 & $+$14:27:37 \\
J1942$+$3941     & [17]  & 1.35329 & 104.5 & 19:42:22.05 & $+$39:41:41.4 & DM        & 1.35329  & 59427.660848 & 95.1(11)   & 19:42:22.27 & $+$39:41:48 \\
J1943$+$2851     & [5]   & 0.736   & 228.5 & 19:43       & $+$28:51      & Posi      & 0.73782  & 60326.134222 & 231.7(23)  & 19:43:23.09 & $+$28:52:28 \\
J1944$+$16       & [8]   & 0.00243 & 170.8 & 19:44       & $+$16         & Posi      & 0.00243  & 59700.920529 & 170.87(3)  & 19:44:15.32 & $+$16:54:44 \\
J1945$+$07       & [2]   & 1.0739  & 62    & 19:45:55    & $+$07:17      & Posi      & 1.07403  & 59413.740988 & 62.5(9)    & 19:45:44.74 & $+$07:20:28 \\
J1945$+$17       & [23]  & 0.6042  & 167.7 & 19:45       & $+$17         & Posi      & 0.60412  & 60513.757082 & 172.4(5)   & 19:45:17.89 & $+$17:40:53 \\
J1947$+$10       & [3,15]& 1.11094 & 128.76& 19:47:36    & $+$10:44      & Posi      & 1.11094  & 59432.673850 & 129.0(9)   & 19:47:20.96 & $+$10:41:60 \\
J1956$+$35       & [18]  & ...     & 153.7 & 19:56:36    & $+$35:36:49   & Posi,P    & 0.87552  & 60488.820901 & 153.1(7)   & 19:56:27.74 & $+$35:44:38 \\  
J1958$+$2213     & [19]  & 1.0502  & 85    & 19:58:34    & $+$22:13      & Posi      & 1.05038  & 59560.385043 & 88.6(10)   & 19:58:45.52 & $+$22:14:27 \\
J1958$+$2332     & [5]   & ...     & 204.5 & 19:58       & $+$23:32      & P         & 8.07375  & 60515.645683 & 201.5(22)  & 19:58:01.19 & $+$23:31:33 \\
J2005$+$14       & [18]  & ...     & 51.2  & 20:05:15	 & $+$14:37:00   & Posi,P    & 2.33214  & 60519.672616 & 50.3(19)   & 20:04:53.97 & $+$14:47:54 \\
J2025$+$2133     & [20]  & 0.6235  & 70.8  & 20:25       & $+$21:33      & Posi      & 0.62348  & 59428.747968 & 71.6(5)    & 20:25:56.55 & $+$21:37:00 \\
J2030$+$31       & [18]  & ...     & 131.4 & 20:30:29 	 & $+$31:29:02   & Posi,P    & 1.01470  & 60516.756135 & 131.6(8)   & 20:30:12.15 & $+$31:36:23 \\
J2041$+$46       & [4]   & 1.160   & 304   & 20:41       & $+$46         & Posi      & 1.15982  & 59701.010270 & 307.7(9)   & 20:41:30.99 & $+$45:51:56 \\
J2102$+$38       & [21]  & 1.19    & 85    & 21:02       & $+$38         & Posi      & 1.18990  & 59370.915960 & 86.3(10)   & 21:02:02.10 & $+$37:58:19 \\
J2125$+$52       & [18]  & ...     & 262.8 & 21:25:32    & $+$52:19:23   & P,Posi    & 5.80150  & 60521.751530 & 261.9(46)  & 21:25:50.78 & $+$52:18:20 \\
J2153$+$44       & [18]  & ...     & 142.7 & 21:53:26    & $+$44:56:43   & P,Posi    & 2.89297  & 60502.875475 & 142.7(23)  & 21:53:18.84 & $+$45:07:08 \\
J2300$+$50       & [18]  & ...     & 58.3  & 23:00:24    & $+$50:21:13   & P         & 3.64840  & 60488.924604 & 59.5(29)   & 23:00:22.28 & $+$50:24:08 \\
J2300$+$52       & [16]  & 0.4265  & 82    & 23:00:30    & $+$52:20      & Posi      & 0.42610  & 60489.909318 & 83.7(13)   & 23:00:05.18 & $+$52:24:34 \\[1mm]
J0928$+$06$^*$   & [2]   & 2.0604  & 50    & 09:28:44    & $+$06:14      & Posi      & 2.06025  & 59485.118013 & 52.0(17)   & 09:28:28.14 & $+$06:14:15 \\
J1656$+$00$^*$   & [22]  & 1.49785 & 46.9  & 16:56:41    & $+$00:26      & Posi      & 1.49796  & 59515.284575 & 44.9(12)   & 16:56:30.17 & $+$00:17:47 \\ 
J1726$-$00$^*$   & [2]   & 1.3086  & 57    & 17:26:23    & $-$00:15      & Posi      & 1.30762  & 59541.278735 & 58.6(20)   & 17:26:28.88 & $-$00:22:37 \\
J1749$+$16$^*$   & [22]  & 2.31165 & 59.6  & 17:49:29    & $+$16:24      & Posi      & 2.31105  & 59517.299127 & 57.7(19)   & 17:49:04.52 & $+$16:29:05 \\
J1802$+$03$^*$   & [2]   & 0.6643  & 77    & 18:02:44    & $+$03:38      & Posi      & 0.66431  & 59517.378403 & 77.7(15)   & 18:02:38.12 & $+$03:45:38 \\
J2057$+$2133$^*$ & [20]  & 1.1667  & 72.2  & 20:57       & $+$21:33      & Posi      & 1.16662  & 59541.416868 & 72.3(28)   & 20:57:41.71 & $+$21:28:33 \\
\hline
\end{tabular}
\tablecomments{Column (1): Pulsar name. The six high latitude pulsars with superscript `*' get parameters measured in a project led by J. Xu; 
References in Column (2): 
[1] \citet{bck+13}; 
[2] \citet{dsm+13}; 
[3] \citet{cns+96}; 
[4] http://astro.phys.wvu.edu/GBNCC/; 
[5] http://groups.bao.ac.cn/ism/CRAFTS/CRAFTS/; 
[6] \citet{lfl+06}; 
[7] \citet{ncb+15}; 
[8] https://palfa.nanograv.org/; 
[9]: \citet{kek+13}; 
[10] \citet{zbd+24}; 
[11] \citet{bri+06}; 
[12] \citet{ebv+01}; 
[13] \citet{lcm+13};
[14] \citet{scb+19};
[15] \citet{dmo+24};
[16] \citet{ttk+24};
[17] \citet{ccl+21};
[18] https://www.chime-frb.ca/galactic;
[19] \citet{tkt+20};
[20] \citet{wzl+21};
[21] \citet{hrk+08};
[22] \citet{dsm+16};
[23] \citet{psf+22};
Columns (3)-(6): parameters in reference; 
Column (7): items updated by FAST; Columns (7) - (12): parameters obtained from the GPPS survey or FAST observations. Newly measured DM has an uncertainty in brackets.
}
\label{updatePara}
\end{table*}


\begin{table*}[ht!]
\centering
\caption{Previosuly known pulsars not detected in the FAST-GPPS survey}
\label{missingPSR}
\setlength{\tabcolsep}{1.5pt} 
\footnotesize
\renewcommand\arraystretch{0.8}
\begin{tabular}{llllll}
\hline\noalign{\smallskip}
PSR Name    & Period  & DM         & RA(2000)     &   DEC(2000)    & Notes and references  \\
            & \hfil (s)  &  (pc~cm$^{-3}$)          & hh:mm:ss     & $\pm$dd:mm:ss  &         \\ 
        (1) & \hfil (2)  & (3)        & \hfil (4)    & \hfil (5)      & (6)     \\
\hline\noalign{\smallskip}             
\multicolumn{6}{c}{High-energy / Radio-quiet pulsars} \\
\hline\noalign{\smallskip}
J0633+0632        &  0.29740   & ...       &   06:33:44.2      &  +06:32:34.9   & \citet{aaa+09,rkp+11} \\
J0633+1746        &  0.23710   & ...       &   06:33:54.2      &  +17:46:12.9   & \citet{hh92,clm+98} \\
J1846+0919        &  0.22555   & ...       &   18:46:26.0      &  +09:19:46     & \citet{sdz+10} \\
J1906+0722        &  0.11152   & ...       &   19:06:31.2      &  +07:22:55.8   & \citet{cpw+15} \\
J1932+1916        &  0.20821   & ...       &   19:32:19.7      &  +19:16:39     & \citet{pga+13} \\
J1954+2836        &  0.09271   & ...       &   19:54:19.2      &  +28:36:06     & \citet{sdz+10} \\
J1958+2846        &  0.29039   & ...       &   19:58:40.1      &  +28:45:54     & \citet{aaa+09,rkp+11}  \\
J2017+3625        &  0.16675   & ...       &   20:17:55.8      &  +36:25:07.9   & \citet{cwp+17} \\
J2021+4026        &  0.26532   & ...       &   20:21:30.0      &  +40:26:45.1   & \citet{aaa+09,rkp+11} \\
J2028+3332        &  0.17671   & ...       &   20:28:19.9      &  +33:32:04.4   & \citet{pga+12} \\
J2030+4415        &  0.22707   & ...       &   20:30:51.4      &  +44:15:38.1   & \citet{pga+12} \\
J2111+4606        &  0.15783   & ...       &   21:11:24.1      &  +46:06:31.3   & \citet{pga+12} \\
J2139+4716        &  0.282849  & ...       &   21:39:56.0      &  +47:16:13     & \citet{pha+12} \\
J2034+3632        &  0.00365   & ...       &   20:34           &  +36:32        & https://einsteinathome.org/gammaraypulsar/FGRP1\_discoveries.html \\
%
J0501+4516        &  5.762097  & ...       &   05:01:06.8      &  +45:16:33.9   & \citet{gwk+08,gwk+10};SGR \\
J0635+0533        &  0.033856  & ...       &   06:35:18        &  +05:33:11     & \citet{cmn+00};X-ray pulsation \\
J1849-0001        &  0.038523  & ...       &   18:49:1.6       &  -00:01:17.4   & \citet{ghtm11};X-ray pulsar \\
J1852+0033        &  11.558713 & ...       &   18:52:46.6      &  +00:33:20.9   & \citet{zcl+14};transient magnetar \\
J1852+0040        &  0.104913  & ...       &   18:52:38.6      &  +00:40:19.8   & \citet{ghs+05};X-ray pulsar \\
J1907+0919        &  5.198346  & ...       &   19:07:14.3      &  +09:19:20.1   & = SGR 1900+14; \citet{ksh+99,fkb+99,met+06} \\
J1935+2154        &  3.24498   & ...       &   19:34:55.6      &  +21:53:48.2   &  = SGR 1935+2154  \\
AX\_J1845.0$-$0258&  6.9712    & ...       &   18:44:54.68     &  $-$02:56:53.1 & \citet{tkk+98,gv98,tkg+06} \\
SGR\_2013+34      &  ...       & ...       &   20:13:56.9      &  +34:19:48       &      \citet{sbb+11}\\
\hline\noalign{\smallskip}             
\multicolumn{6}{c}{Weak/Scatted/Binary pulsars} \\
\hline\noalign{\smallskip}
J0454+4529        &  1.389137  & 20.8      &   04:54:59.3      &  +45:29:46.7   & \citet{mhl+18,tbc+20}\\
J1848-0129A       &  0.01978   & 491.1     &   18:48           &  -01:29        & https://www.trapum.org/discoveries/ \\
J1848-0129C       &  0.00644   & 489.2     &   18:48           &  -01:29        & https://www.trapum.org/discoveries/ \\
J1848-0129D       &  0.0171    & 458.0     &   18:48           &  -01:29        & https://www.trapum.org/discoveries/ \\
J1848-0129E       &  0.00454   & 479.9     &   18:48           &  -01:29        & https://www.trapum.org/discoveries/ \\
J1848-0129F       &  0.00417   & 520.1     &   18:48           &  -01:29        & https://www.trapum.org/discoveries/ \\
J1907+0602        &  0.106633  & 82.1      &   19:07:54.7      &  +06:02:16.9   & \citet{aaa+09,aaa+10}\\
J1953+1846C       &  0.02893   & 116.2     &   19:53:46.5      &  +18:46:45     & \citet{pqm+21} \\
J1953+1846D       &  0.10067   & 119.0     &   19:53:46.5      &  +18:46:45     & \citet{pqm+21} \\
J2016+3711        &  0.050806  & 429.5     &   20:16:9.14      &  +37:11:10.4   & \citet{lzc+24};15.5uJy \\
J2022+3842        &  0.048579  & 429.1     &   20:22:21.6      &  +38:42:14.8   & Scatter.       \\
\hline\noalign{\smallskip}             
\multicolumn{6}{c}{Radio transients /Nulling pulsars} \\
\hline\noalign{\smallskip}
J1855+0626        &  0.528832  & 253.8     &   18:55:25        &  +06:26:53     & \citet{psf+22};98.4\% nulling. \\
J1905+09          &  3.48784   & 288.0     &   19:05:33.3      &  +08:57:30.2   & https://palfa.nanograv.org/ (posi by Ryan) \\ 
J1910+0517        &  0.308048  & 300       &   19:10:37.9      &  +05:17:56.1   & \citet{lsf+17};intermittent \\
J1911+00          &  6.94      & 100       &   19:11:48        &  +00:37        & \citet{mll+06,kkl+11} \\
J1905+0414        &  0.15648   & 381.1     &   19:05           &  +04:14        &  \citet{pab+18}. P?  \\
J1928+15          &  0.403     & 242       &   19:28:20        &  +15:13        & \citet{dcm+09};intermittent \\
J1929+1357        &  0.866927  & 150.7     &   19:29:10.6      &  +13:57:35.9   & \citet{lsf+17};intermittent \\
J1933+2421        &  0.81369   & 106.0     &   19:33:37.8      &  +24:36:39.6   & \citet{stw+85,klo+06}\\
J2108+4516        &  0.577231  & 83.5      &   21:08:23.3      &  +45:16:24.9   & \citet{gac+21} \\
\hline\noalign{\smallskip}             
\multicolumn{6}{c}{Other reasons for no detection.} \\
\hline\noalign{\smallskip}
J0324+5239        &  0.33662   & 115.5     &   03:24:55.4      &  +52:39:31.3   & \citet{bck+13} \\
J1914+0805        &  0.455499  & 344.4     &   19:14:05.5      &  +08:05:12.7   & \citet{psf+22} \\
J0553+4111        &  0.559493  & 37.9      &   05:53:23.8      &  +41:11:38.9   & \citet{wyw+23} \\
J0611+04          &  1.67443   & 69.9      &   06:11:18        &  +04:06        & \citet{dsm+16} \\
J0534+2200        &  0.033392  & 56.8      &   05:34:31.9      &  +22:00:52.0   &     Crab pulsar / Saturated   \\
J1905+0154B       &  0.004968  & 192.0     &   19:05           &  +01:54        & \citet{hrs+07} 
\\
\hline
\end{tabular}
\end{table*}

For any candidates standing out from these searches, we make the verification observations with the central beam M01 targeting the best-estimated position (occasionally a bright pulsar shown in a few adjacent beams) for 15 minutes, and data from the other 18 beams are also recorded for pulsar searching. The package {\sc  PSRCHIVE} \citep{hvm04} has been extensively used for folding pulsar data and polarization analysis. 

\begin{figure*}
  \centering
  \includegraphics[height=0.45\textwidth]{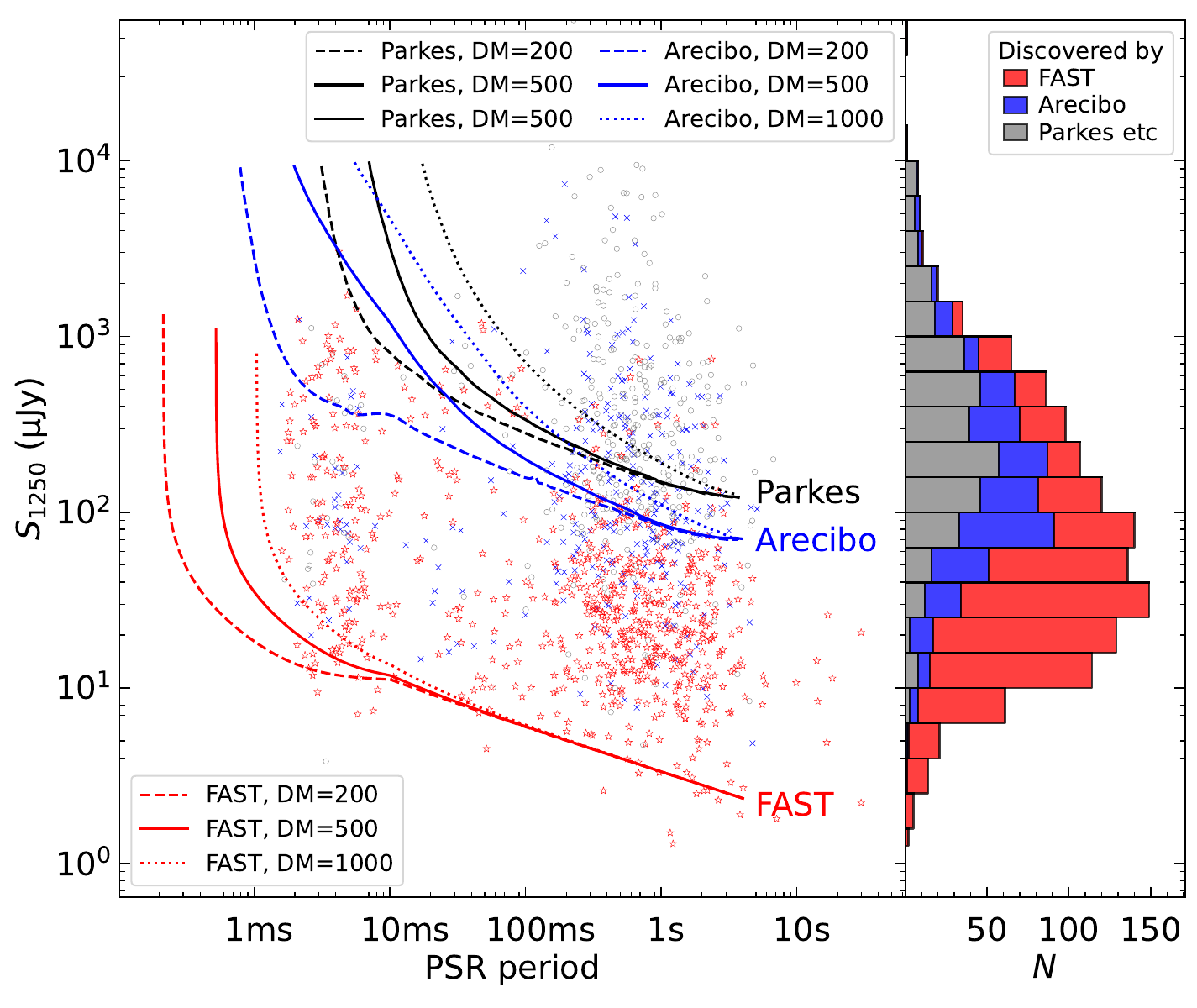}
  \includegraphics[height=0.45\textwidth]{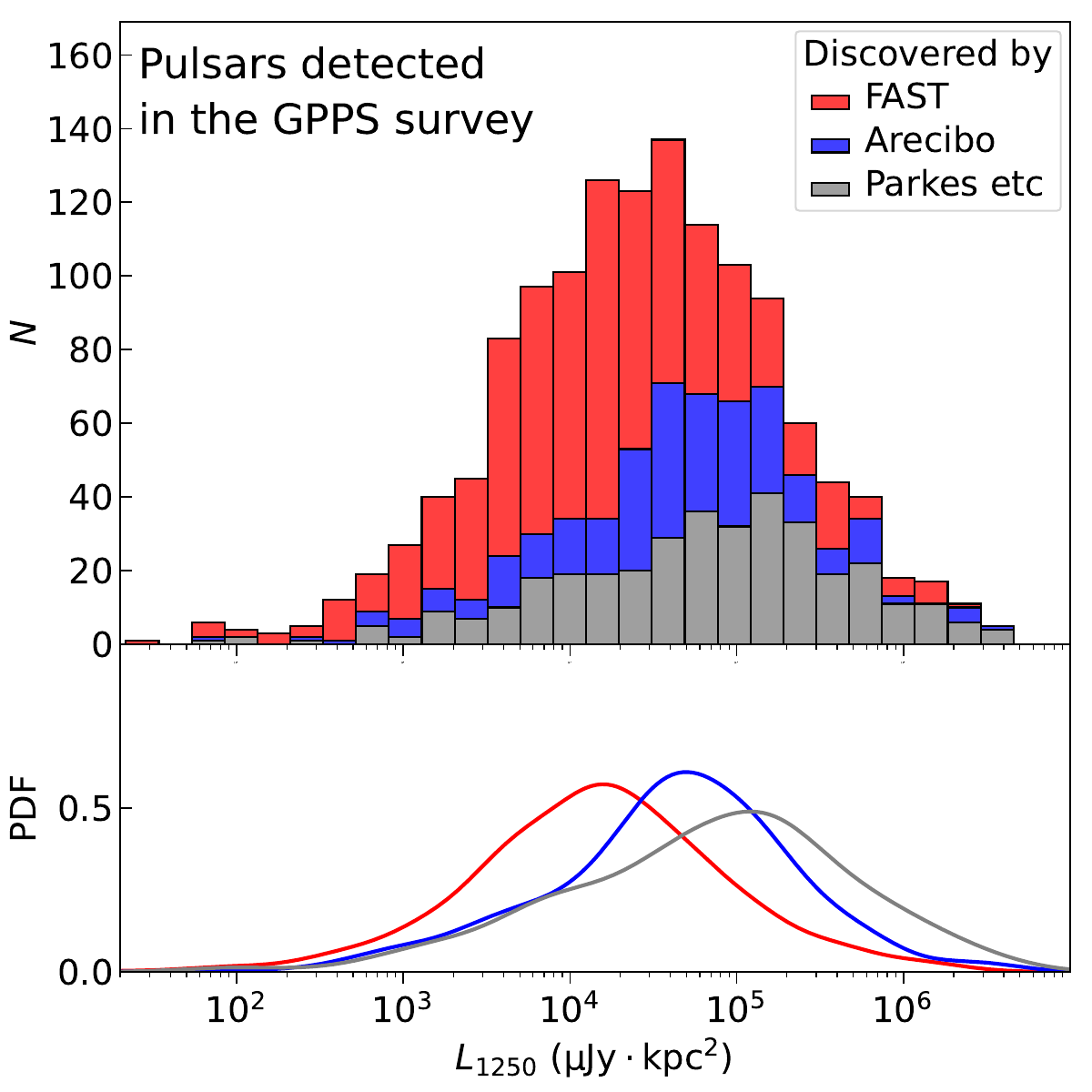}
  \caption{The distributions of flux densities and luminosities of pulsars observed by FAST. In the left panel, the sensitivity curves for different DMs of the FAST GPPS survey are given, compared to those for the Arecibo and Parkes surveys \citep{cfl+06,mld+96}, together with the histogram of the three surveys. The right panel shows that the FAST GPPS pulsars have much lower flux densities and dominate at the lower end of luminosity distributions. The probability distribution functions (PDFs) are smoothed curves for the normalized fraction distribution. }
  \label{flux_lum}
\end{figure*}

\section{New survey results}

Up to now, the FAST GPPS survey has uncovered 751 pulsars, as listed on the webpage\footnote{\label{gppsnewpsr}http://zmtt.bao.ac.cn/GPPS/GPPSnewPSR.html}. heir periods and DM values have been well-determined, except for some RRATs whose periods cannot be determined from only a few pulses detected by FAST \citep{zhx+23}. In addition to the publication of the massive pulsar sample in the first GPPS paper, we here publish the second massive sample in Table~A\ref{gppsPSRtab1} in the Appendix and several examples in Table~\ref{gppstab9exam} in the main text. The profiles are plotted for examples in Fig.~\ref{prof20in_text} and for all pulsars in Figure~A\ref{gppsPSRprof} in the Appendix. 

In the sky regions within $\pm10\degr$ from the Galactic plane where our FAST GPPS survey is observing, there are 800 previously known pulsars in 600 covers (there are more than one pulsar in some covers). To verify our FAST observation system the covers with known pulsars have always been observed during some survey sessions.

As shown in Figure~\ref{pdm}, the newly discovered pulsars found in the FAST GPPS survey generally have larger DMs than previously known pulsars. Though the number of newly discovered normal pulsars is still less than that of the previously known pulsars, we discovered twice the number of pulsars with a period of less than 100 ms compared to the previously known cases, which demonstrates the superior sensitivity of the FAST GPPS survey compared to previous experiments.

One can estimate the distances of all newly discovered pulsars by using the Galactic electron density distribution models, the NE2001 model \citep{ne2001} and the YMW16 model \cite{ymw17}, as listed in Table~\ref{gppsPSRtab1}. With these distances, we get the pulsar distribution projected onto the Galactic plane, as shown in Figure~\ref{dis-disk}. We discover many pulsars around the Scutum Arm (the most inner arm in the figure), the Sagittarius Arm (the first interior to the Sun), and also some in the Perseus arm (the first outside the Sun). For almost all of these pulsars, we recorded polarization data in our FAST verification observations. We will present their polarization profiles and Faraday rotation measures (RMs) in another paper by Jun Xu et al. (2025, in preparation).

With these estimated distances, one can derive the luminosity of pulsars based on the measured flux density at 1250 MHz. The pulsars discovered by FAST in the GPPS survey dominate in the flux-density range of 10 -- 100 $\mu$Jy and many in the level of several $\mu$Jy, see Fig.~\ref{flux_lum}. The results of the FAST GPPS survey contribute decisively to the lower end of pulsar luminosity distribution.

\subsection{Verification of known pulsars}

Up to now, we have observed 600 covers with previously known pulsars and detected 670 pulsars. Those known pulsars are generally very strong when using FAST and are often detected in several beams of the L-band 19 beam receivers. Some pulsars were published in some papers or made public on some web pages without an accurate position. Our FAST observations can give the position of a detected pulsar accurate to around 1' \citep{hww+21}, depending on how strong the pulsar is and in how many of the 19 beams the pulsar appears. For pulsars without a period or a DM (e.g., high energy pulsars) and with a very coarse position, FAST observations detect them and give the updated parameters, see 52 pulsars listed in Table~\ref{updatePara}. 

Note, however, that some previously known pulsars were not detected during the FAST GPPS survey for several reasons. First, some pulsars with very coarse positions have not been detected, mainly because the covers we observed probably have not yet been on the real position of pulsars. Second, some pulsars in globular clusters \citep{pqm+21} are weak and could not be detected by a 5 minute FAST observation during the GPPS survey. Third, some pulsars are $\gamma$-ray bright and radio- quiet and hence not detectable even by FAST. Fourth, it is also possible that some pulsars are just nulling or eclipsed during
the 5 minute observation for the GPPS survey. See the list of FAST undetected pulsars in Table~\ref{missingPSR}.

\subsection{Pulsars independently discovered by others}

After we detect any pulsar, we immediately compare its period and DM with the values of all known pulsars which we collected for all known pulsars. The database of known pulsars is updated every month or two. All new pulsars have to be verified by further FAST observations. After we get some tens of new pulsars, we often update our GPPS web page by adding these newly discovered pulsars.

For a few cases, we published our discoveries on the web page, but later we found someone published an object on their web page or paper earlier than ours, showing a similar DM without a period and/or a very coarse position, or a similar period without DM and/or position. Often hard to recognize, we withdrew and replaced the pulsar in question with a newer discovered one in a new version of the web page with a mark ``NewItem4misID”. All previous versions of the web page are kept there for tracking.

\begin{figure}
  \centering
  \includegraphics[width=0.98\columnwidth]{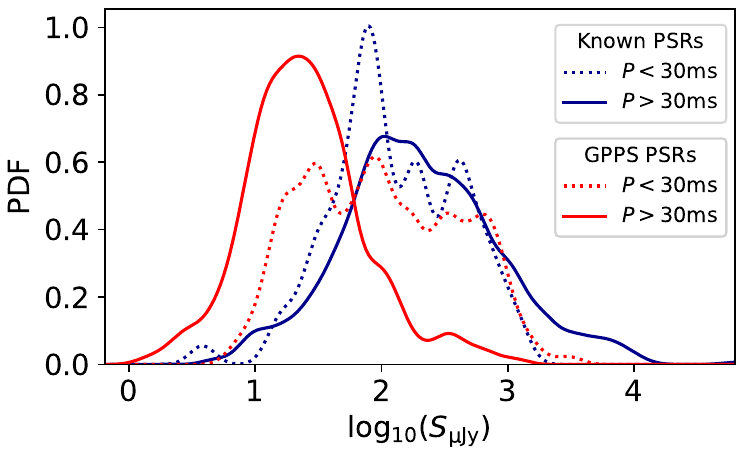}
  \caption{Comparison of the PDFs for millisecond pulsars and normal pulsars discovered in the FAST GPPS survey and others, represented by the smoothed curves for normalized fraction along the flux density.}
  \label{fig:pdf-s}
\end{figure}

\begin{table}[t]
\centering
\footnotesize
\renewcommand\arraystretch{0.85}
\caption{A list of 34 Millisecond Pulsars with TOA Accuracy Better than 3 $\mu$s by FAST 15 minute observations.}
\label{msptoa}
\begin{tabular}{lclr}
\hline
Name &  GPPS No. &  Period & $\sigma_{\rm TOA}$ \\
     &           &   (s)   &  ($\mu$s) \\
\hline
J2023+2853 & gpps0201 & 0.01132890 & 0.18 \\
J1918+0621 & gpps0494 & 0.002104 & 0.31 \\
J1955+3114g & gpps0692 & 0.003357 & 0.56 \\
J1821+0007g & gpps0613 & 0.004222 & 0.58 \\
J1859-0224g & gpps0466 & 0.006170 & 0.82 \\[1mm]
J1857+0642 & gpps0236 & 0.00353 & 0.84 \\
J1930+1708g & gpps0274 & 0.00228 & 0.84 \\
J1814+0045g & gpps0549 & 0.002309 & 0.95 \\
J1939+1848g & gpps0500 & 0.003357 & 1.0 \\
J1924+2027g & gpps0138 & 0.00195 & 1.1 \\[1mm]
J1845+0201g & gpps0547 & 0.004309 & 1.2 \\
J0622+0338g & gpps0388 & 0.008771 & 1.3 \\
J1846+0507g & gpps0614 & 0.003073 & 1.4 \\
J0520+3722g & gpps0538 & 0.007915 & 1.5 \\
J1913+0152g & gpps0702 & 0.003227 & 1.5 \\[1mm]
J2027+2837g & gpps0700 & 0.004794 & 1.5 \\
J1908+0949g & gpps0128 & 0.00905 & 1.8 \\
J1835+0158g & gpps0636 & 0.003320 & 2.0 \\
J1906-0200g & gpps0584 & 0.002530 & 2.0 \\
J1943+2210 & gpps0227 & 0.01287 & 2.0 \\[1mm]
J1904+0553A & gpps0039 & 0.0049080 & 2.3 \\
J1912+1416 & gpps0169 & 0.00317 & 2.3 \\
J1915+0601g & gpps0683 & 0.004011 & 2.5 \\
J1832+0113g & gpps0665 & 0.006345 & 2.6 \\
J1929+1259g & gpps0345 & 0.00285 & 2.6 \\[1mm]
J1957+2754g & gpps0437 & 0.003313 & 2.6 \\
J0408+4955g & gpps0638 & 0.011442 & 2.7 \\
J1930+1403 & gpps0013 & 0.00321 & 2.7 \\
J1908+1036 & gpps0114 & 0.01069 & 2.8 \\
J1920+0129g & gpps0230 & 0.00358 & 2.8 \\[1mm]
J2007+3343g & gpps0604 & 0.002681 & 2.8 \\
J1903+0839 & gpps0100 & 0.0046212 & 2.9 \\
J1931+2333g & gpps0675 & 0.003860 & 2.9 \\
J1908+0705g & gpps0278 & 0.00199 & 3.0 \\
\hline 
\end{tabular}
\end{table}

\begin{figure}
  \centering
  \includegraphics[width=0.9\columnwidth]{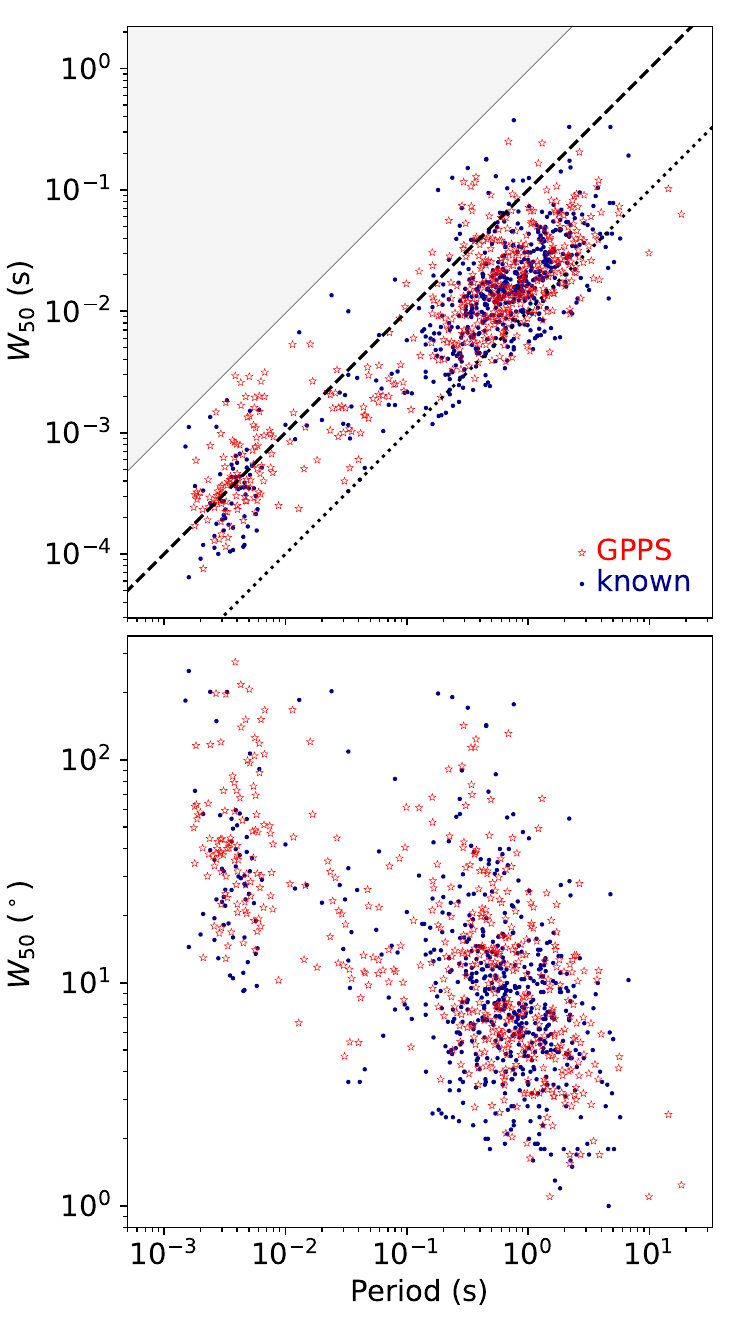}
  \caption{The pulse width $W_{50}$ at the 50\% level of the peak against the pulsar period. The top panel is for $W_{50}$ in units of second, and the lower panel is for units of rotation phase in degree. The light gray area is the impossible area with $W_{50}>P$. The dashed and dotted lines are drawn for $W_{50}=P/10$ and $W_{50}=P/100$, respectively. The open stars are for the newly discovered GPPS pulsars, and the dots are the known pulsars with pulse width measured in \citet{whx+23}. We have checked these outstanding dots with low $W_{50}$ and found that these pulsars have many components but only one is very sharp and dominates the $W_{50}$.
    }\label{fig:w50-p}
\end{figure}

Some pulsars are independent discoveries that should be shared with other authors. Generally several months after we published some new pulsars on the web page, we got an email telling us that some pulsars had been found by using data from other telescopes, but no public information was ever available, and their team indeed had even made some follow-up observations. We understand these pulsars are in fact independent discoveries. For example, PSR J1851+1021g (gpps0258) and J1901+1316g (gpps0426) were also detected by  \citet{sbb+23} from the archive Parkes pulsar survey data. PSR J1845+0200 (gpps0547) is a bright millisecond pulsar discovered on July 8th, 2022 and later confirmed by FAST. We put on the web on April 11th, 2023. After we completed eight follow-up observations and determined the binary orbit had a period of 5.3 days, we received an email notice from Dr. H. T. Cromartie on April 11th, 2024 that this pulsar was discovered by Dr. J. Deneva in 2016 in a Fermi unassociated source using Arecibo, and Dr. Cromartie had some follow-up timing and presented in her PhD thesis \citep{cro20}. The relevant information on this pulsar was first published by \citet{saa+23} on December 1st, 2023. We understand that this is a perfectly independent discovery, and we now have got its timing solution as shown in the GPPS binary pulsar paper by \citet{why+24}.

\subsection{Millisecond pulsars discovered by the GPPS survey}

We have discovered 177 millisecond pulsars in the FAST GPPS survey. In this new release of the major pulsar sample of the FAST GPPS survey, there are 133 new millisecond pulsars, see Table~A\ref{gppsPSRtab1}. Here we define the millisecond pulsars by a simple criterion of the spin period being less than 30 ms. As shown in Figure~\ref{fig:pdf-s}, the GPPS survey discovers weak pulsars, an order of magnitude weaker than previous surveys. For the millisecond pulsars, however, the GPPS survey detects them in a wide range of flux densities, which implies that many strong millisecond pulsars were missed in previous surveys probably due to their high DMs. 

In general, bright short-period millisecond pulsars with sharp profiles can join Pulsar Timing Arrays (PTAs), which are the best probes for gravitational waves in a PTA because they can be observed with a high signal-to-noise ratio (S/N) and hence the time of arrival (TOA) can be measured with high accuracy. We measure the TOAs of the newly discovered millisecond pulsars using FAST 15 minute observations and find that 34 millisecond pulsars in Table~\ref{msptoa} have a TOA accuracy better than 3$\mu$s and they are suitable to be included in PTAs.

We here compare the pulse widths of normal pulsars and millisecond pulsars. \citet{whx+23} measured the full widths at half maximum (FWHMs, $W_{50}$) of known pulsars and the GPPS pulsars detected in the FAST GPPS survey and other FAST projects. The measurements were made for profiles with the best S/N ever. We also measured the $W_{50}$ of the newly discovered pulsars in the FAST GPPS survey, as listed in Table~A\ref{gppsPSRtab1}. As seen in Figure \ref{fig:w50-p}, the millisecond pulsars always have relatively wider profiles than normal pulsars. Normal pulsars often have FWHMs ($W_{50}$) of about a few percent of the period, a few larger than 10\% of the periods, while millisecond pulsars often have width greater than $20\degr$, many $50\degr$. outstanding small-width dots are these pulsars with multiple components but one sharp tall component gives $W_{50}$so that they appear unusually narrow. 

\begin{table}[th]
\centering
\caption{Newly discovered pulsars from the FAST GPPS survey showing nulling, mode-changing or driftinging behaviors.}\label{Tab:phenom}
\setlength{\tabcolsep}{2.2pt} 
\footnotesize
\renewcommand\arraystretch{0.75}
\begin{tabular}{lclrl}
\hline \noalign{\smallskip}
PSR Jname      & GPPS No. &  \hfill Period     & DM    & Features\\
           &     &  \hfill (s)     &  (*)     &        \\
\hline\noalign{\smallskip}
J0514$+$4009g & gpps0556 & 2.6255 & 106.5 & drifting \\
J0630$+$1002g & gpps0649 & 2.8815 & 149.3 & drifting \\
J1810$-$0100g & gpps0570 & 1.1062 & 78.1  & nulling; drifting \\
J1819$-$0040g & gpps0597 & 0.1601 & 57.9  & modulation \\
J1820$+$0006g & gpps0629 & 2.4050 & 187.9 & nulling; drifting \\[1mm]
J1822$+$0044g & gpps0674 & 0.9989 & 54.0  & nulling; drifting \\
J1825$-$0208g & gpps0483 & 3.3074 & 192.9 & nulling; drifting \\
J1827$-$0125g & gpps0397 & 0.3380 & 98.0  & modulation \\
J1832$+$0203g & gpps0664 & 1.2128 & 226.4 & drifting \\
J1835$-$0112g & gpps0253 & 0.5996 & 316.7 & drifting \\[1mm]
J1835$-$0149g & gpps0395 & 1.2776 & 102.3 & drifting \\
J1838$+$0044g & gpps0005 & 2.2032 & 229.6 & nulling; drifting \\
J1838$-$0014g & gpps0445 & 0.3605 & 282.8 & modulation \\
J1840$-$0141g & gpps0374 & 1.9268 & 279.7 & nulling; drifting \\
J1842$+$0131g & gpps0438 & 1.5903 & 115.4 & modulation \\[1mm]
J1843$-$0127g & gpps0363 & 2.1649 & 450.8 & nulling \\
J1845$-$0229Ag& gpps0493 & 0.6577 & 832.0 & modulation \\
J1846$-$0211g & gpps0456 & 0.7882 & 838.0 & modulation \\
J1847$+$0614g & gpps0165 & 1.6631 & 273.0 & drifting \\
J1847$-$0048g & gpps0337 & 0.5825 & 667.0 & modulation \\[1mm]
J1848$+$0127g & gpps0035 & 0.5340 & 77.0  & mode-changing \\
J1848$+$1245g & gpps0223 & 0.2482 & 100.0 & drifting \\
J1849$+$0225g & gpps0116 & 1.4745 & 259.9 & mode-changing \\ 
J1849$+$0340g & gpps0184 & 1.6667 & 349.5 & nulling; drifting \\
J1851$+$0501g & gpps0408 & 2.3274 & 341.3 & drifting \\ [1mm]
J1852$+$0421g & gpps0364 & 3.1609 & 294.3 & nulling \\ 
J1852$-$0033g & gpps0173 & 1.3690 & 321.6 & drifting \\
J1854$-$0230g & gpps0574 & 0.6875 & 550.5 & drifting;mode-changing \\
J1855$-$0115g & gpps0360 & 2.5623 & 261.7 & nulling \\ 
J1856$-$0134g & gpps0402 & 0.3819 & 236.2 & mode-changing \\ [1mm]
J1857$-$0120g & gpps0384 & 1.2226 & 353.9 & drifting \\
J1858$+$0026g & gpps0094 & 4.7147 & 415.3 & nulling; drifting \\  
J1901$-$0104g & gpps0401 & 0.7395 & 261.2 & modulation \\
J1903$+$1728g & gpps0669 & 1.7166 & 150.3 & drifting \\
J1905$+$0935g & gpps0153 & 1.6345 & 418.0 & nulling; drifting \\[1mm]
J1906$+$1211g & gpps0458 & 3.8050 & 292.4 & nulling  \\
J1909$+$0310g & gpps0447 & 1.9721 & 111.0 & modulation \\
J1909$+$0423g & gpps0436 & 0.5116 & 254.9 & drifting \\
J1919$+$1527g & gpps0130 & 1.3715 & 697.5 & mode-changing \\
J1921$+$0851g & gpps0234 & 0.9567 & 101.0 & nulling; drifting \\[1mm]
J1921$+$1341g & gpps0088 & 4.6035 & 754.9 & nulling \\
J1922$+$1512g & gpps0385 & 2.3572 & 395.0 & nulling \\
J1924$+$1923g & gpps0002 & 0.6893 & 384.9 & mode-changing \\
J1924$+$2037g & gpps0192 & 0.6848 & 82.3  & nulling \\
J1926$+$1631g & gpps0053 & 0.6784 & 196.7 & drifting \\[1mm]
J1935$+$1901g & gpps0407 & 0.8973 & 362.3 & nulling \\
J1937$+$1358g & gpps0240 & 2.6454 & 174.8 & nulling; drifting \\
J1939$+$2352g & gpps0150 & 2.1455 & 415.5 & nulling \\
J1939$+$2453g & gpps0208 & 2.4025 & 240.5 & nulling \\ 
J1944$+$1934g & gpps0678 & 3.4453 & 241.4 & nulling; drifting \\[1mm]
J1945$+$2410g & gpps0225 & 2.3776 & 478.4 & drifting \\
J1959$+$3141g & gpps0617 & 0.5145 & 340.6 & drifting \\
J2001$+$2856g & gpps0486 & 1.4456 & 233.1 & drifting \\
J2011$+$3006g & gpps0136 & 2.5057 & 12.0  & drifting \\
J2011$+$3521g & gpps0610 & 0.9432 & 438.1 & nulling; drifting \\[1mm]
J2026$+$3656g & gpps0344 & 1.7855 & 280.7 & nulling \\
J2029$+$4453g & gpps0422 & 1.3614 & 333.2 & drifting \\
J2051$+$4434g & gpps0085 & 1.3031 & 616.0 & drifting;mode-changing \\
\hline\noalign{\smallskip}
\end{tabular}
\tablecomments{*: DM values have a unit of pc~cm$^{-3}$.} 
\end{table}

\begin{figure*}
\centering
\includegraphics[width=0.25\textwidth]{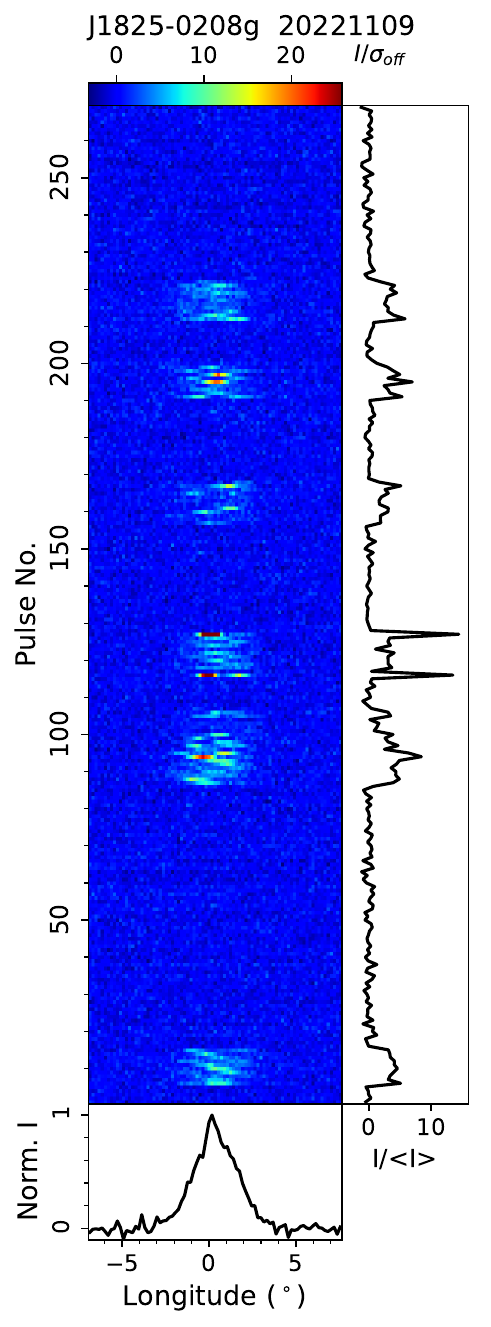}
\includegraphics[width=0.25\textwidth]{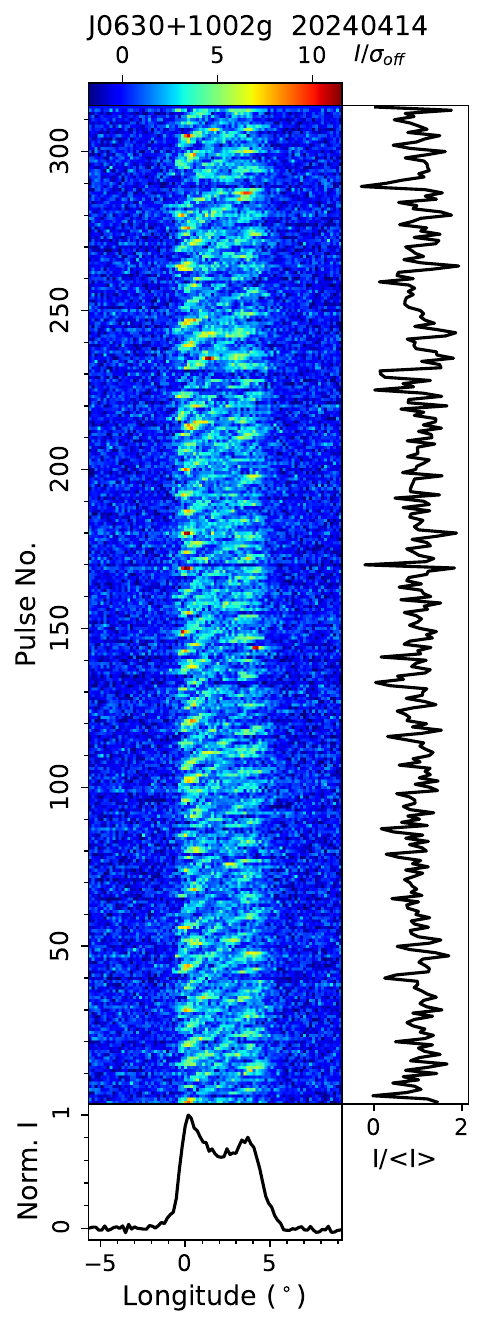}
\includegraphics[width=0.25\textwidth]{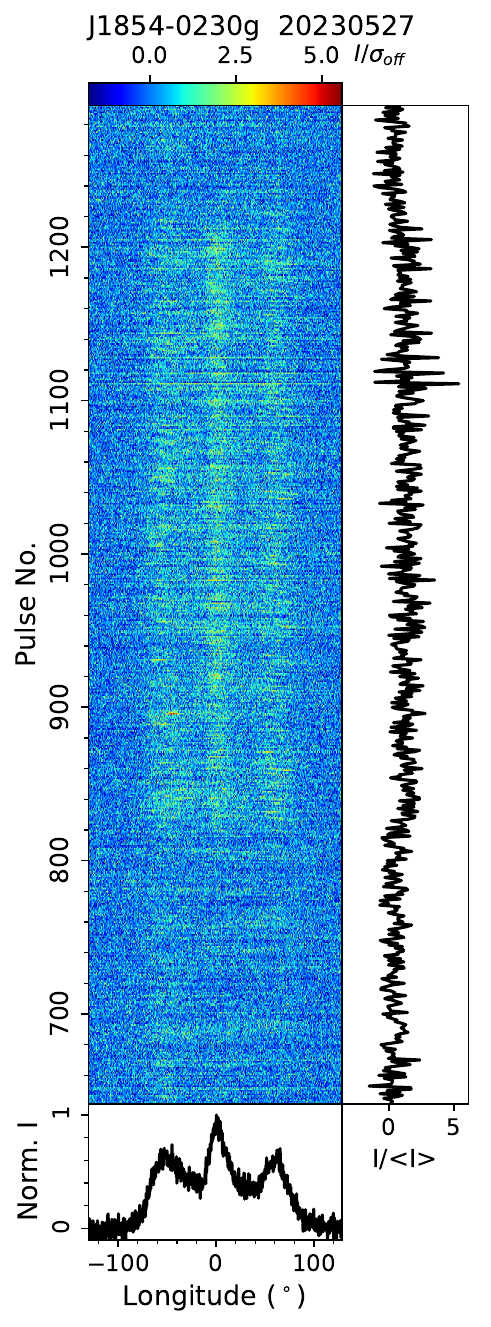}
\caption{The pulse sequences of three example pulsars, PSRs J1825-0208g (gpps0483), J0630+1002g (gpps0649) and J1854-0230g (gpps0574) exhibiting nulling, subpulse drifting, or mode-changing behavior, respectively. The integrated profiles are plotted in the bottom subpanels, and variations of pulse energy along the pulse period are shown in the right subpanels.
}\label{fig:phenom}
\end{figure*}

\subsection{Binary pulsars discovered by the GPPS survey}

During the FAST GPPS survey, we have found about 160 binary pulsars \citep{hww+21,shw+23,why+24}. According to the indications for binaries, we extract the barycenter spin periods and accelerations by using the package PRESTO \citep{ran01} or PDMP in PSRCHIVE \citep{hvm04}, and do more follow-up observations. and do more follow-up observations. We plot the barycenter spin periods and accelerations in two dimensions \citep{fkl01} and find the orbital period and the projected semimajor axis directly by fitting these data by an ellipse. After more observations are made, the phase-connected timing solution can be found. We have derived the timing solutions for a quarter of binary pulsars discovered by the GPPS survey, see details in \citet{why+24}.

A few interesting binaries are worth mentioning here. The first pulsar, PSR J1901+0658, discovered by the FAST GPPS survey is a double neutron star system with a total mass of 2.79(7) $M_\odot$ \citep{shw+23}. The millisecond pulsar PSR J1953+1844 was discovered by the GPPS survey (gpps0190) and has an orbital period of 53 minutes \citep{plj+23}, which
should be a descendant from ultracompact X-ray binaries \citep{yhj+23}. Careful timing of PSR J1928+1815(gpps0121) revealed its compact orbit of 3.6 hr with a heavy companion eclipsing the pulsar signal for about 17\% of the orbit \citep{yhz+24}. From timing observations, we find six millisecond pulsars having a massive white dwarf companion, and four of them have a detectable Shapiro delay  \citep{yhw+24}, from which the mass of PSR J1943+2210 has been determined to be   1.84$^{+0.11}_{-0.09}$~M$_\odot$.

\subsection{Newly discovered pulsars with nulling, mode-changing, and subpulse drifting phenomena}

Among the pulsars discovered by the FAST GPPS survey, some 58 bright pulsars show nulling, mode-changing, and subpulse drifting phenomena, as listed in Table~\ref{Tab:phenom}. Three examples are shown in Figure~\ref{fig:phenom}.

Detailed analyses of these single pulse behaviors for the GPPS pulsars and also known pulsars have been carried out by Yi Yan et al. (2025, in preparation) using the FAST observation data.

\subsection{Long period pulsars}

Long-period pulsars with a period of, e.g., longer than 10 s have been difficult to discover because they have much fewer pulses in a given observation time, and hence do not appear clearly in the Fourier spectrum. The system variations over scales of a few seconds lead to large signal amplitude fluctuations which finally appear as red noise. The single pulse search approach we developed \citep{zhj+23} can be efficiently used for the detection of long-period pulsars. 

During the FAST GPPS survey, we detected 8 pulsars with periods larger than 10~s:  
PSRs 
J2053+4455g (gpps0645, $P$ = 10.72~s, DM = 440.6 pc~cm$^{-3}$), 
J1940+2203 (gpps0312, $P$ = 11.91~s, DM = 59.4 pc~cm$^{-3}$), 
J1903+0433 (gpps0090, $P$ = 14.05~s, DM = 200.6 pc~cm$^{-3}$), 
J1902$-$0012g (gpps0488, $P$ = 14.31~s, DM = 126.6  pc~cm$^{-3}$), 
J2044+4331g (gpps0658, $P$ = 16.61~s, DM = 483.2 pc~cm$^{-3}$), 
J1911+0906g (gpps0285, $P$ = 16.93~s, DM = 24.5 pc~cm$^{-3}$), 
J1920+0941g (gpps0569, $P$ = 18.29~s, DM = 54.2 pc~cm$^{-3}$), 
J1847$-$0308g (gpps0730, $P$ = 29.77~s, DM = 150.3 pc~cm$^{-3}$). 
Most of them were detected first by the single pulse searching module. PSRs J1940 +2203, J2044+4331g and J1911+0906g are RRATs, and only a few pulses have been detected.

Currently the longest period pulsar discovered in the FAST GPPS survey is PSR J1847$-$0308g, which is located in the smallest Galactic longitude ($l = 29.6312 \degr$, $b = -0.4558\degr$) that FAST can survey (see Fig.\ref{gppsSky}). Only 4 pulses were detected in the survey data obtained at 20210427. From the follow-up tracking observation for 15 minutes on 20240917, we detect many harmonics of this pulsar (3, 4, 6, 7, 10, 11, 13, 15, 16, 17). In reality, the longest period we detect in the period search is 9.923090~s, but its 1/3, 3/4, 3/7, 3/11 harmonics directly tell the real period of 29.63~s.

Follow-up timing observations are the key to revealing their natures. Are they RRATs or normal pulsars and do they have ultra-strong magnetic fields? The timing result of PSR J1856+0211 (gpps0158, $P$ = 9.89s) has already shown that a long-period pulsar can be a normal pulsar near the death line \citep{shw+23}.

\begin{table}
\centering
\caption{Properties of single pulses for newly discovered Galactic transient sources by FAST.}
\label{tab:fewPulses}
\footnotesize
\renewcommand\arraystretch{0.85}
\setlength{\tabcolsep}{4.0pt}
\begin{tabular}{crcrrr}
\hline
ObsDate & No. & TOA            & $\rm S/N$  & $W_{\rm 50}$    & $F_{\rm 1250MHz}$      \\
        &     & (MJD)          &      & (ms)   & (mJy ms)   \\
\hline
\multicolumn{6}{c}{{J1905$+$1200g (gpps0712)}}          \\ 
    20230727 & 1   & 60151.69484912 & 12.7 & 25.3  & 94.4  \\
    20230727 & 2   & 60151.69504097 & 7.3  & 20.6  & 31.2  \\
    20240125 & 1   & 60334.09010890 & 6.2  & 67.4  & 37.5  \\
    20240125 & 2   & 60334.09164371 & 13.1 & 83.3  & 159.2 \\
\hline
\multicolumn{6}{c}{{J1914$+$1053g (gpps0713)}}          \\ 
    20191123 & 1   & 58810.34124023 & 16.6 & 92.5  & 182.9 \\
    20191123 & 2   & 58810.35130488 & 17.6 & 111.1 & 165.4 \\
    20191123 & 3   & 58810.38946132 & 6.5  & 44.4  & 26.4  \\
    20191124 & 1   & 58811.37380287 & 24.8 & 58.5  & 274.3 \\
\hline
\multicolumn{6}{c}{{J1913$+$1058g (gpps0714)}}          \\ 
    20191123 & 1   & 58810.35640275 & 11.2 & 19.0  & 52.6  \\
    20191124 & 1   & 58811.35026513 & 15.2 & 11.6  & 56.4  \\
\hline
\multicolumn{6}{c}{{J1930$+$1713g (gpps0715)}}          \\ 
    20230904 & 1   & 60191.60094709 & 27.4 & 340.3 & 604.4 \\
    20240413 & 1   & 60412.97178654 & 11.7 & 79.8  & 258.6 \\
\hline
\multicolumn{6}{c}{{J1859$-$0233g (gpps0716)}}          \\ 
    20240528 & 1   & 60457.82530378 & 10.6 & 31.0 & 40.1   \\
    20240528 & 2   & 60457.82625133 & 12.3 & 19.7 & 57.7   \\
    20240528 & 3   & 60457.82786567 & 9.0  & 17.1 & 29.8   \\
    20240627 & 1   & 60487.73142270 & 16.7 & 11.0 & 55.5   \\
\hline
\multicolumn{6}{c}{{J0639$+$0828g (gpps0717)}}          \\ 
    20240613 & 1   & 60474.27783075 & 47.5 & 7.6  & 117.8  \\
    20240613 & 2   & 60474.28532394 & 61.7 & 45.3 & 666.4  \\
\hline
\multicolumn{6}{c}{{J2044$+$3843g (gpps0727)}}          \\ 
    20240212 & 1   & 60352.23971594 & 18.4 &  4.4 & 54.2   \\
    20240917 & 1   & 60570.64981733 & 15.5 & 20.1 & 68.2   \\
    20240917 & 2   & 60570.65101100 &  9.7 & 44.2 & 36.0   \\
    20240917 & 3   & 60570.65166538 & 13.3 & 19.9 & 58.6   \\
\hline
\multicolumn{6}{c}{{J1906$+$0310g (gpps0728)}}          \\ 
    20201117 & 1   & 59170.34129234 & 14.4 & 5.3  & 42.4   \\
    20240927 & 1   & 60580.47441629 & 10.4 & 1.4  & 19.4   \\
\hline
\multicolumn{6}{c}{{J1830$-$0231g (gpps0729)}}          \\ 
    20230627 & 1   & 60121.72042370 & 35.6 & 12.8 & 201.1  \\
    20230627 & 2   & 60121.72461619 & 29.2 & 21.1 & 185.0  \\
    20240920 & 1   & 60573.48722729 & 48.2 & 16.9 & 354.8  \\
\hline
\multicolumn{6}{c}{{J1843$-$0147g (gpps0731)}}          \\ 
    20231102 & 1   & 60250.35810808 & 17.2 & 50.8 & 156.6  \\
    20241001 & 2   & 60584.45125998 & 19.8 & 92.2 & 213.4  \\
\hline
\end{tabular}

\end{table}

\begin{figure*}[!t]
\centering
\includegraphics[width=0.195\textwidth]{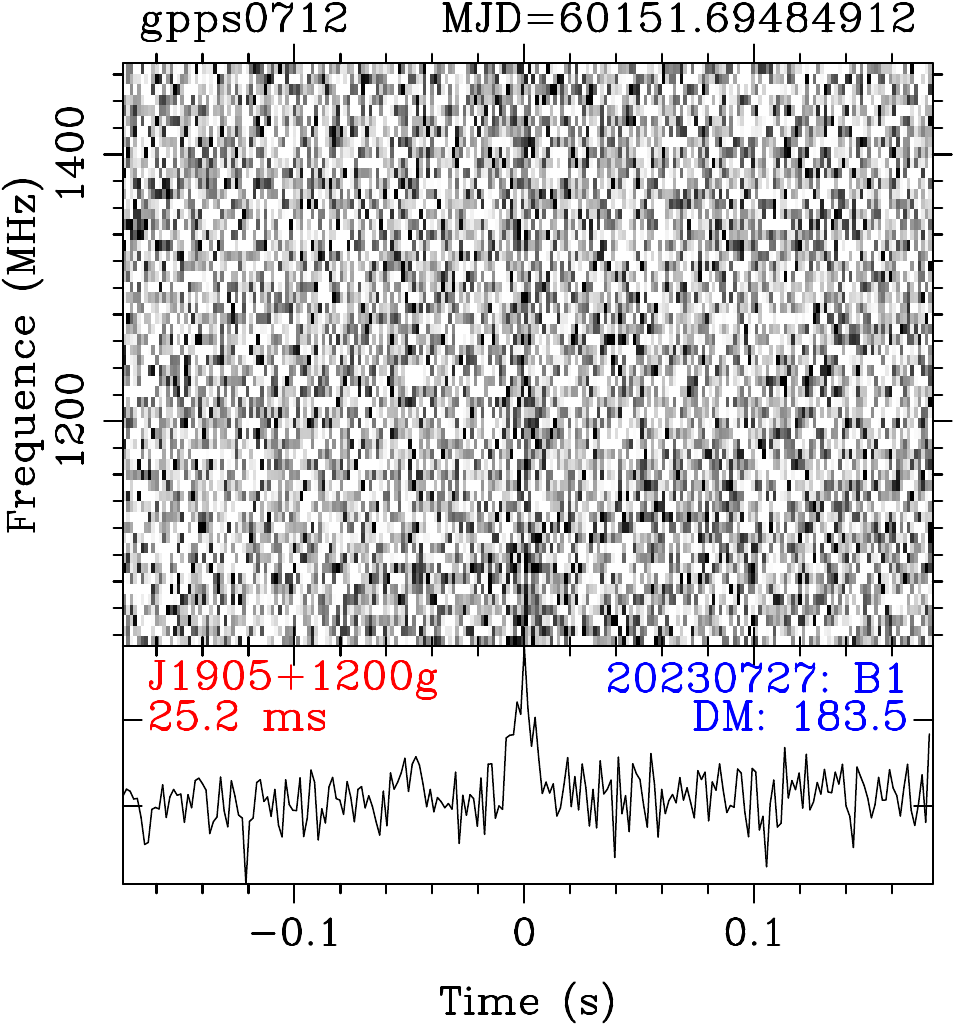}
\includegraphics[width=0.195\textwidth]{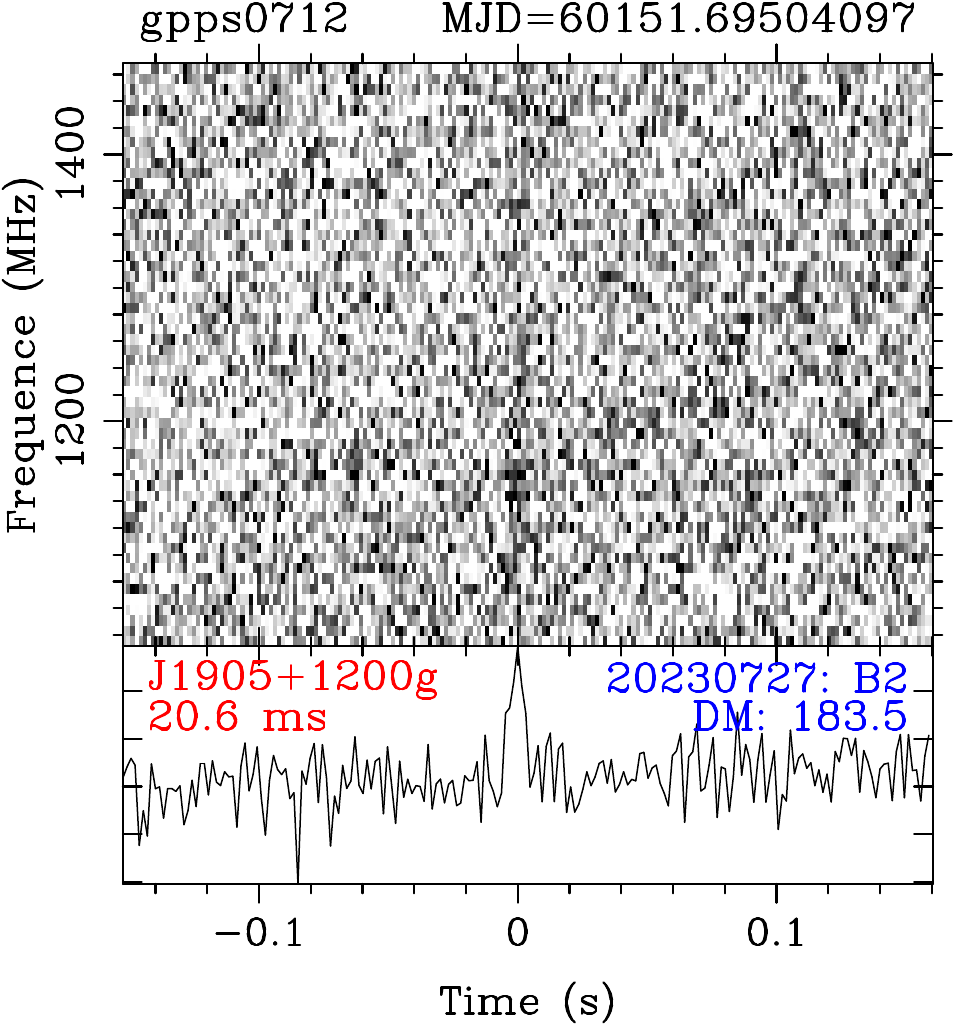}
\includegraphics[width=0.195\textwidth]{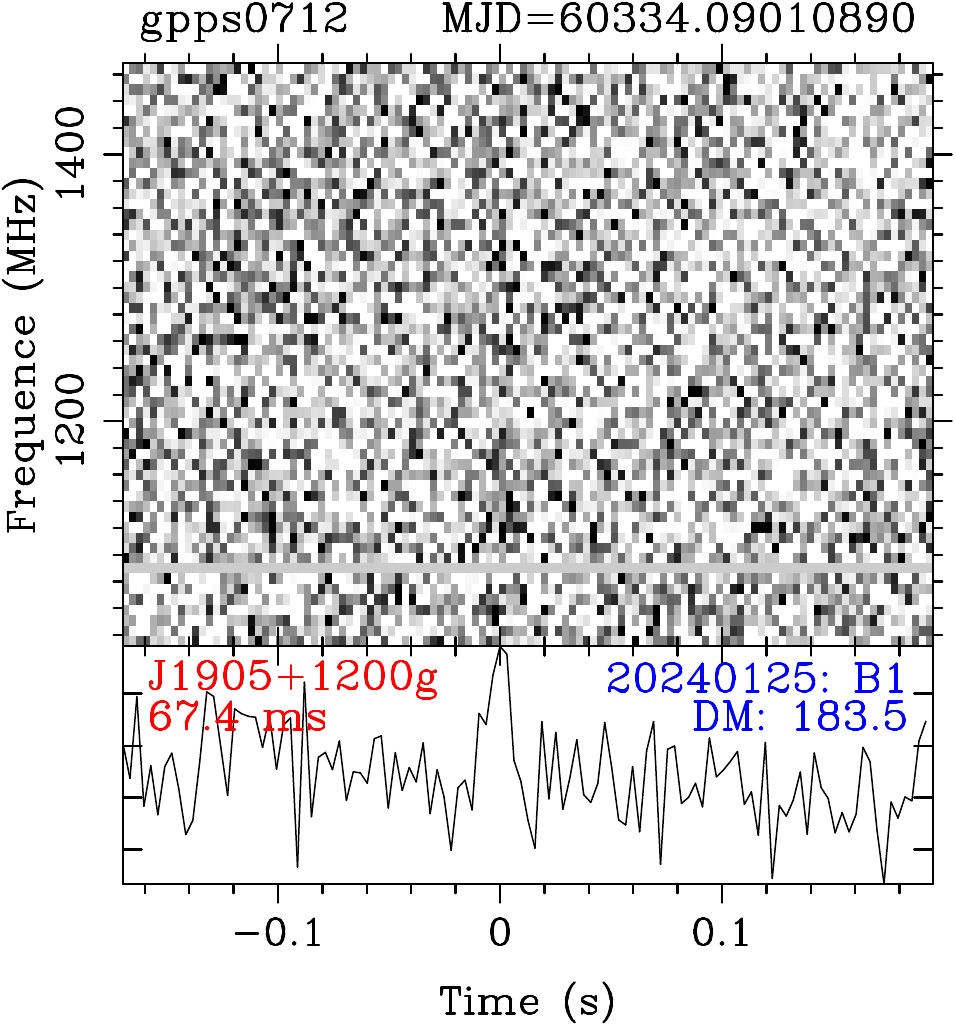}
\includegraphics[width=0.195\textwidth]{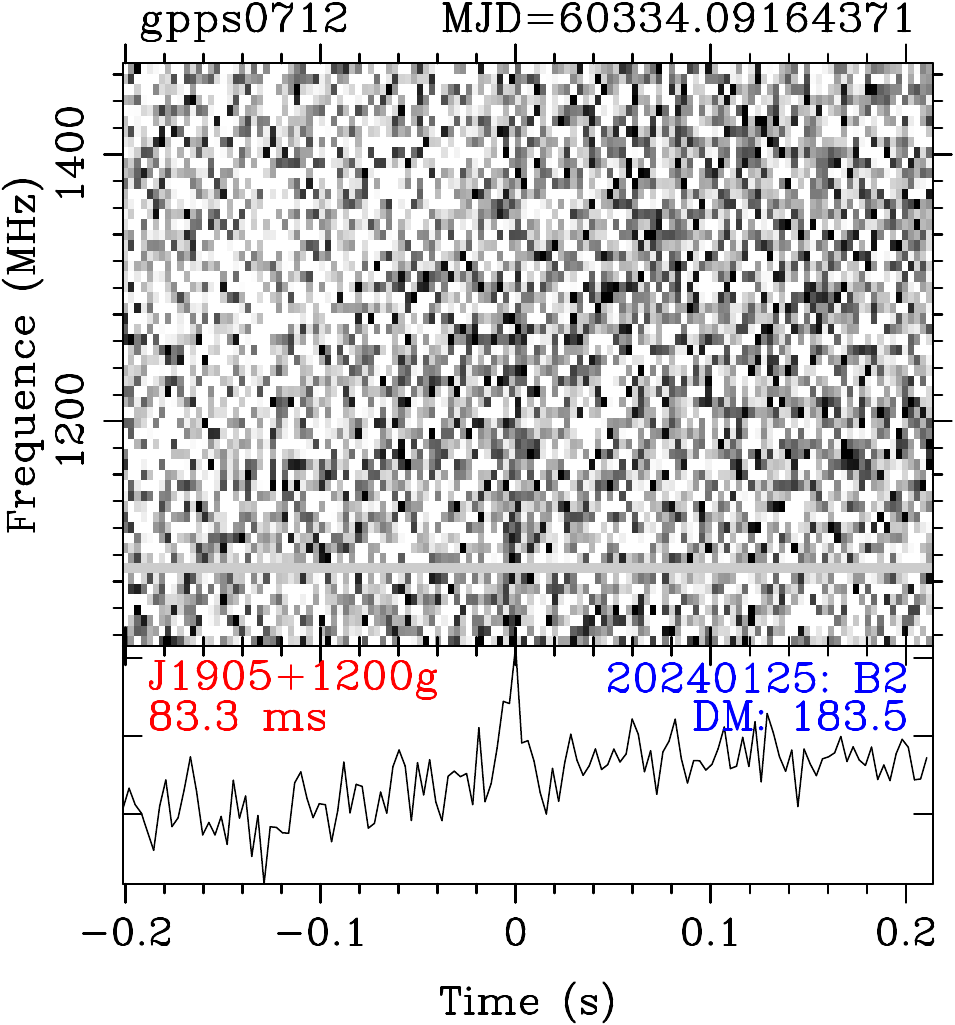}
\includegraphics[width=0.195\textwidth]{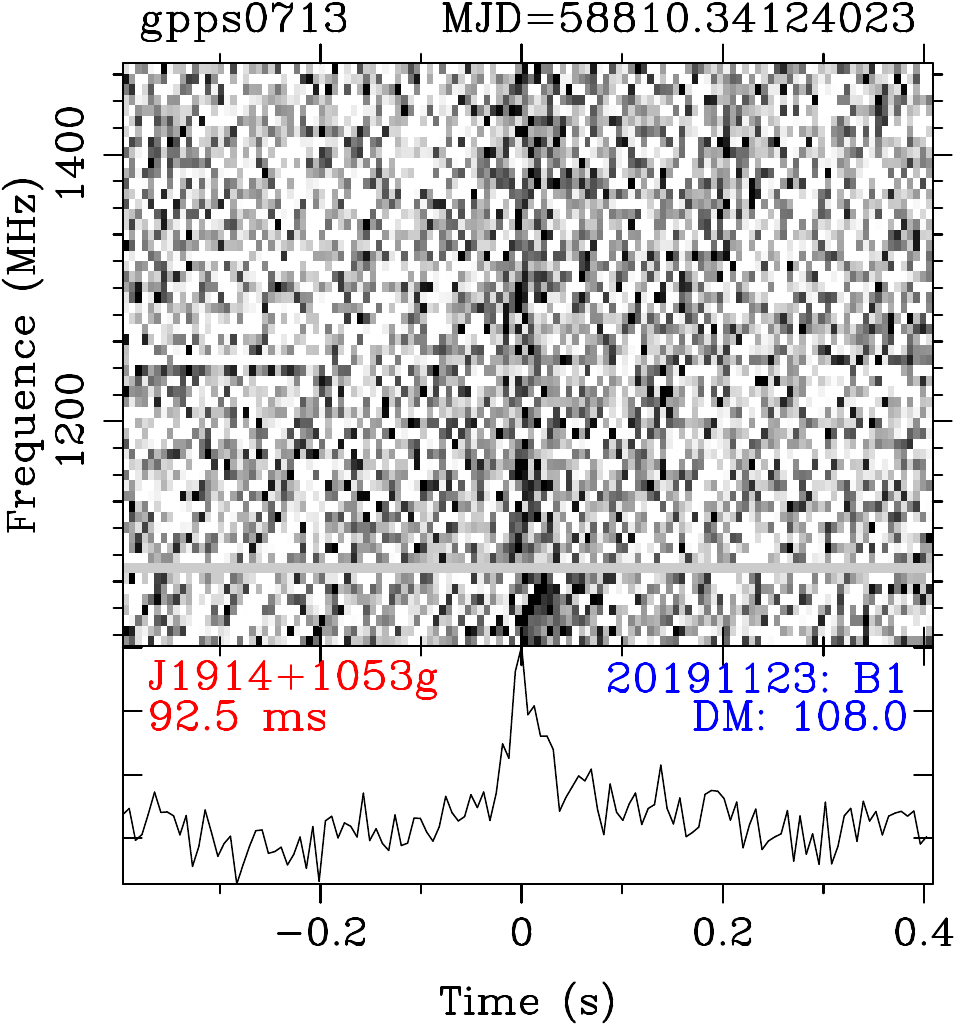}
\includegraphics[width=0.195\textwidth]{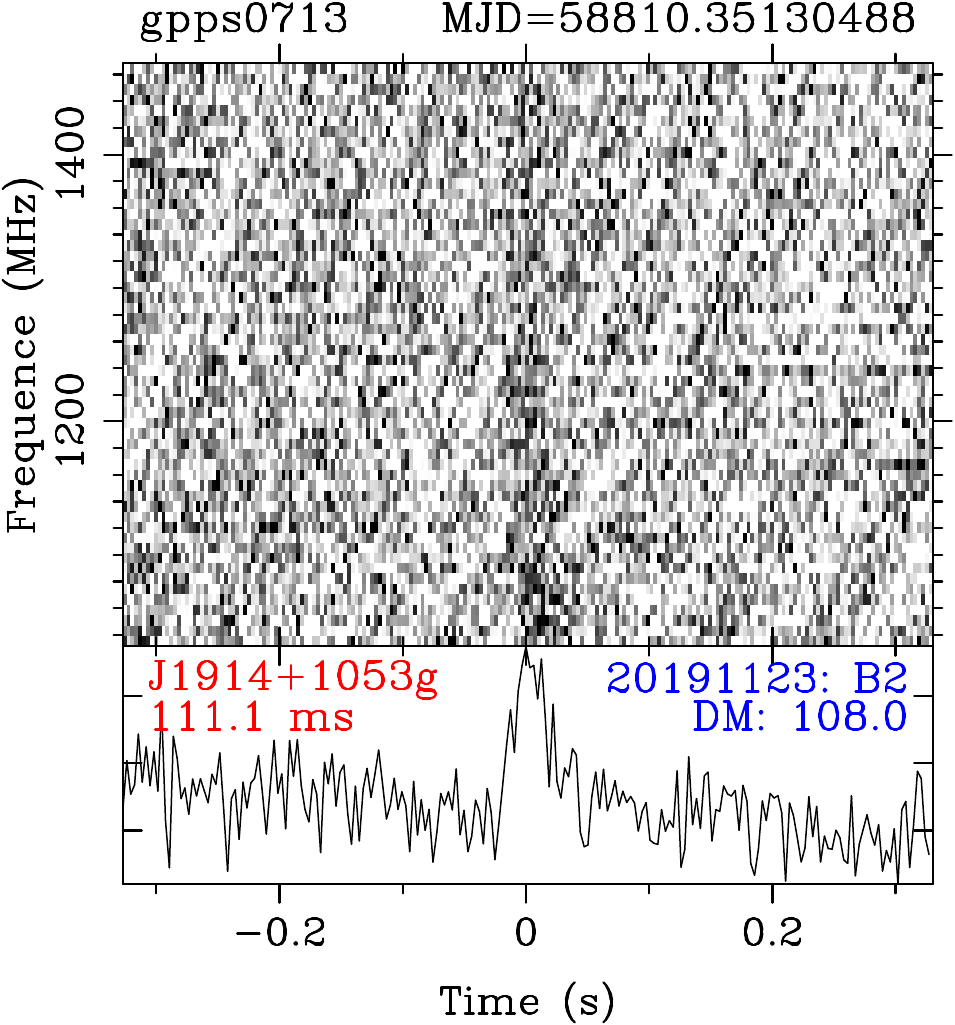}
\includegraphics[width=0.195\textwidth]{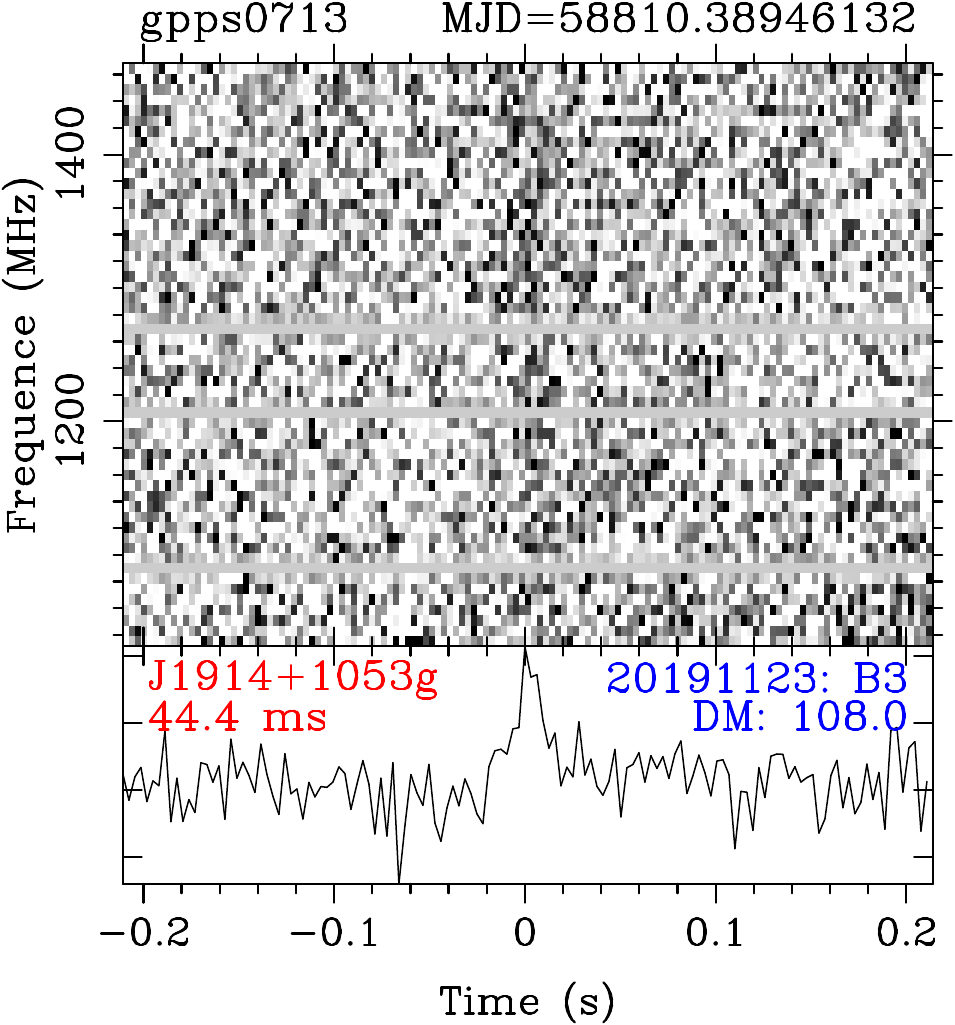}
\includegraphics[width=0.195\textwidth]{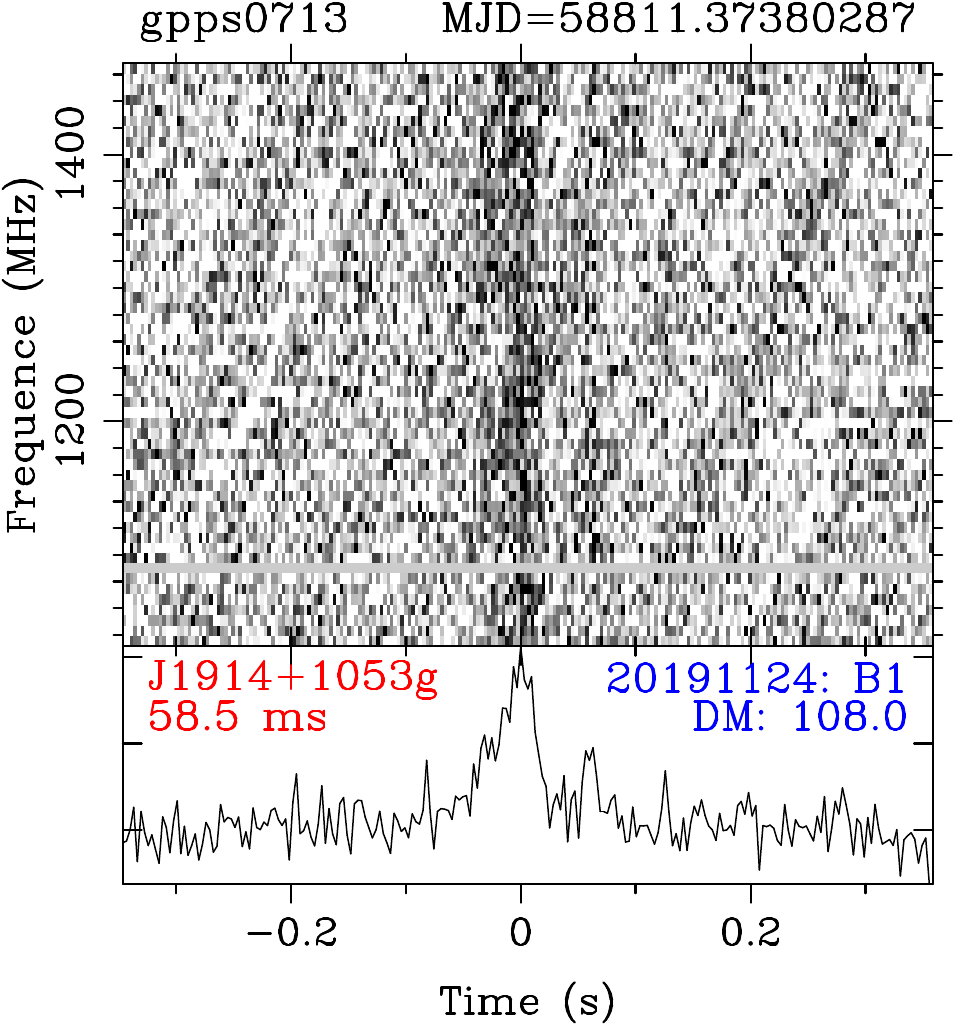}
\includegraphics[width=0.195\textwidth]{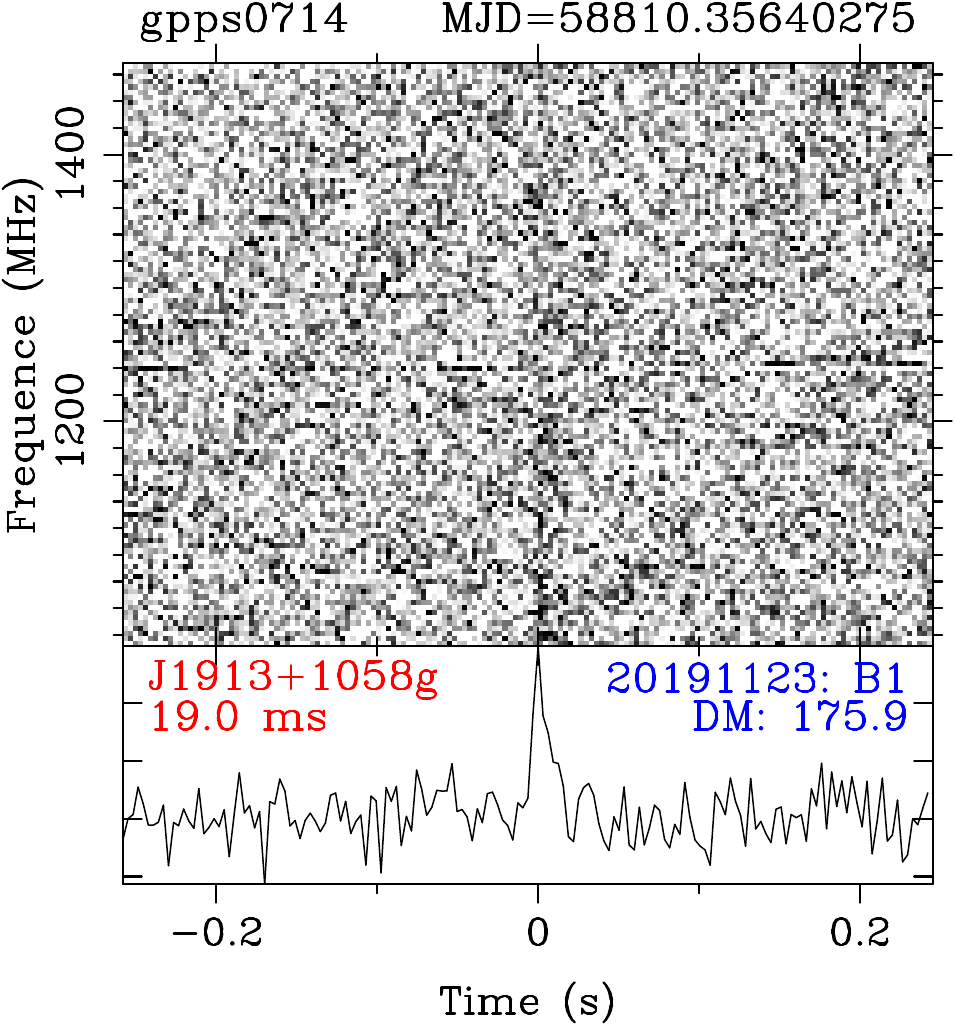}
\includegraphics[width=0.195\textwidth]{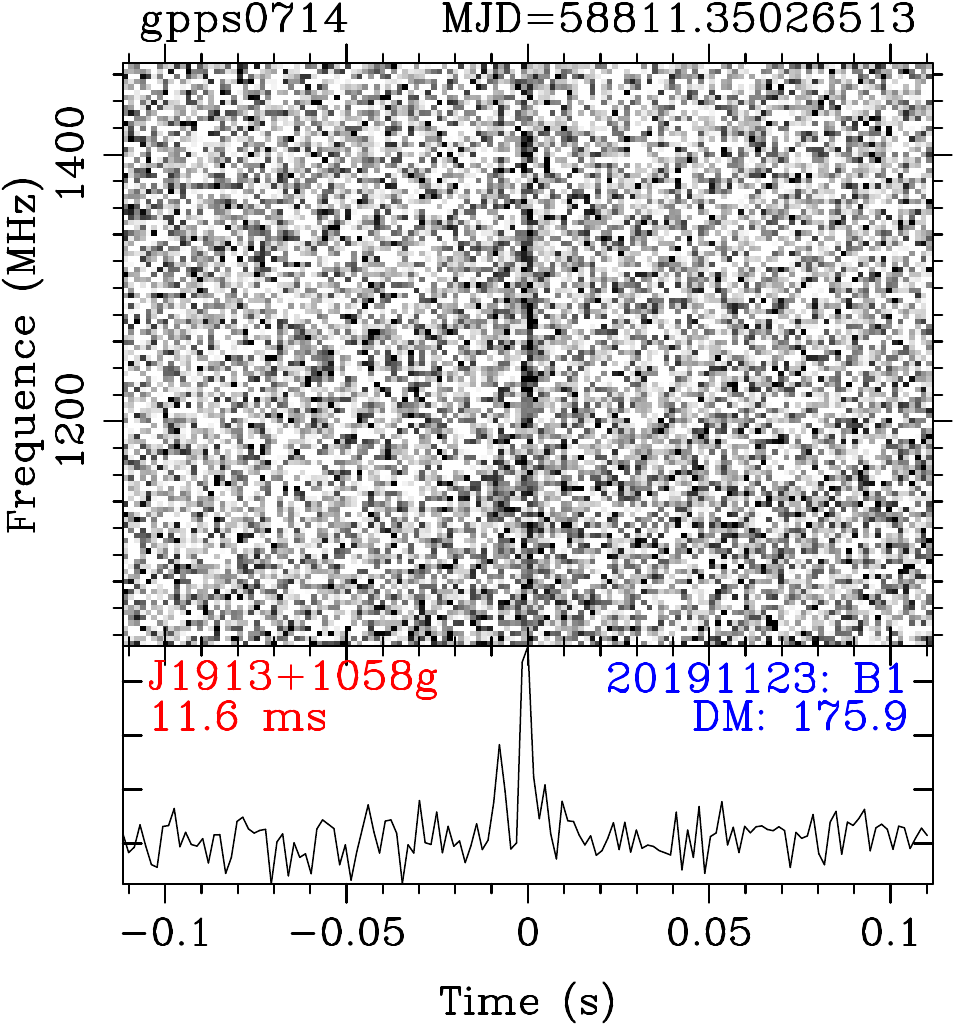}
\includegraphics[width=0.195\textwidth]{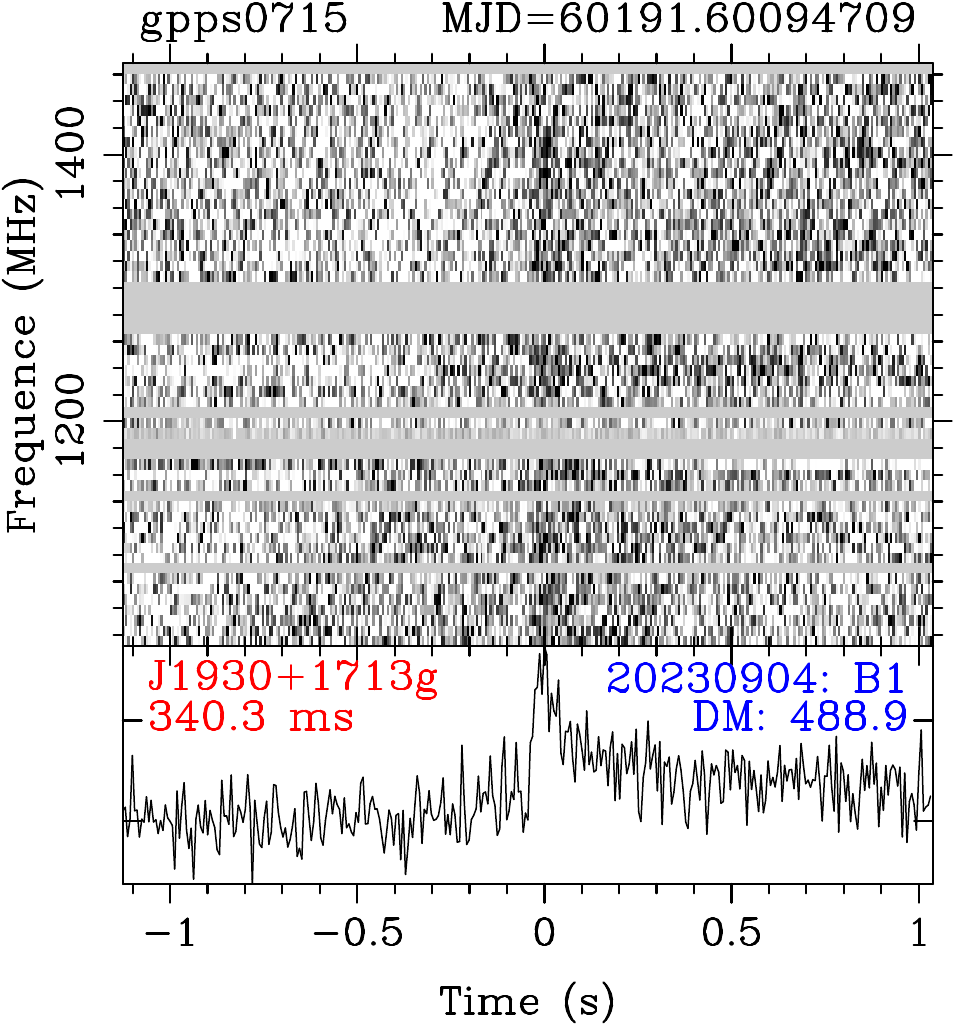}
\includegraphics[width=0.195\textwidth]{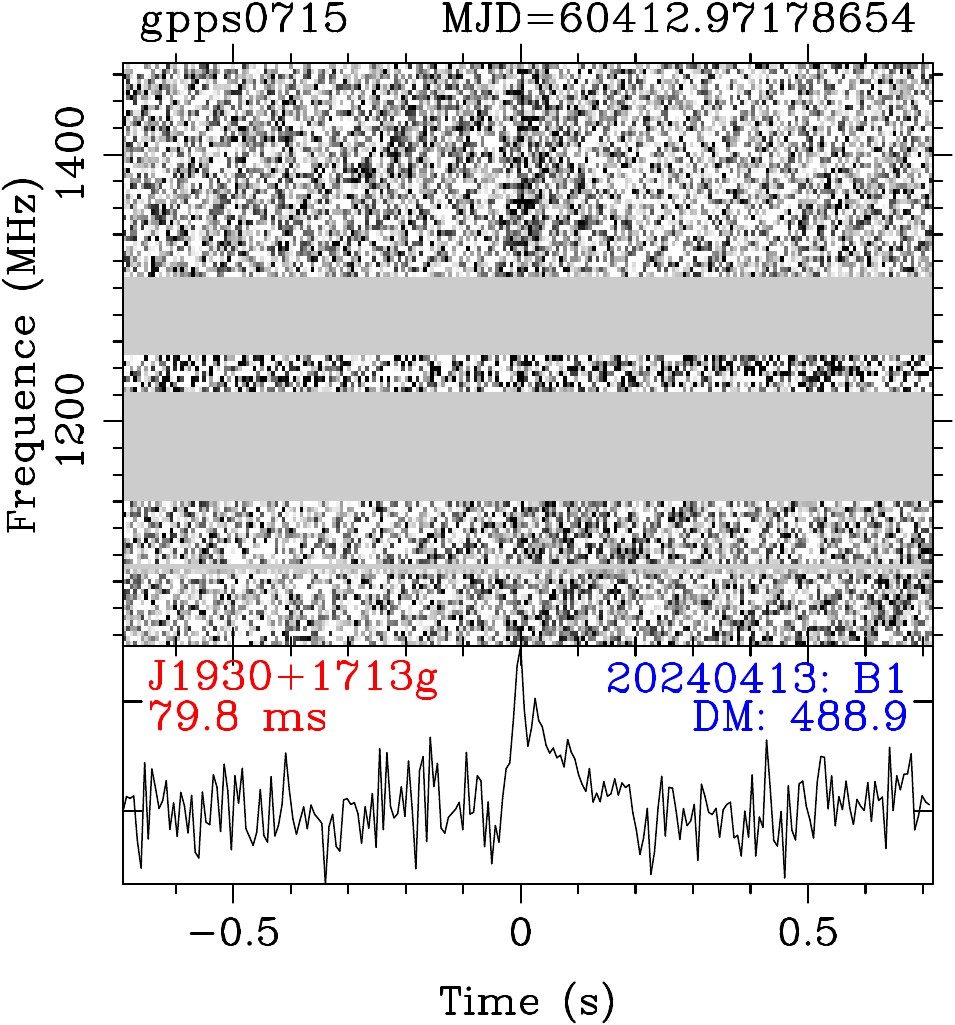}
\includegraphics[width=0.195\textwidth]{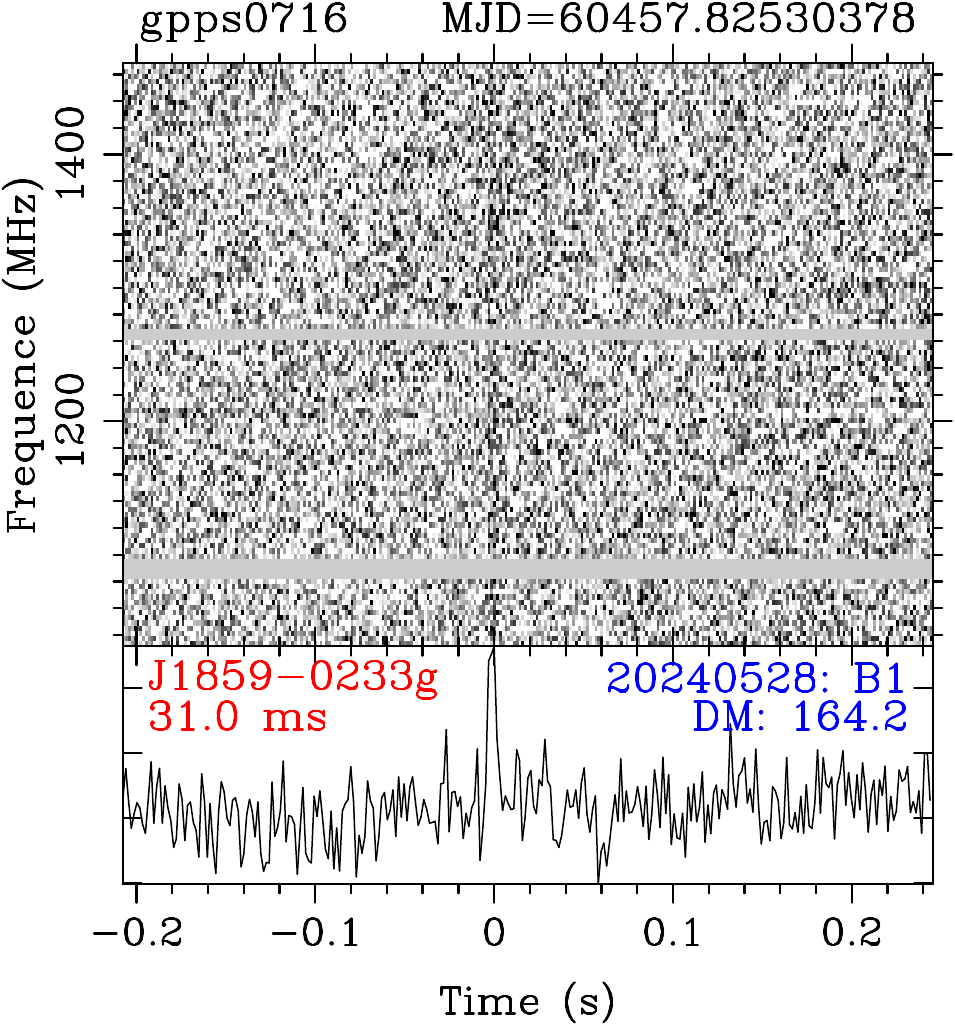}
\includegraphics[width=0.195\textwidth]{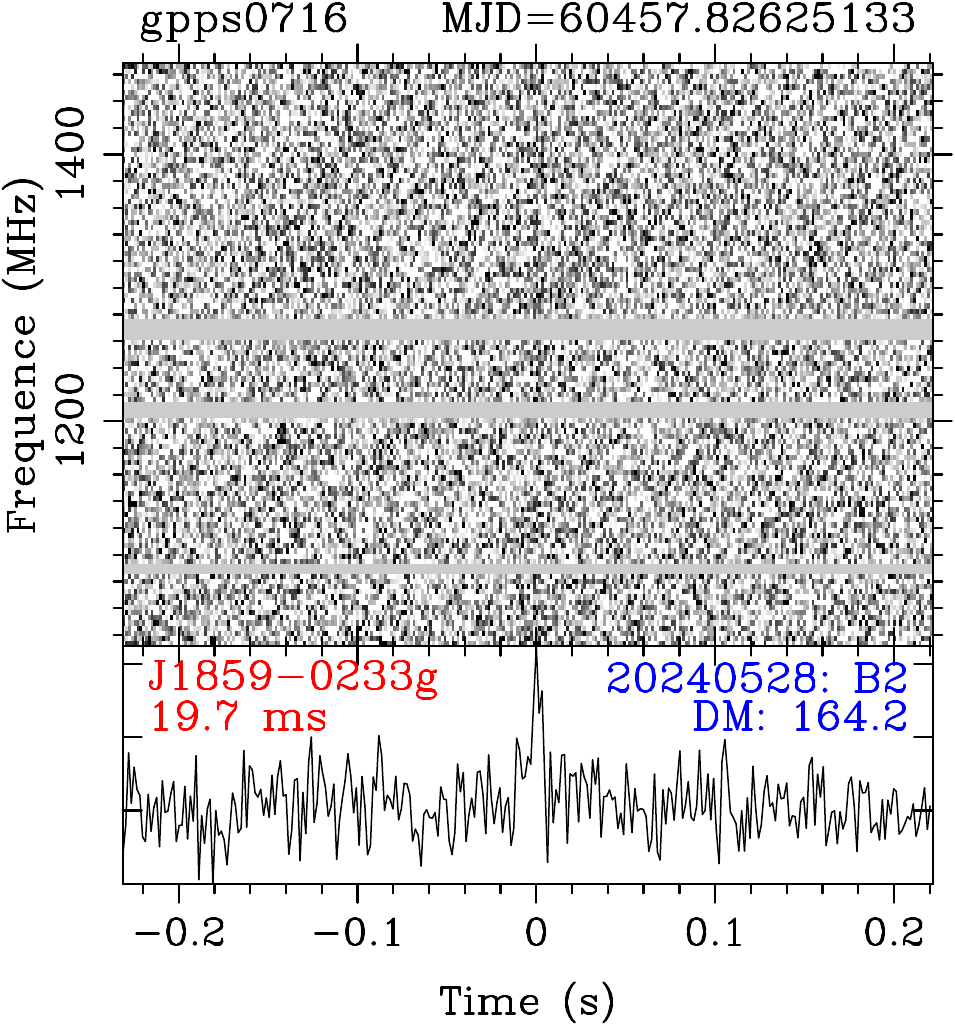}
\includegraphics[width=0.195\textwidth]{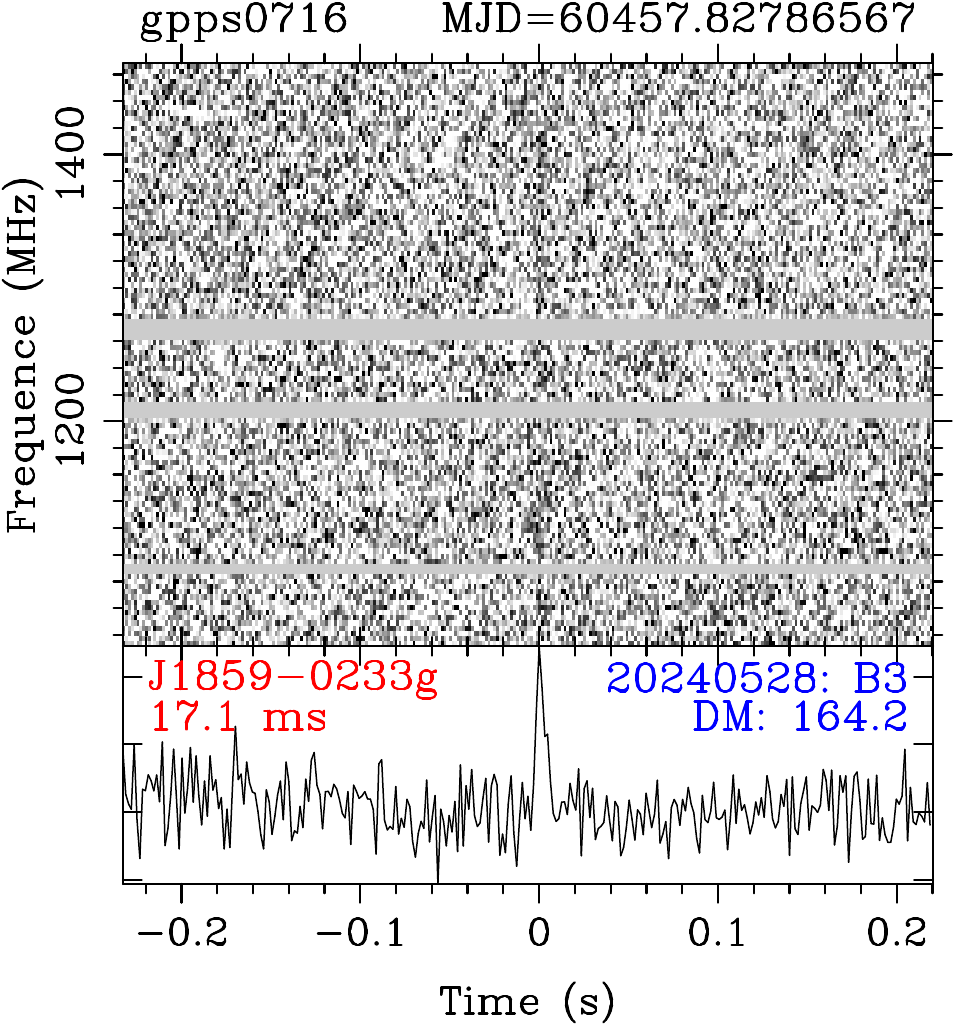}
\includegraphics[width=0.195\textwidth]{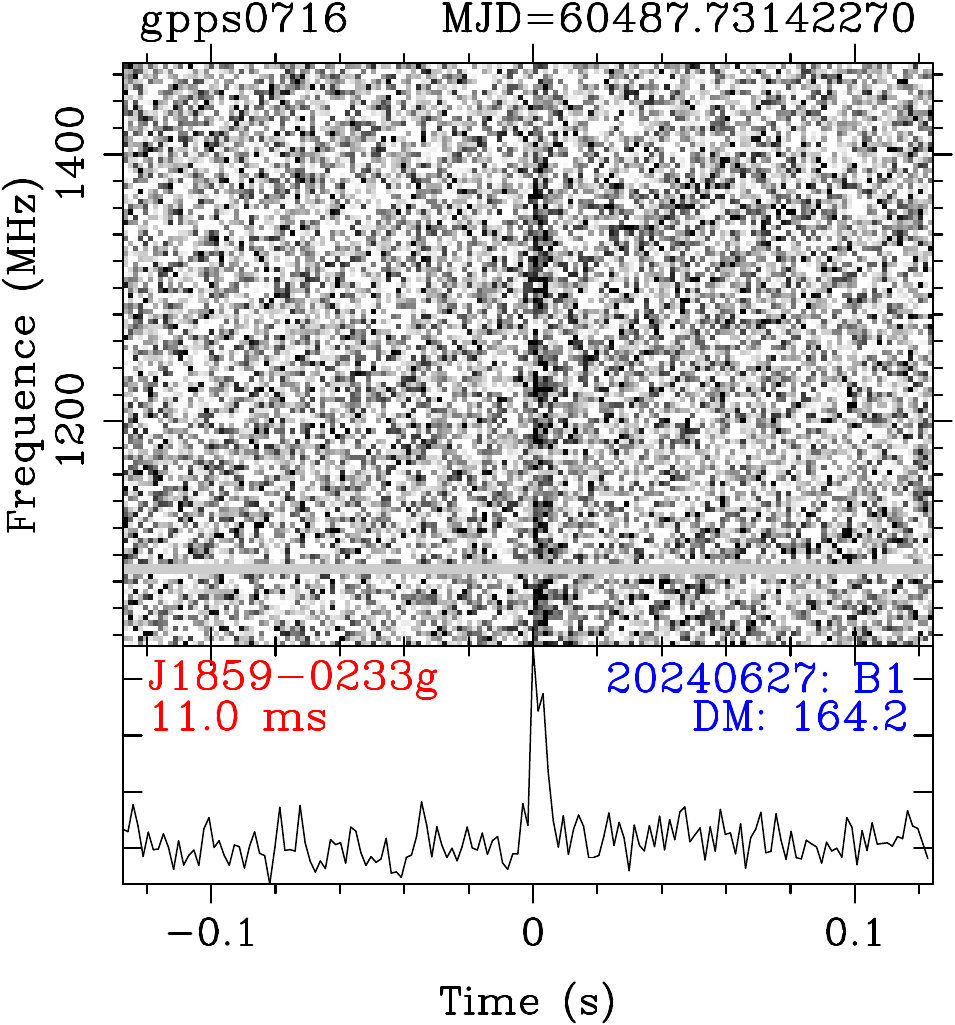}
\includegraphics[width=0.195\textwidth]{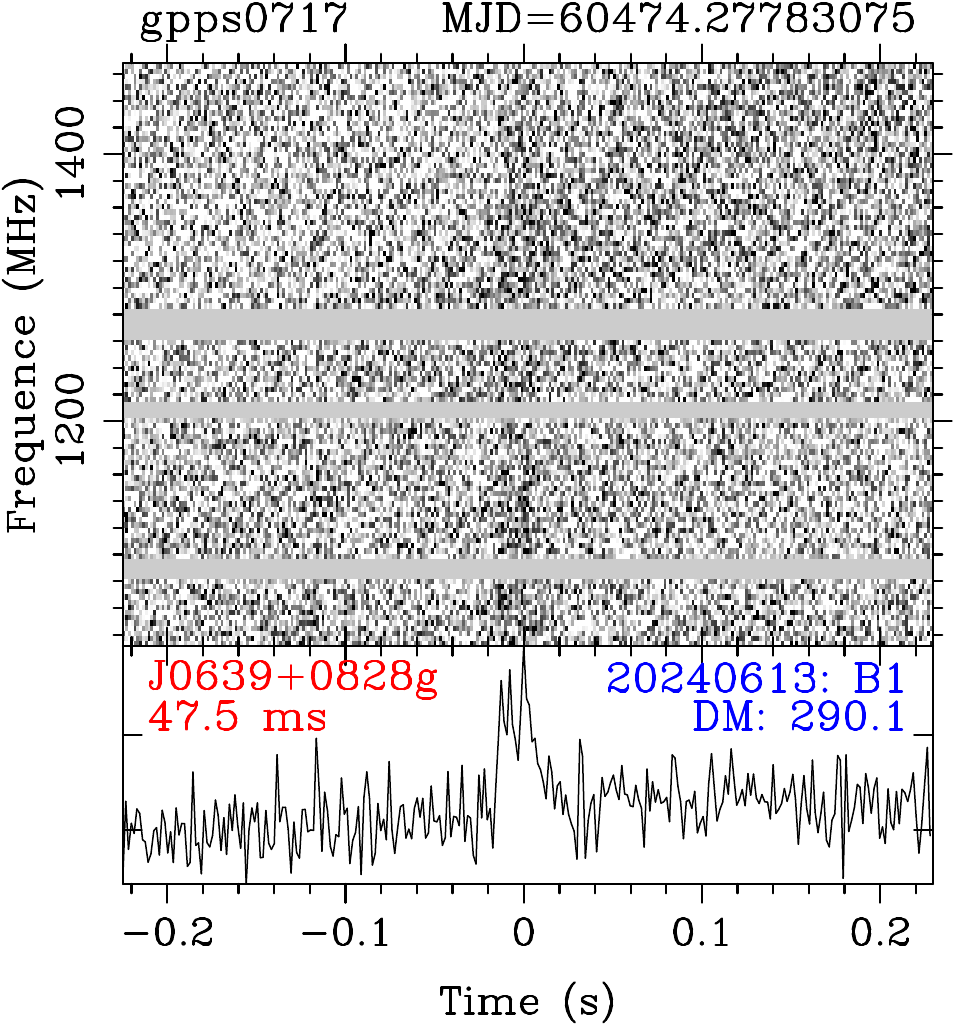}
\includegraphics[width=0.195\textwidth]{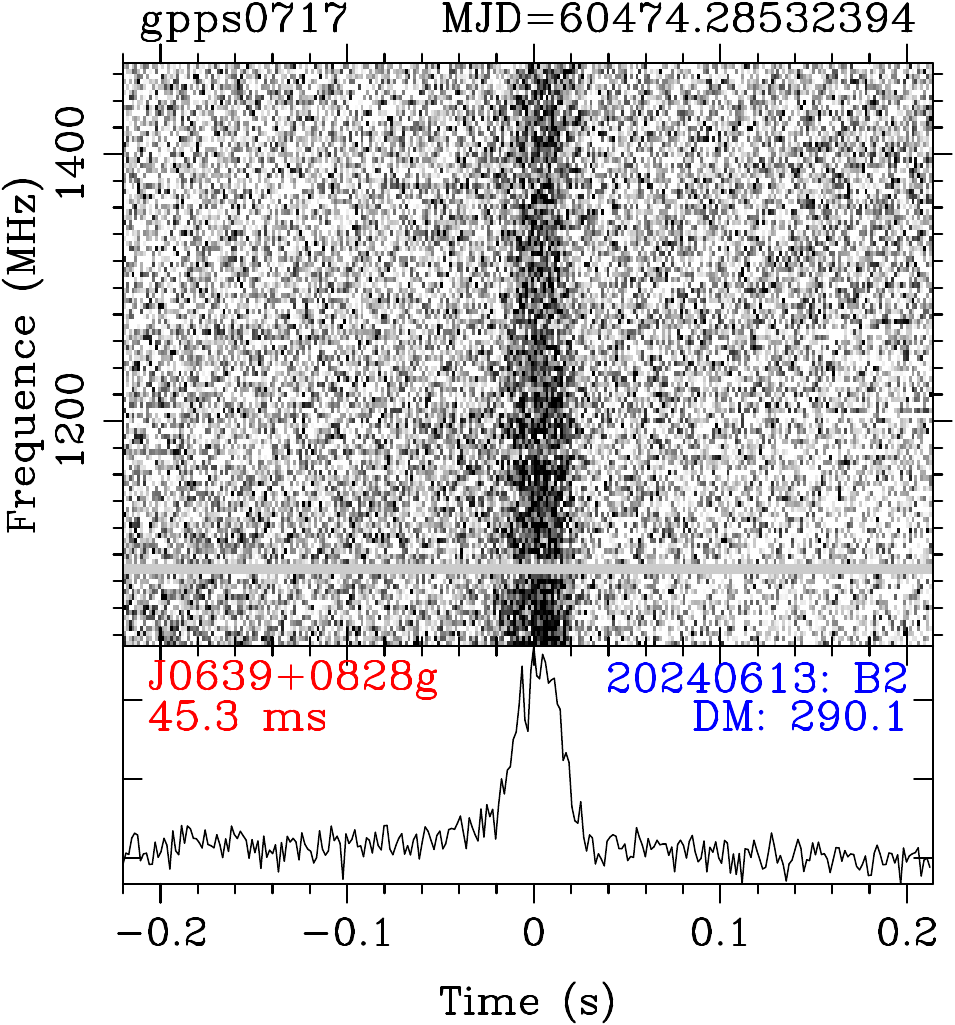}
\includegraphics[width=0.195\textwidth]{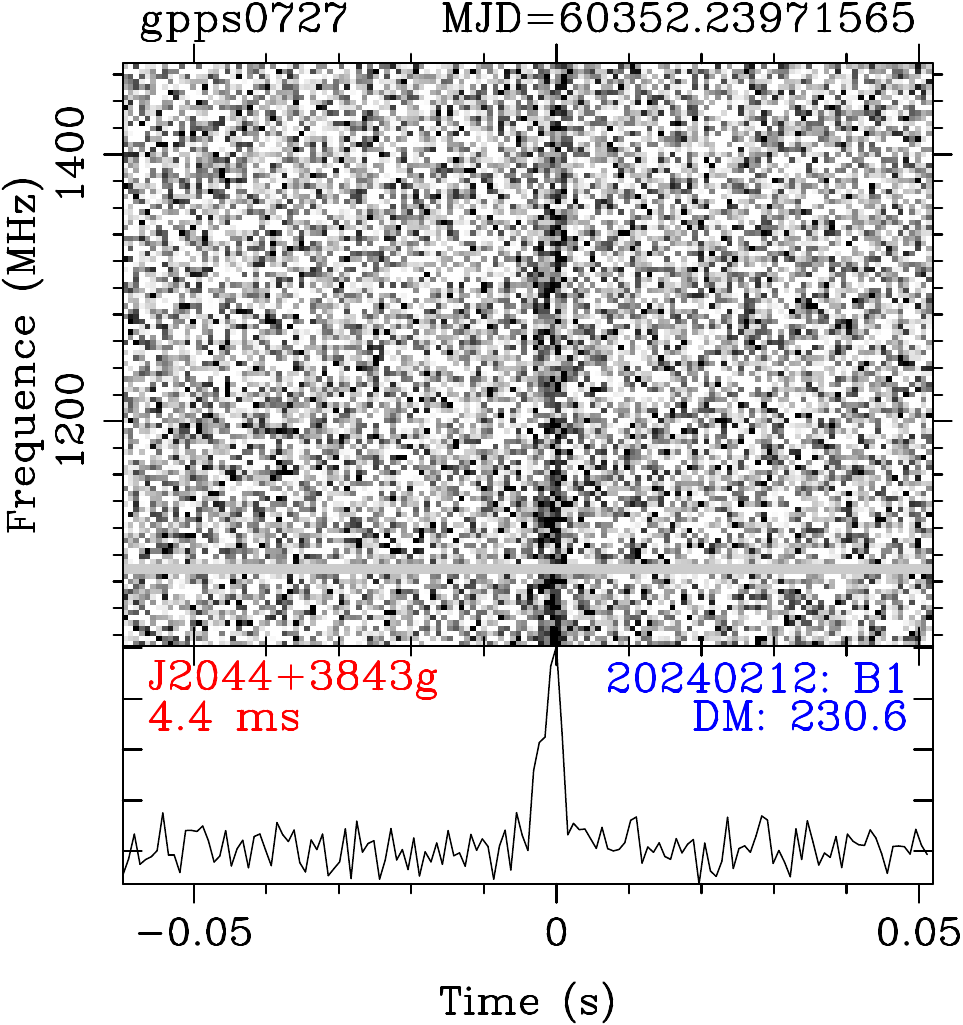}
\includegraphics[width=0.195\textwidth]{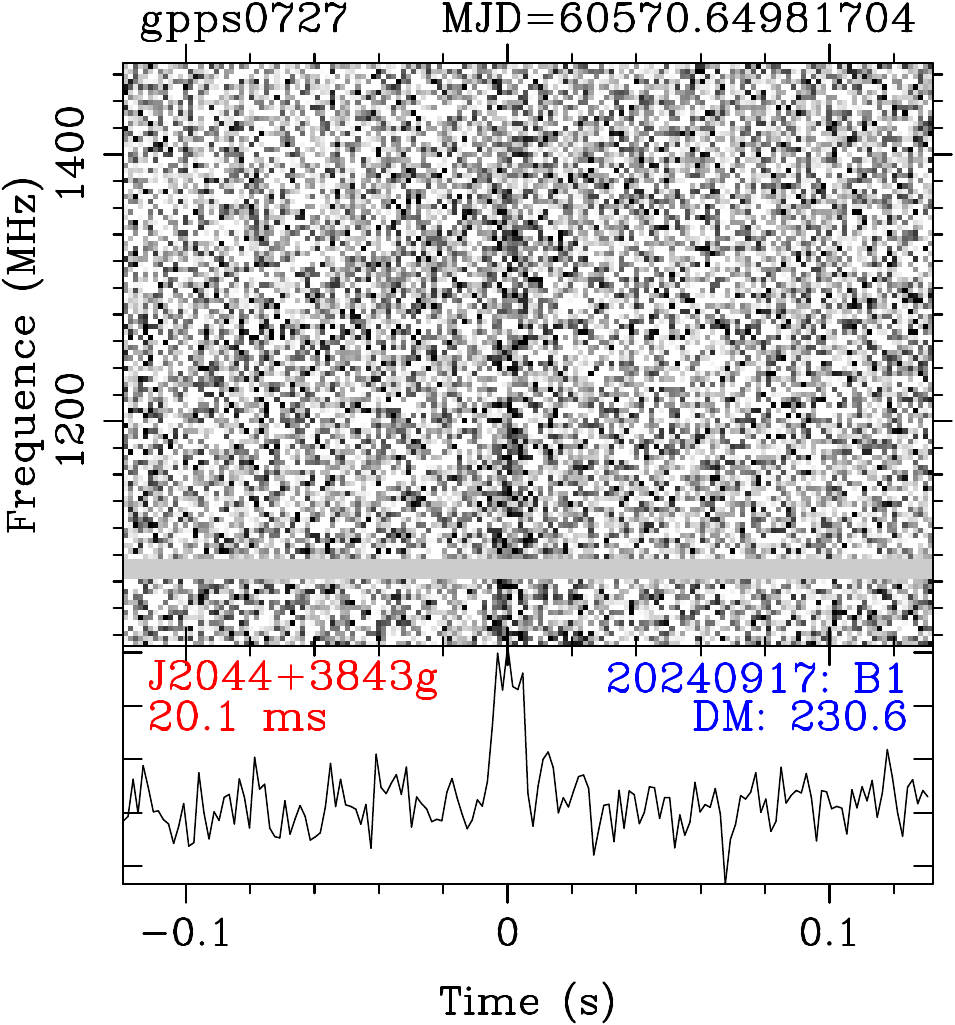}
%
\includegraphics[width=0.195\textwidth]{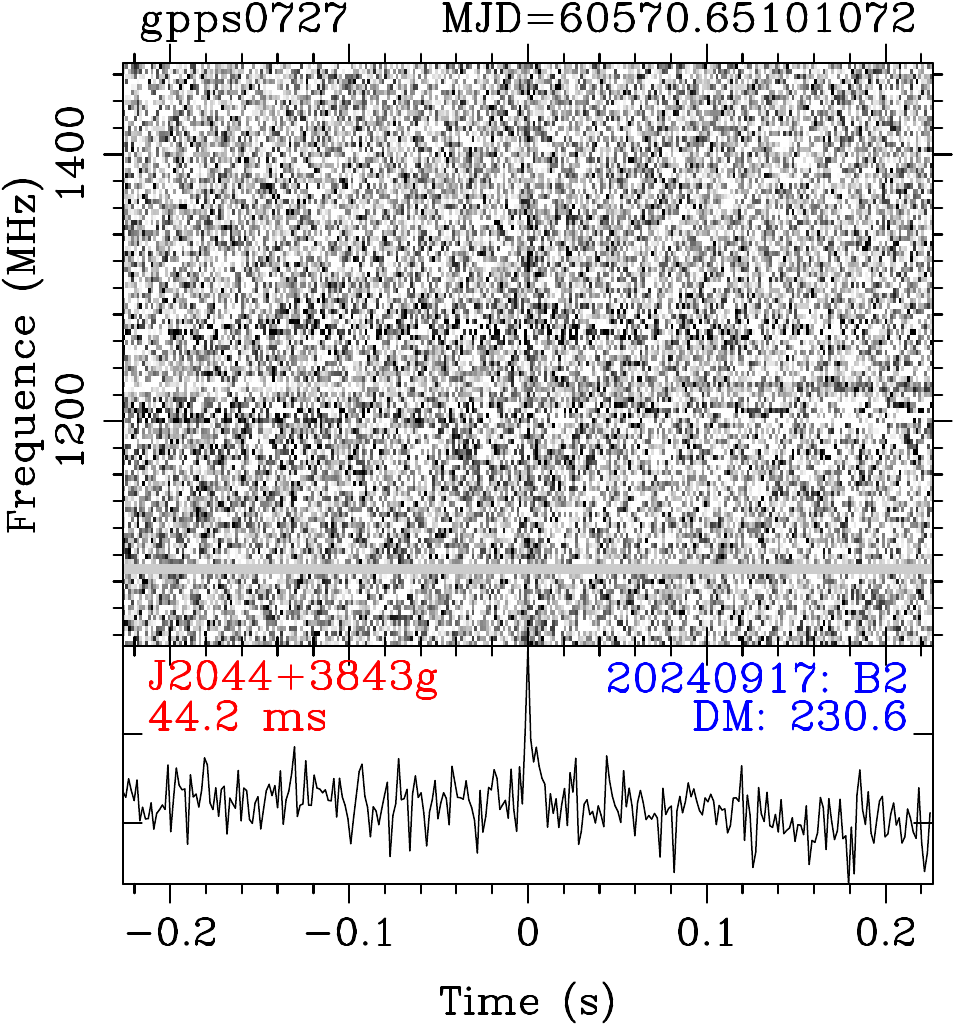}
\includegraphics[width=0.195\textwidth]{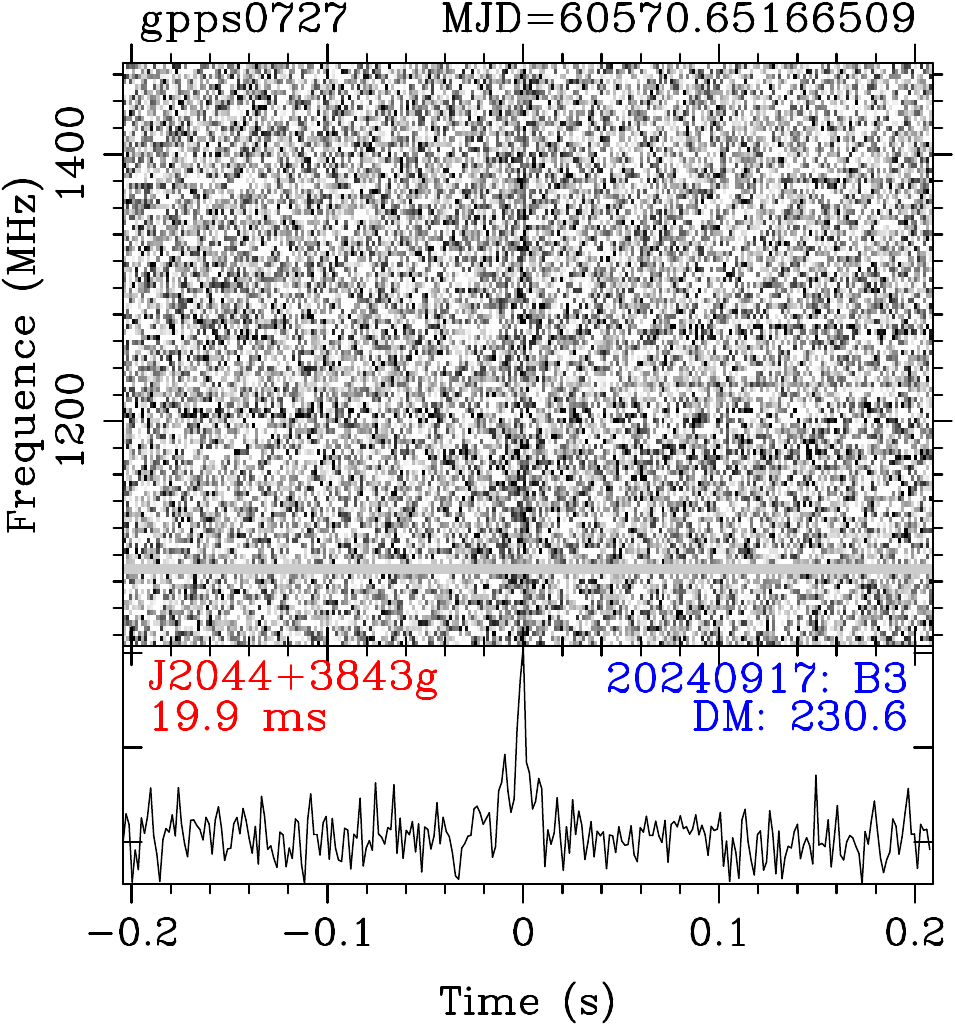}
\includegraphics[width=0.195\textwidth]{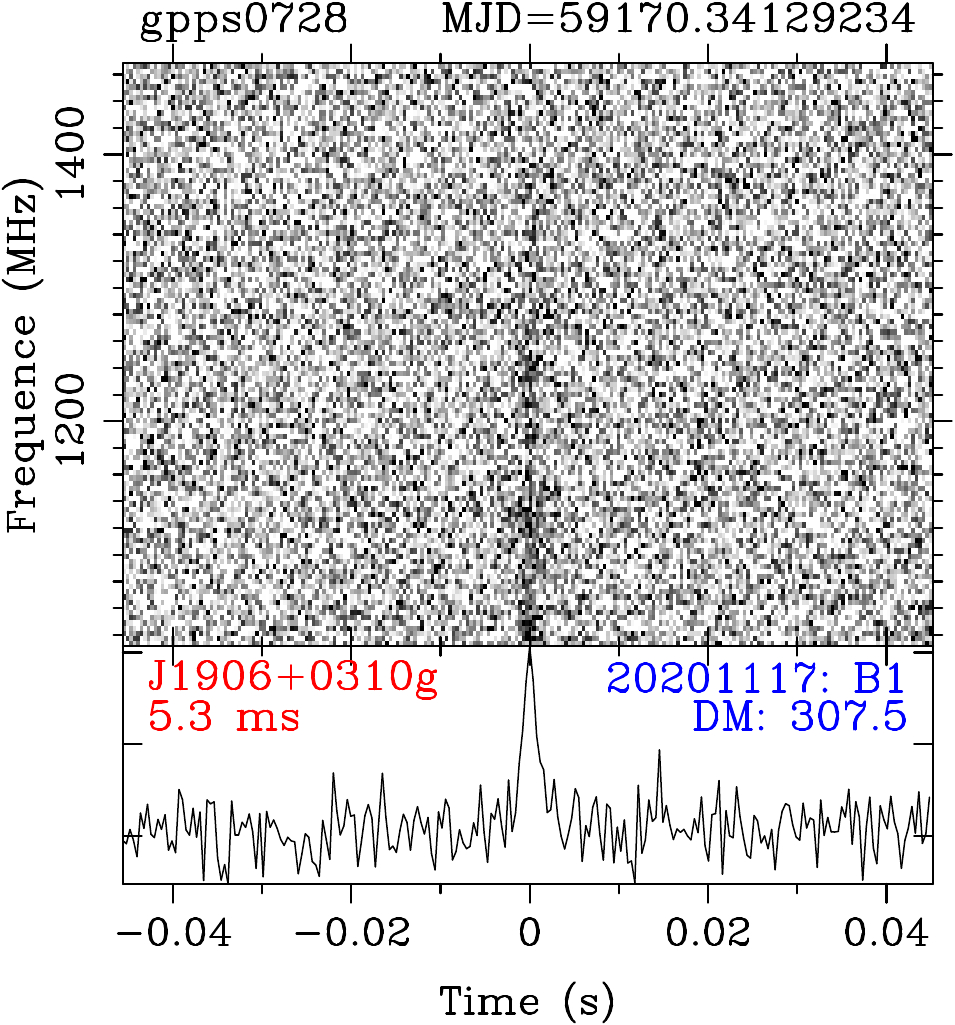}
\includegraphics[width=0.195\textwidth]{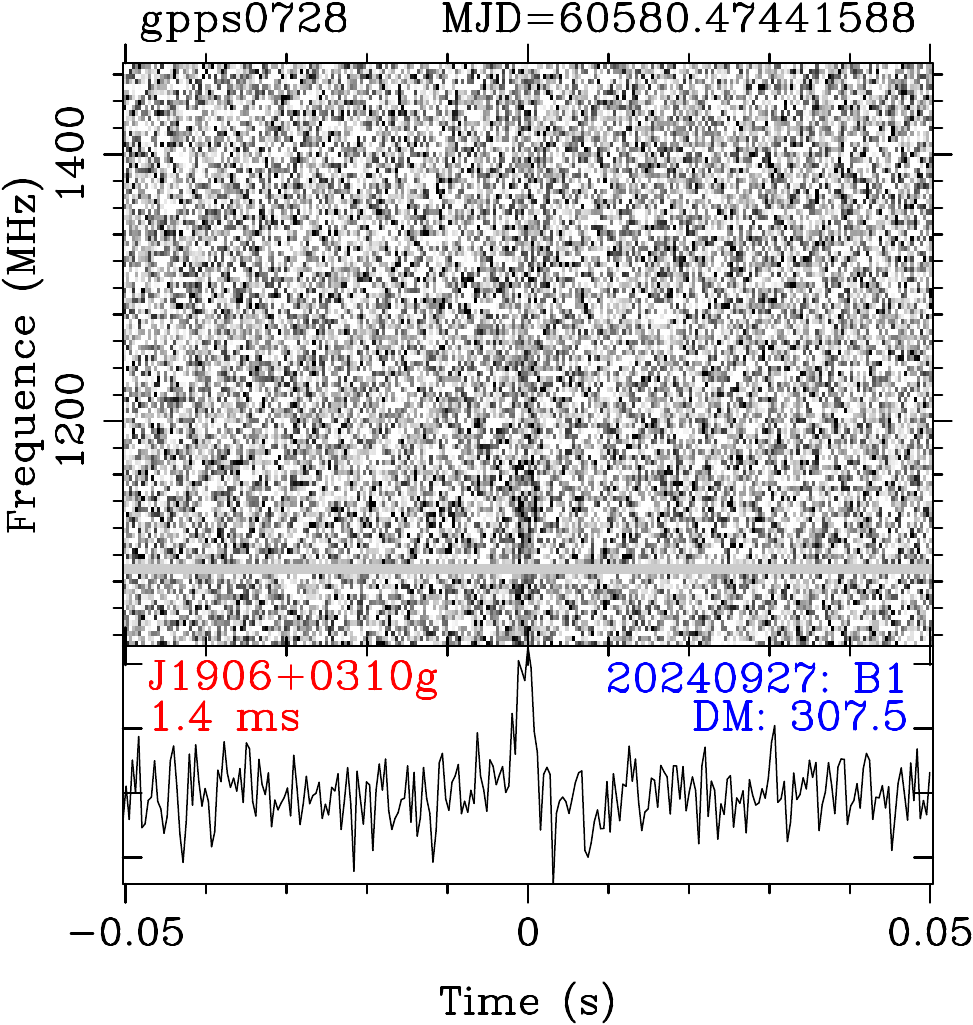}
\includegraphics[width=0.195\textwidth]{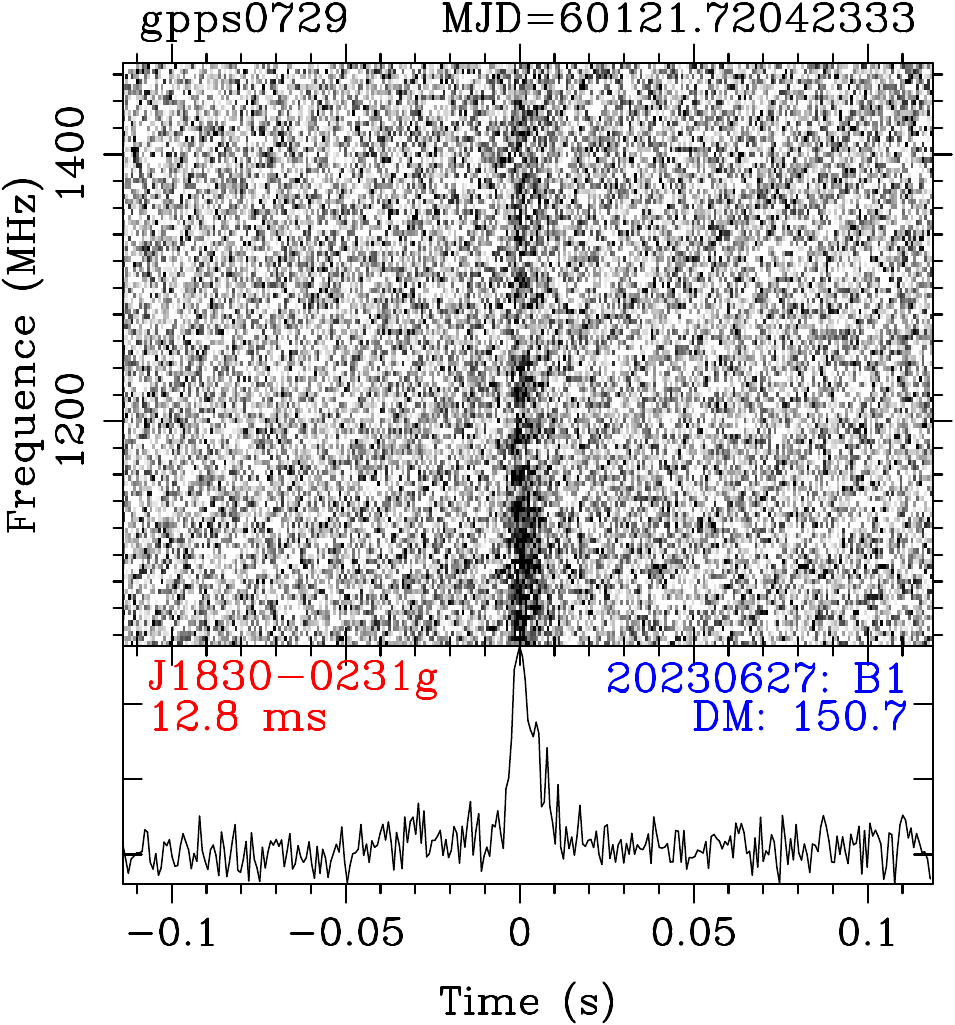}
\includegraphics[width=0.195\textwidth]{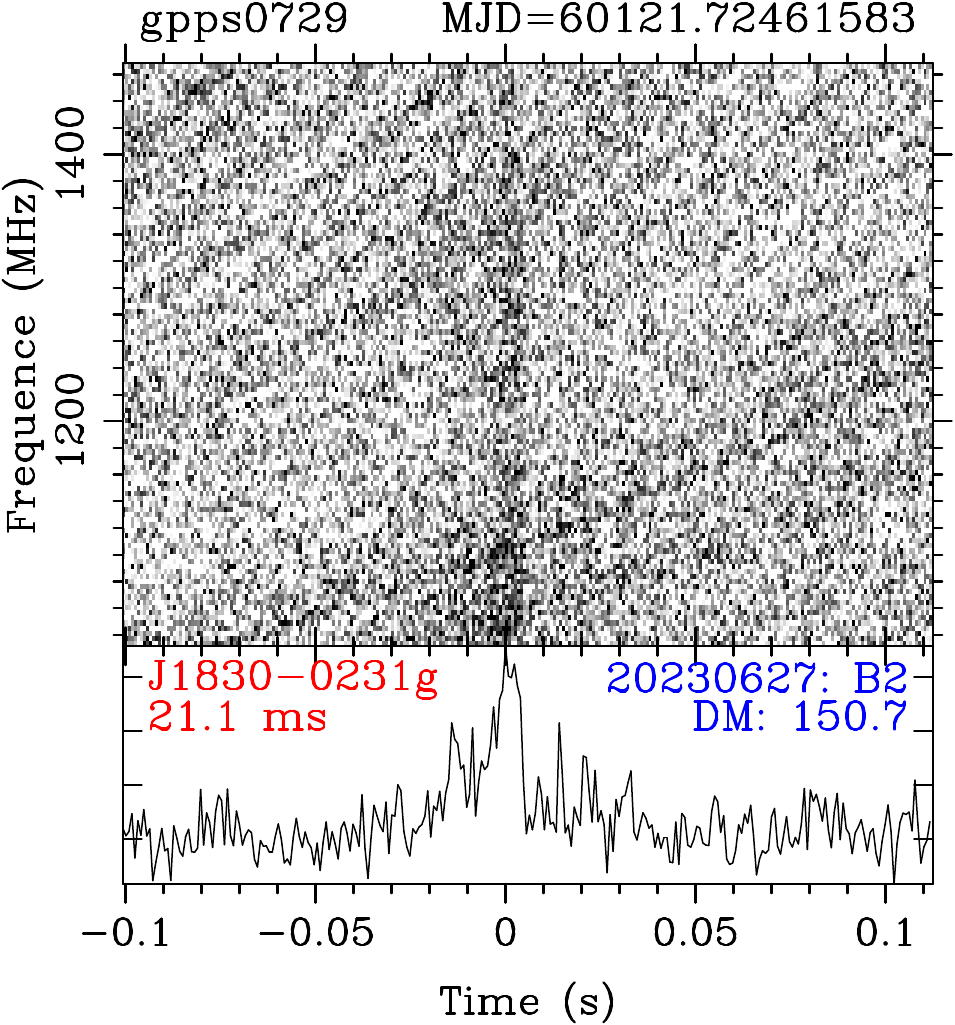}
\includegraphics[width=0.195\textwidth]{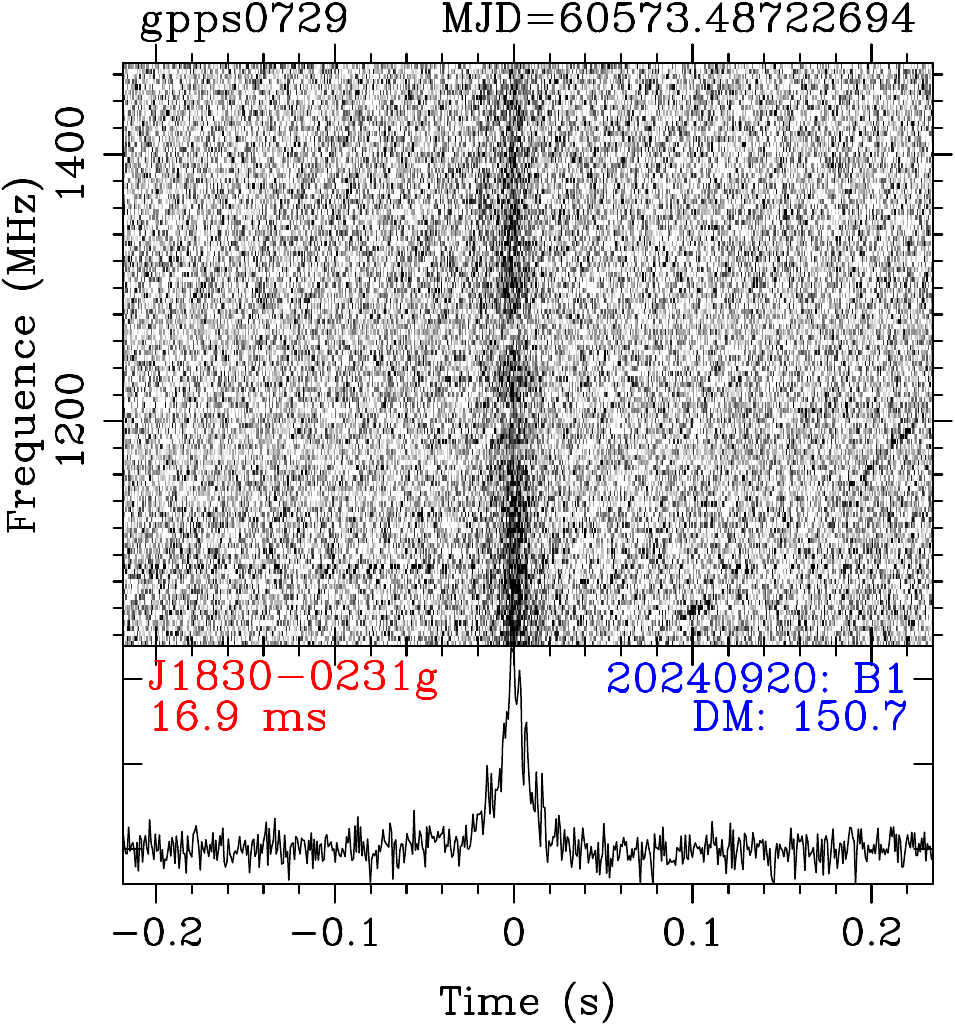}
\includegraphics[width=0.195\textwidth]{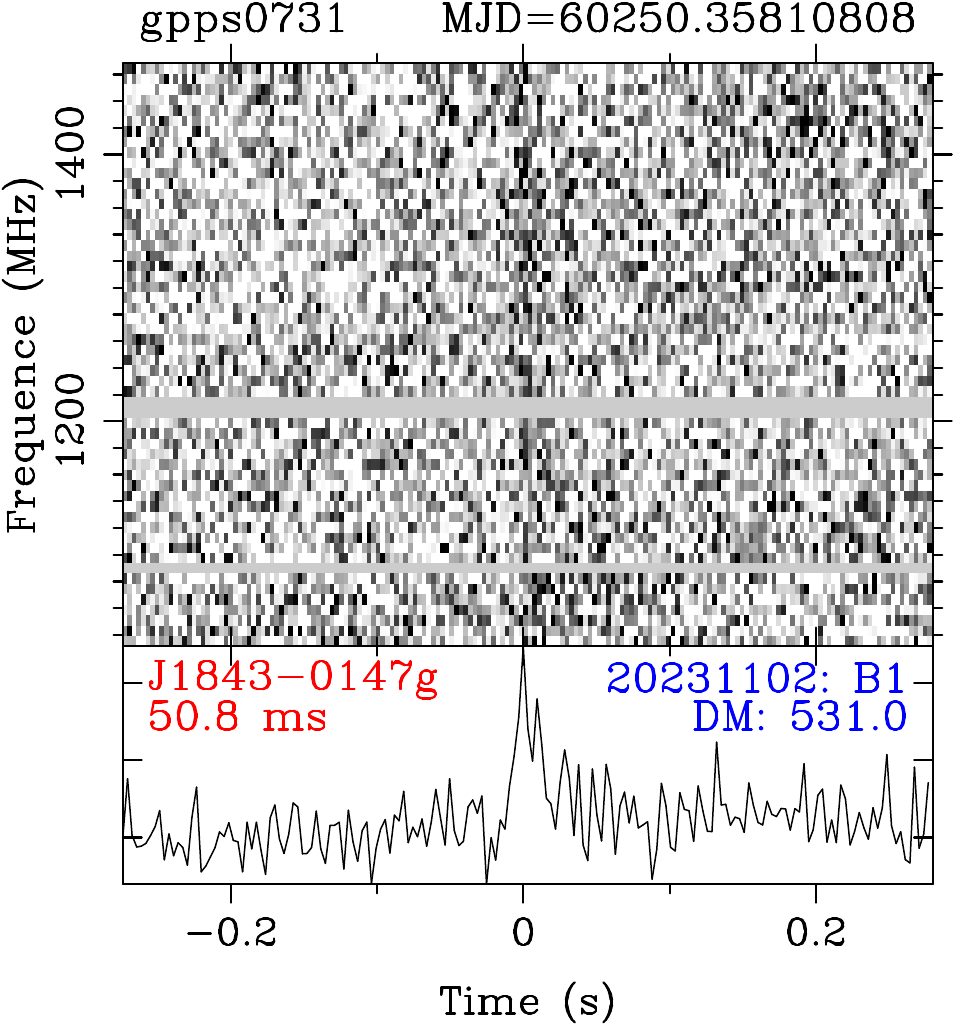}
\includegraphics[width=0.195\textwidth]{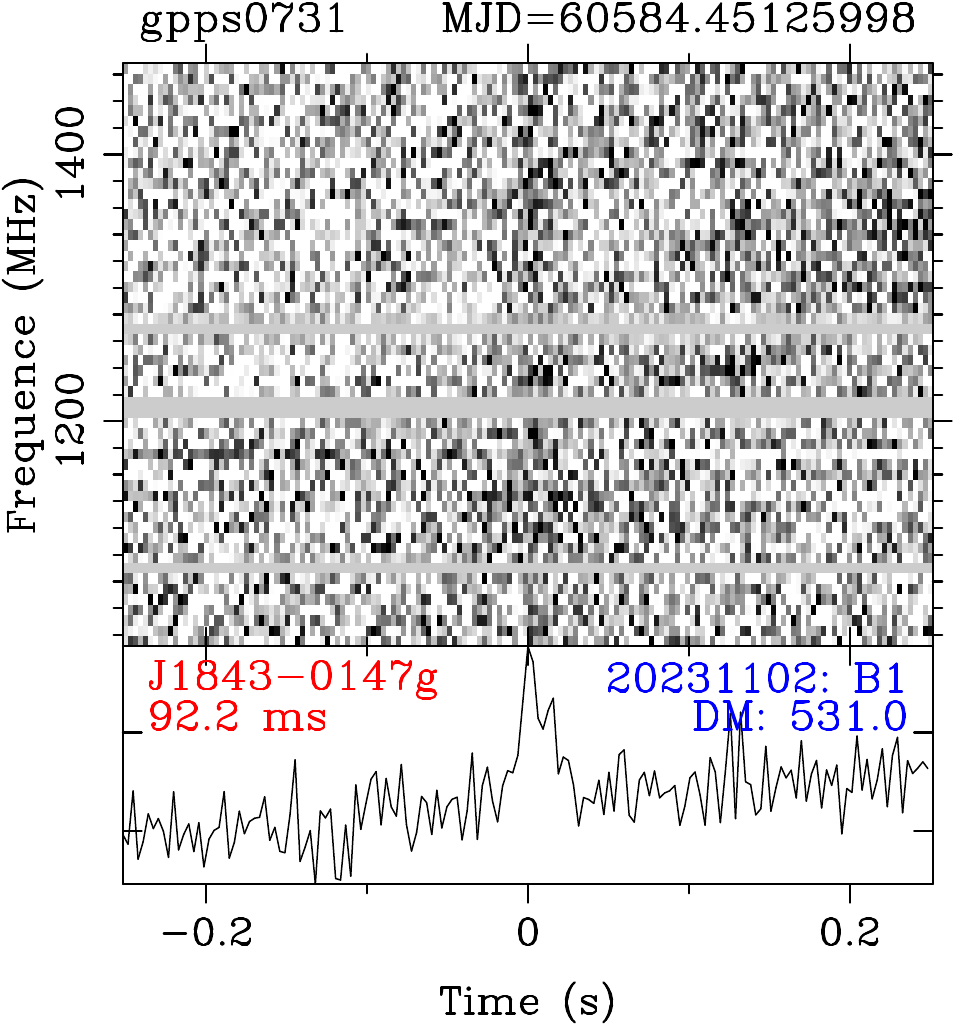}
\caption{The waterfall plots in the frequency–time dimension for pulses of the Galactic radio transients detected by FAST. The lower subpanel is the integrated pulse profile over all frequency channels.
}
\label{fig:AppfewPulses}
\end{figure*}

\begin{figure}
\centering
\includegraphics[width=0.48\columnwidth]{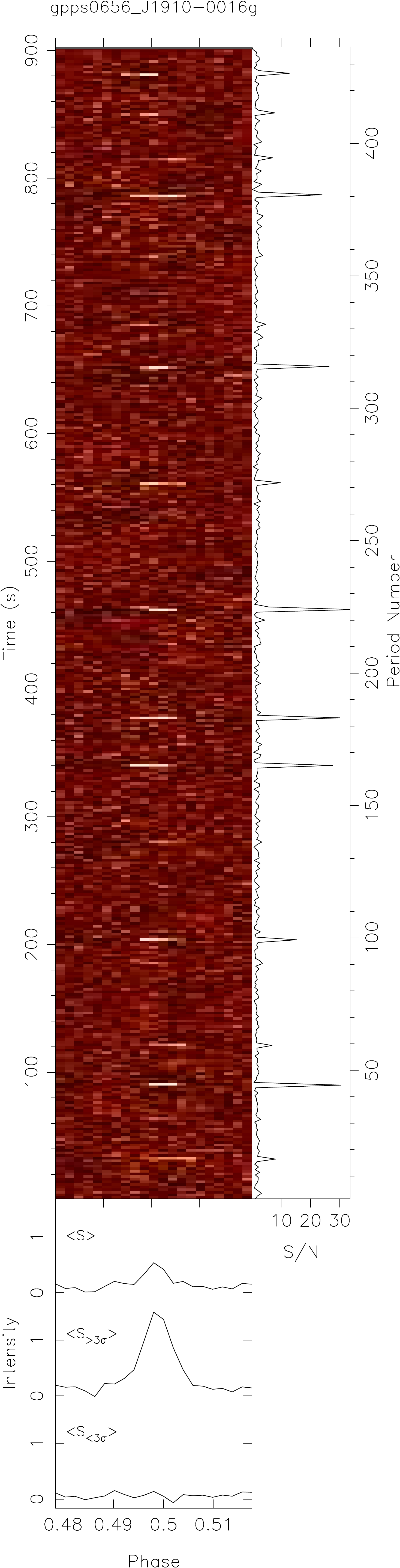}
\includegraphics[width=0.48\columnwidth]{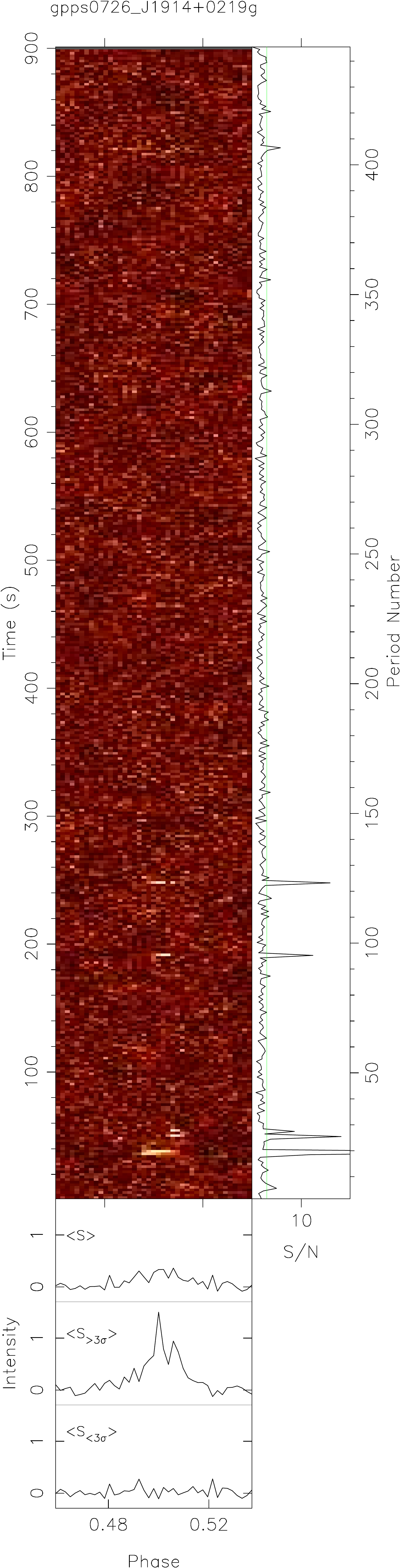}
\caption{he pulse-stacks of two RRATs newly discovered in the GPPS survey, Plots for 20 newly discovered RRATs are presented in Figure~A\ref{19RRATs} in the Appendix. The pulse-stack is shown in the main left panel, where only a few pulses occasionally emit. The right panel displays the curve of S/N over pulse number, with the sigma calculated from a given width of off-pulse phase range. Three subpanels below the main panel are the averaged profiles of all periods, and of single pulses with the S/N $>3$ and $<3$. 
}\label{exam2RRAT}
\end{figure}

\subsection{Newly discovered RRATs}

By using the specially designed single pulse searching module, we discovered and published 76 RRATs in \citep{zhx+23}. For completeness, we include them in Table~A\ref{gppsPSRtab1} with a superscript mark ``s" after the names. We noticed that the period of PSR J1828$-$0003 (gpps0501) has been updated by \citet{zbd+24} with a value of 3.8~s.

We have improved and updated the module after  \citet{zhx+23}, and processed the newly obtained survey data. We discovered 10 radio transients with only a few pulses detected, as listed in Table~\ref{tab:fewPulses} and shown in Figure~\ref{fig:AppfewPulses}. Their DM values signify that they are Galactic RRATs though the periods cannot be derived from the detection of limited number of pulses. We measure the pulse width $W_{\rm 50}$ in ms, and their fluence $F_{\rm \nu}$ in units of mJy\,ms at the central frequency $\nu=\rm1.25\,GHz$ by integrating the previously defined flux density $S_{\rm i}$ over the sampling time $t_{\rm bin}$ (measured in ms). Mathematically, it can be expressed as $F_{\rm\nu} = \sum S_{\rm i} \cdot~t_{\rm bin}$. The results for fluence are listed in Table~\ref{tab:fewPulses}.

We have discovered 20 new conventional RRATs, and present the pulse-stacks and their mean profiles for all observed periods, radiated periods and not significantly radiated periods, as shown for two examples in Figure~\ref{exam2RRAT}. The plots for all 20 RRATs are shown in Figure~A\ref{19RRATs} in the Appendix.

\begin{figure}
  \centering
 \includegraphics[width=0.7\columnwidth]{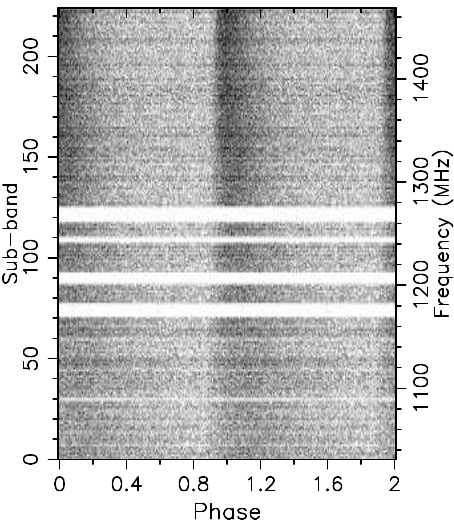}
  \caption{The water-fall plots in the frequency-phase dimension for a scattered pulsar PSR J1850$-$0050g (gpps0257, P = 221~ms, DM = 1072 pc~cm$^{-3}$).}
  \label{wf2psr_scatter}
\end{figure}

\begin{figure}
  \centering
  \includegraphics[width=0.8\columnwidth]{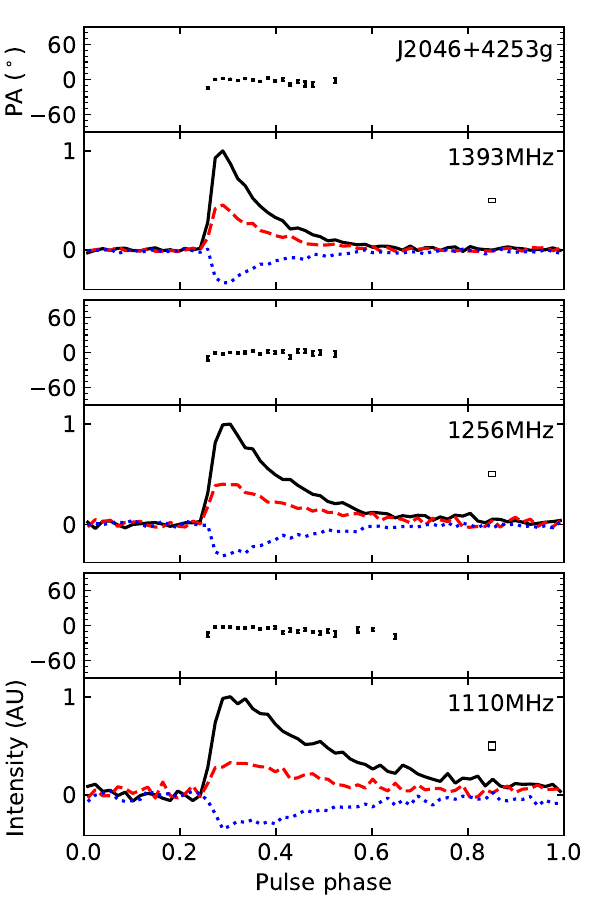}
  \caption{Polarization profiles of PSR J2046+4253g (gpps0464) at three subbands of FAST $L$-band observations. The lower subpanel shows the intensity profile in the solid line, linear polarization in the dashed line, and circular polarization in the dotted line. The linear polarization position angles (PAs) with error bars are displayed in the upper subpanel. The central frequency is marked in the lower subpanel, together with a box for the bin size and 1$\sigma$ of profiles. } 
\label{sub4scat}
\end{figure}

\subsection{Pulsars with wide and scattered profiles}

Looking at Figure~A\ref{gppsPSRprof}, one may find that many pulsars have very wide profiles, for example, PSRs J1908+0705g (gpps0278), J1905$-$0048 (gpps0367), and J2007+3343g (gpps0604). Some profiles are not only wide but also have many components, such as J1835$-$0011g (gpps0221) and J1931+2333 (gpps0675). Such wide profiles, according to the pulsar emission geometry \citep{lm88}, should be produced by viewing the radiation beam almost aligned to the rotation axis. Such a geometry can be verified by the polarization profiles.

Also impressive in Figure~A\ref{gppsPSRprof} are the scattered profiles of 64 GPPS pulsars, all of which have high DMs. We show the waterfall plot of the mean pulses in the frequency-phase dimension for two pulsars in Figure~\ref{wf2psr_scatter}, which demonstrates that pulsar searches at lower frequencies are not good for detecting these distant high-DM and highly scattered pulsars \citep{xwhh11} aand that a survey at 2–3 GHz is probably the best. Another issue is the polarization angle curves of the scattered pulsars, which are always very flat as discussed and predicated by \cite{lh03}.

FAST has great sensitivity and can detect the different tails of the scattered pulsars in the subbands of the 1.0–1.5 GHz band, as affirmed in Figure~\ref{sub4scat}. Conventionally, pulsar DM is determined by aligning the pulse peak, which is not good for the largely scattered profiles in subbands or channels. We developed a new method to determine DM, which is to align the front edge of subband profiles at the 1/4 level of the peak. See Jing et al. (2025, in preparation) for the FAST pulsar database for the scattered profiles of the 64 newly discovered GPPS pulsars and 61 previously known pulsars and determined scattering parameters.

\begin{figure*}
    \centering
    \includegraphics[width=0.7\columnwidth]{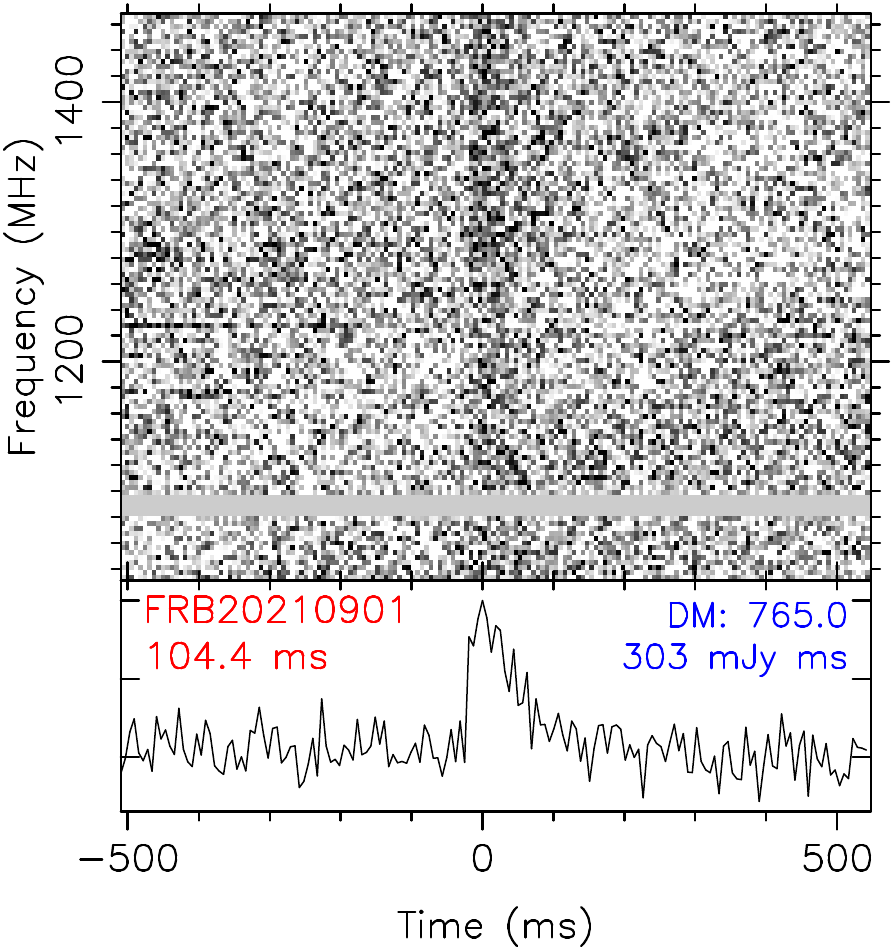}
    \includegraphics[width=0.7\columnwidth]{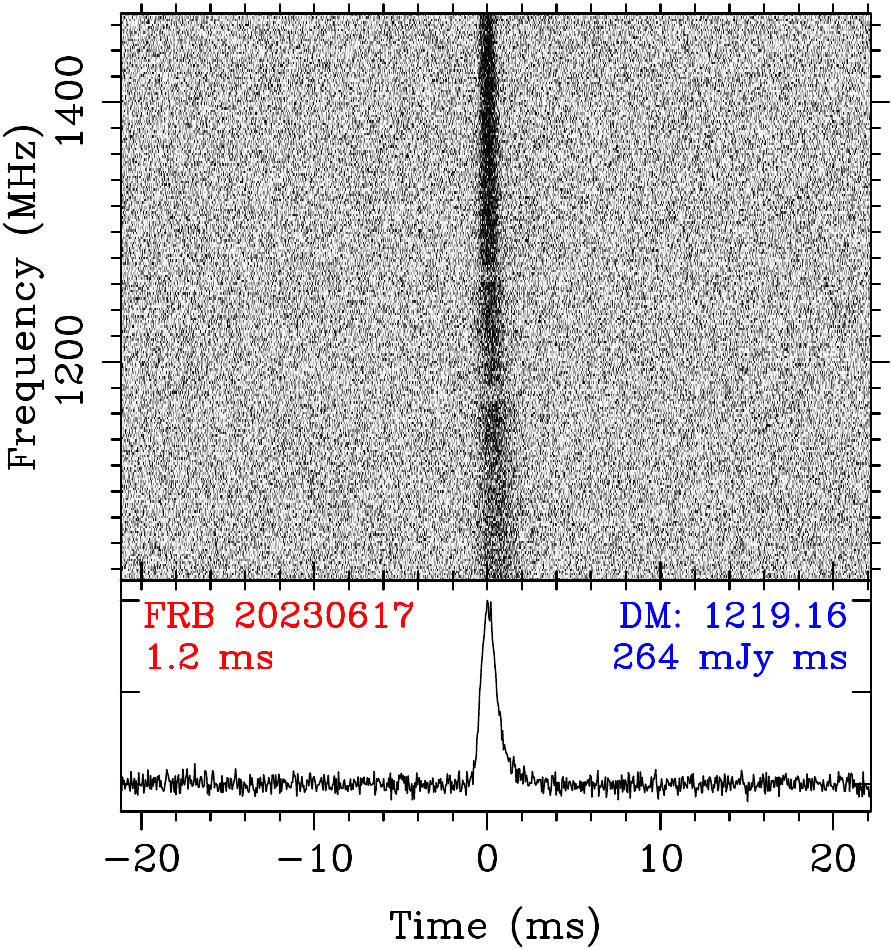}
    \includegraphics[width=0.7\columnwidth]{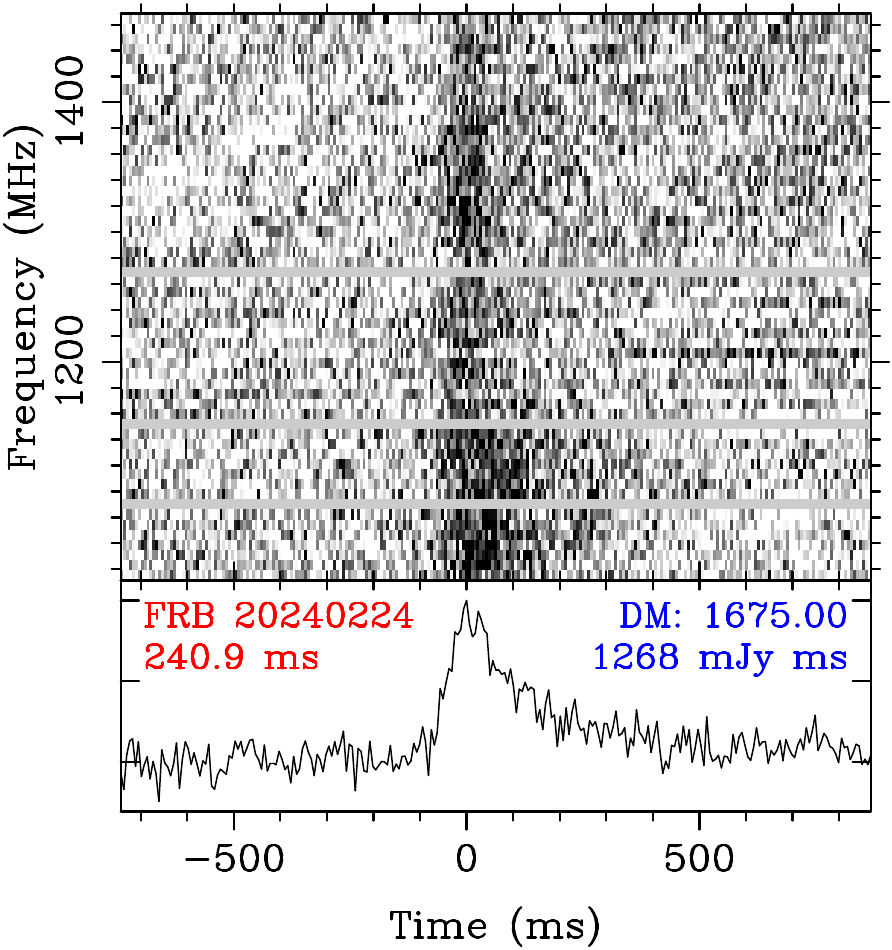}
    \caption{The dynamic spectra and profile for three new FRBs discovered in the FAST GPPS survey.}
    \label{3frb}
\end{figure*}

\begin{table}
\centering
\caption{Basic Parameters of the Newly Discovered Probable FRBs from the FAST GPPS Survey}
\label{tab3frb}
\setlength{\tabcolsep}{2pt}
\footnotesize
\begin{tabular}{lccc}
\hline
Name                         & FRB\,20210901       & FRB\,20230617       & FRB\,20240224     \\
                             & (J2045$+$4213)      & (J1839$+$0203)      & (J2102$+$4030)    \\
\hline
R.A. (hh:mm:ss)              & 20:45:09.3          & 18:39:39.7          & 21:02:31.2        \\
Dec ($\pm$dd:mm:ss)          & $+$42:13:03         & $+$02:03:33         & $+$43:30:23       \\
Galactic: $l$ ($^\circ$)     & 82.2919             & 33.4241             & 85.3245           \\
Galactic: $b$ ($^\circ$)     & -0.4236             & 3.5584              & -2.0334           \\
DM  (pc~cm$^{-3}$)           & 765$\pm$17          & 1219.16$\pm$0.78    & 1675$\pm$12       \\
DM$\rm_{MW}$ (pc~cm$^{-3}$)* & 376 / 422           & 528 / 459           & 326 / 335         \\
DM$\rm _{Ex}$ (pc~cm$^{-3}$)* & 389 / 343           & 691 / 760           & 1349 / 1340       \\
TOA (MJD)-59000              & 458.673323046       & 1111.742909304      & 1364.175237223    \\
W$_{\rm 50}$ (ms)           & 92.3                & 1.2                 & 240.9             \\
$S_{\rm peak}$ (mJy)         & 5.5                 & 238.2               & 7.8               \\
Fluence $F_\nu$ (mJy ms)     & 278.9               & 263.6               & 1268.3            \\
$\tau_{\rm 1GHz}$ (ms)       & $187\pm19$          & $1.07\pm0.04$       & $217\pm15$        \\
\hline
\multicolumn{4}{l}{Note: $^a$ Two values estimated by using the NE2001/YMW16 models.}
\end{tabular}
\end{table}

\begin{table}
\centering
\caption{Observation Sessions for the Two Newly Discovered FRBs and One Probable Case.}
\label{tab:frbobs}
\setlength{\tabcolsep}{9.0pt}
\footnotesize
\begin{tabular}{lccrr}
 \hline\noalign{\smallskip}
 Name        & Date     & MJD   & T$_{\rm obs}$    & Burst \\
              &          &       & (min)           & No.   \\
 \hline  
 FRB\,20210901& 20210901 & 59458 & 5              & 1     \\
              & 20211004 & 59491 & 15              & 0     \\
              & 20220903 & 59825 & 40              & 0     \\
              & 20230105 & 59949 & 40              & 0     \\
              & 20230117 & 59961 & 40              & 0     \\
              & 20230307 & 60010 & 25              & 0     \\
              & 20230409 & 60043 & 40              & 0     \\
              & 20230423 & 60057 & 40              & 0     \\
              & 20230514 & 60078 & 40              & 0     \\
              & 20240814 & 60536 & 50              & 0     \\
 FRB\,20230617& 20230617 & 59240 & 5               & 1     \\
              & 20241110 & 60624 & 50              & 0     \\
 FRB\,20230224& 20230224 & 59240 & 5               & 1     \\
\hline
\end{tabular} 
\end{table}


\begin{figure*}
    \centering
    \includegraphics[width=0.9\textwidth]{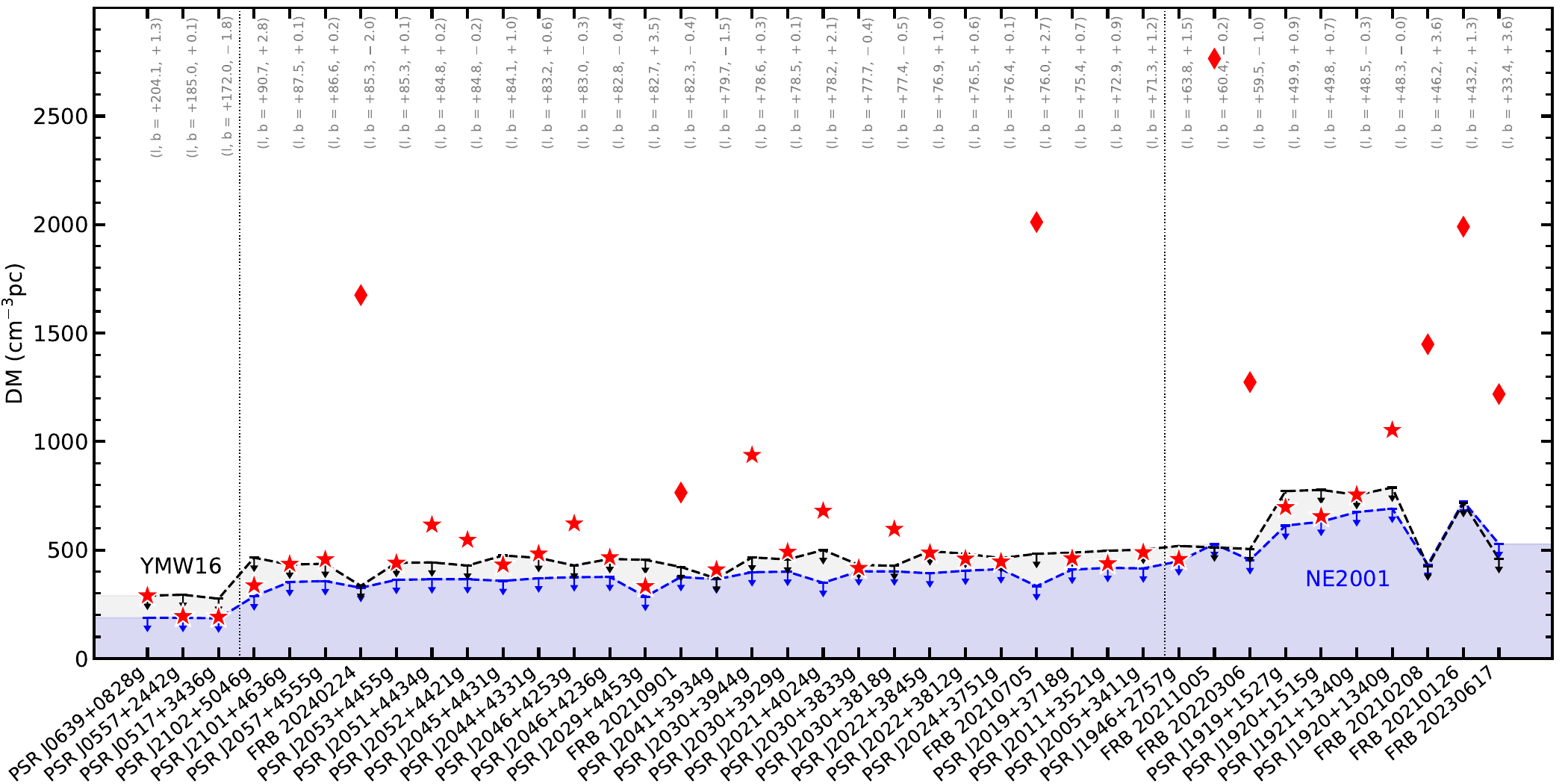}
    \caption{Pulsars (stars) and FRBs (diamonds) discovered by the FAST GPPS survey with DM excesses from the model predictions by NE2001 \citep{cl02} {\it or} YMW16 \citep{ymw17}.}
    \label{fig:dmexcess}
\end{figure*}

\begin{figure*}
    \centering
    \includegraphics[width=0.9\textwidth]{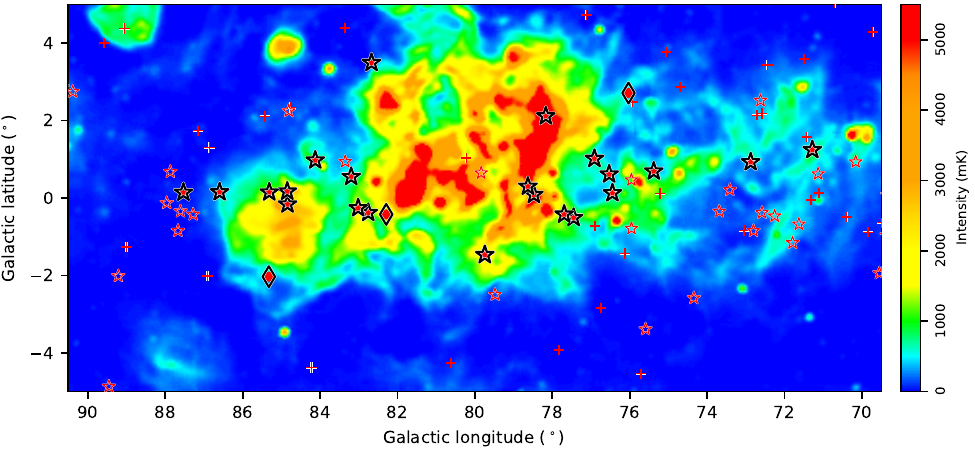}
    \caption{The radio continuum maps of the Galactic plane at the 6~cm wavelength from \citet{gsh+11} around the local arm with newly discovered FRBs (in diamonds) and pulsars (in stars). Previously known pulsars are plotted as ``+''. Filled symbols with a black frame stand for those with excessive DMs from the model predictions by NE2001 \citep{cl02} and YMW16 \citep{ymw17}.}
    \label{fig:localarm}
\end{figure*}

\subsection{Weakest pulsars}

Looking at Figure~\ref{flux_lum} and Table~A\ref{gppsPSRtab1}, we find that many pulsars we have discovered are among the weakest, i.e., very low flux densities. Note, however, that the flux density is the averaged value of the detected pulse energy over observation time. Therefore, RRATs and very nulling pulsars, as discussed above, deliver a few pulses over many minutes and are the weakest pulsars.

If talking about normal pulsars, not as extreme as RRATs, we find that a longer observation time is important for detecting weak pulsars. For example, PSR J1857+0249g (gpps0276) and PSR J1903+0830 (gpps0531) were first detected from another beam of these 15 minute tracking verification observations, not from the general 5 minute snapshot observations. Some weaker pulsars, such as PSR J1840$-$0245g (gpps0313), were first detected via the single pulse search module and then from a longer verification observation. PSR J1943+2205g (gpps0514) was discovered during the 1.5 hr tracking of a binary pulsar.

There is No doubt that there are {\it more} weak pulsars in the Milky Way than we have detected, so a longer integration time is needed to uncover them for a given telescope. 

\subsection{Two FRBs with one probable case}

FRBs are generally detected as dispersed radio pulses \citep{lbm+07}, and they appear not much different from single pulses of pulsars \citep{zhz+22}. Some FRBs are repeating, and most detected FRBs are one-off events \citep{cc19, pgk+21}. There have been 10 FRBs discovered by FAST \citep{zll+20, nll+21, nal+22, zhj+23}, and 5 of them from the FAST GPPS survey data at the low Galactic latitudes \citet{zhj+23}.

Processing newly observed FAST GPPS survey data by using the single pulse-searching module \citep{zhj+23}, we have found three new one-off pulses with DMs much larger than the prediction of maximum DMs by the Galactic electron density distribution models \citep{cl02,ymw17}. Two of them are located at a slightly high Galactic latitude, with large excess of DM, and hence are extragalactic FRBs, and we name them FRB 20230617, and FRB 20240224. One probable case is named FRB 20210901. Their waterfall plots in the frequency–time dimensions are presented in Figure~\ref{3frb}, and their physical parameters are given in Table~\ref{tab3frb}. 

FRB20210901 has a DM of 765 pc~cm$^{-3}$ located at about the direction of the Galactic coordinates of (gl, gb) = (82\degr.3, -0\degr.4), where the predicated DM from the medium in our Galaxy is 376 or 422 pc~cm$^{-3}$ according to the Galactic electron density distribution models NE2001 \citep{cl02} or YMW16 \citep{ymw17}. In general, any pulses with a DM of 765 pc~cm$^{-3}$, with an excess DM of 389 or 343 pc~cm$^{-3}$ should be the extragalactic origin. However, it
is located near the Galactic plane where the local arm is laid. 
We have discovered pulsars in the nearby directions with excessive DMs \citep{hww+21}, see also Fig.~\ref{fig:dmexcess}. For example PSR J2046+4236g (gpps0625, GL=82.7557\degr, GB=0.3808\degr) with a DM of about 466 pc~cm$^{-3}$, PSR J2046+4253g (gpps0464, GL=83\degr.0161, GB = -0\degr.2593) with a DM of 622 pc~cm$^{-3}$. Therefore, we are not confident in claiming this burst is extragalactic origin. Therefore, we are not confident in claiming this burst is of extragalactic origin. Nevertheless, given the non-detection in so many hours of follow-up observations as listed in Table~\ref{tab:frbobs}, we believe it is probably an FRB.

We noticed that all three FRBs show the scattering effect, i.e., the more extended tails at lower frequencies (see Figure~\ref{3frb}). The scattering parameters are listed in Table~\ref{tab3frb}. 
We concurrently fit for the scattering times $\tau$ of different subbands utilizing the relationship $\tau_\nu = \tau_{\rm1GHz} \nu^{-4}$ \citep{nab+13}. The scattering times at 1~GHz are listed in Table~\ref{tab3frb}.

\subsection{Pulsars with excessive DMs from the models}

One can see that in Fig.~\ref{fig:dmexcess} there are a number of pulsars that have DM values exceeding the model predictions by NE2001 \citep{cl02} or YMW16 \citep{ymw17}. Those pulsars are inside the Milky Way, and most of them have slightly larger DM than the maximum DMs given by NE2001 but smaller than those given by YMW16. Two pulsars, PSRs J2046+4253g and J1920+1340g, have DMs of 622 and 1053 pc~cm$^{-3}$ respectively, much larger than the model values caused by the unpredicted clouds in the spiral arms. 
PSR J1920+1340 is at the Galactic coordinates of ($l,b$) = ($48\degr.3345$, $-0\degr.0051$), probably behind the Sagittarius Arm and the Perseus arm. 
PSR J2046+4253g at ($l,b$) = ($83\degr.0161$, $-0\degr.2593$) as several pulsars in \citet{hww+21}, is behind the local arm and has an excessive DM due to uncounted clouds (see Fig.\ref{fig:localarm}).

\section{Summary and perspectives}

The super-sensitive FAST, mounted with the $L$-band 19-beam receiver, is currently the most powerful tool to discover pulsars. We have conducted the FAST-GPPS survey for about 5yr and finished observations of 4359 covers among the planned 18,413 covers. In total, we have found 751 pulsars, including 157 binary pulsars and 107 RRATs. In addition to previous publications by \citet{hww+21} and \citet{zhx+23}, in this paper, we present data on 473 new pulsars, including 137 MPSs and 30 new RRATs. In addition to five FRB published by \citet{zhj+23}, we here publish two FRBs with one probable case, whose DM values exceed the upper limits given by the Galactic electron density distribution models \citep{ymw17, ne2001},

Most pulsars we have discovered are fairly weak, including many weakest ones, with a flux density down to the $\mu$Jy level. During some verification observations with longer integration time, some sub-$\mu$Jy have also been discovered. With the help of the estimated distances from the Galactic electron density distribution models \citep{ymw17, ne2001}, we found that our newly discovered pulsars make a decisive contribution to the lower end of the luminosity distribution of pulsars.

Our survey detects 177 millisecond pulsars, nearly 20\% of the total number, which is a much higher fraction of millisecond pulsars in the Galactic fields, partially because of the improved sensitivity and partially due to the lack of millisecond pulsars with high flux densities.

Because of using the independent single pulse searching module, we found eight long-period pulsars with a period greater than 10 s, and also 100 RRATs. These new pulsars are hard to detect during normal periodical signal searches. The results suggest that there are many more slow neutron stars missing in previous surveys. There are many RRATs in our Milky Way, about 15\%, which occasionally emit strong pulses and are missed also in previous surveys.

With polarization measurements of these newly discovered pulsars, including RRATs, one can get their RMs. Together with DMs, the details of the Galactic magnetic fields can be better revealed, as shown by \citep{xhw+22}. 

n the sky region where the FAST-GPPS survey or verification observations have been observed, there are 1288 known pulsars. We detected most of them, but some have not been detected due to various reasons. The parameters of 46 previously known pulsars have been improved based on the FAST observations. 

In the future, we have to finish the planned FAST-GPPS survey. We understand that the pulsar discovery rate, i.e., the number of new pulsars discovered per 100 observation hours, becomes smaller when the observation pointings go away from the Galactic plane. The survey results should be fundamental to outlining the distribution of generated neutron stars along the Galactocentric radius.

\section*{Acknowledgements}
We are grateful to the referee, Prof. Duncan Lorimer, for the careful reading and helpful comments. This work made use of the data from FAST (https://cstr.cn/31116.02.FAST). FAST is a Chinese national mega-science facility, operated by National Astronomical Observatories, Chinese Academy of Sciences. We all appreciate the excellent performance of FAST and the operation team. The authors are supported by the National Natural Science Foundation of China (NSFC, Grant Nos. 11988101, 12133004 and 11833009) and also the National SKA Program of China 2020SKA0120100. The pulsar searching team is specially supported by the Chinese Academy of Sciences via the project JZHKYPT-2021-06.

\section*{Authors contributions}  
The FAST GPPS survey is a key FAST science project led by J.~L. Han. He organized the teamwork for the survey and follow-up observations, processed the survey data, and discovered these pulsars, supervised all the work;   
D.~J. Zhou developed the single pulse searching module, and processed all GPPS survey data, and found RRATs and FRBs; he also updated the core of searching software to speed up calculations;  
W.~Q. Su patiently processed prepared pulsar profiles from FAST observation data, and prepared the big pulsar table published in this paper; 
C. Wang initialized the FAST snapshot observations, feeds all targets for the FAST GPPS observations, and worked all survey plane; 
Y. Yan worked on known pulsars and updated their parameters; 
P.~F. Wang initialized the pulsar search package and polarization calibration procedures; 
T. Wang worked on the data preparation procedures for pulsar searching; 
W. C. Jing jointly verified many new GPPS pulsars, prepared the web-page, and also made plots of parameter distribution published in this paper; 
P.~F. Wang and Jun Xu made fundamental contributions to the construction and maintenance of the computer clusters for pulsar searching; 
Z. L. Yang jointly verified many newly discovered GPPS pulsars;
J.~H. Sun and Q.~L. Yang realized the snapshotZ observation mode together with Chen Wang with the design given by J.~L. Han; 
Other people jointly propose or contribute to the FAST key project.
All authors contributed to the finalization of this paper.  \\

\section*{Data Availability}
 
Original FAST observational data will be open resources according to the FAST data 1-year protection policy. The folded and calibrated pulsar profiles presented in this paper can be found on the webpage: \url{http://zmtt.bao.ac.cn/GPPS/}.

\bibliographystyle{aasjournal}

\begin{thebibliography}{}
\expandafter\ifx\csname natexlab\endcsname\relax\def\natexlab#1{#1}\fi
\providecommand{\url}[1]{\href{#1}{#1}}
\providecommand{\dodoi}[1]{doi:~\href{http://doi.org/#1}{\nolinkurl{#1}}}
\providecommand{\doeprint}[1]{\href{http://ascl.net/#1}{\nolinkurl{http://ascl.net/#1}}}
\providecommand{\doarXiv}[1]{\href{https://arxiv.org/abs/#1}{\nolinkurl{https://arxiv.org/abs/#1}}}

\bibitem[{{Abdo} {et~al.}(2009){Abdo}, {Ackermann}, {Ajello}, {Atwood},
  {Axelsson}, {Baldini}, {Ballet}, {Band}, {Barbiellini}, {Bastieri},
  {Battelino}, {Baughman}, {Bechtol}, {Bellazzini}, {Berenji}, {Bignami},
  {Blandford}, {Bloom}, {Bonamente}, {Borgland}, {Bouvier}, {Bregeon}, {Brez},
  {Brigida}, {Bruel}, {Burnett}, {Caliandro}, {Cameron}, {Caraveo},
  {Casandjian}, {Cavazzuti}, {Cecchi}, {Charles}, {Chekhtman}, {Cheung},
  {Chiang}, {Ciprini}, {Claus}, {Cohen-Tanugi}, {Cominsky}, {Conrad}, {Corbet},
  {Costamante}, {Cutini}, {Davis}, {Dermer}, {de Angelis}, {de Luca}, {de
  Palma}, {Digel}, {Dormody}, {do Couto e Silva}, {Drell}, {Dubois}, {Dumora},
  {Farnier}, {Favuzzi}, {Fegan}, {Ferrara}, {Focke}, {Frailis}, {Fukazawa},
  {Funk}, {Fusco}, {Gargano}, {Gasparrini}, {Gehrels}, {Germani}, {Giebels},
  {Giglietto}, {Giommi}, {Giordano}, {Glanzman}, {Godfrey}, {Grenier},
  {Grondin}, {Grove}, {Guillemot}, {Guiriec}, {Hanabata}, {Harding}, {Hartman},
  {Hayashida}, {Hays}, {Healey}, {Horan}, {Hughes}, {J{\'o}hannesson},
  {Johnson}, {Johnson}, {Johnson}, {Johnson}, {Kamae}, {Katagiri}, {Kataoka},
  {Kawai}, {Kerr}, {Kn{\"o}dlseder}, {Kocevski}, {Kocian}, {Komin}, {Kuehn},
  {Kuss}, {Lande}, {Latronico}, {Lee}, {Lemoine-Goumard}, {Longo}, {Loparco},
  {Lott}, {Lovellette}, {Lubrano}, {Madejski}, {Makeev}, {Marelli},
  {Mazziotta}, {McConville}, {McEnery}, {McGlynn}, {Meurer}, {Michelson},
  {Mitthumsiri}, {Mizuno}, {Moiseev}, {Monte}, {Monzani}, {Moretti},
  {Morselli}, {Moskalenko}, {Murgia}, {Nakamori}, {Nolan}, {Norris}, {Nuss},
  {Ohno}, {Ohsugi}, {Omodei}, {Orlando}, {Ormes}, {Ozaki}, {Paneque},
  {Panetta}, {Parent}, {Pelassa}, {Pepe}, {Pesce-Rollins}, {Piron}, {Porter},
  {Poupard}, {Rain{\`o}}, {Rando}, {Ray}, {Razzano}, {Rea}, {Reimer}, {Reimer},
  {Reposeur}, {Ritz}, {Rochester}, {Rodriguez}, {Romani}, {Roth}, {Ryde},
  {Sadrozinski}, {Sanchez}, {Sander}, {Saz Parkinson}, {Scargle}, {Schalk},
  {Sellerholm}, {Sgr{\`o}}, {Shaw}, {Shrader}, {Sierpowska-Bartosik},
  {Siskind}, {Smith}, {Smith}, {Spandre}, {Spinelli}, {Starck}, {Stephens},
  {Strickman}, {Strong}, {Suson}, {Tajima}, {Takahashi}, {Takahashi}, {Tanaka},
  {Thayer}, {Thayer}, {Thompson}, {Tibaldo}, {Tibolla}, {Torres}, {Tosti},
  {Tramacere}, {Uchiyama}, {Usher}, {Van Etten}, {Vilchez}, {Vitale}, {Waite},
  {Wallace}, {Wang}, {Watters}, {Winer}, {Wood}, {Ylinen}, {Ziegler}, \&
  {Fermi/LAT Collaboration}}]{aaa+09}
{Abdo}, A.~A., {Ackermann}, M., {Ajello}, M., {et~al.} 2009, \apjs, 183, 46,
  \dodoi{10.1088/0067-0049/183/1/46}

\bibitem[{{Abdo} {et~al.}(2010){Abdo}, {Ackermann}, {Ajello}, {Baldini},
  {Ballet}, {Barbiellini}, {Bastieri}, {Baughman}, {Bechtol}, {Bellazzini},
  {Berenji}, {Blandford}, {Bloom}, {Bonamente}, {Borgland}, {Bregeon}, {Brez},
  {Brigida}, {Bruel}, {Burnett}, {Buson}, {Caliandro}, {Cameron}, {Camilo},
  {Caraveo}, {Casandjian}, {Cecchi}, {{\c{C}}elik}, {Chekhtman}, {Cheung},
  {Chiang}, {Ciprini}, {Claus}, {Cognard}, {Cohen-Tanugi}, {Cominsky},
  {Conrad}, {Cutini}, {de Angelis}, {de Palma}, {Digel}, {Dingus}, {Dormody},
  {Silva}, {Drell}, {Dubois}, {Dumora}, {Farnier}, {Favuzzi}, {Fegan}, {Focke},
  {Fortin}, {Frailis}, {Freire}, {Fukazawa}, {Funk}, {Fusco}, {Gargano},
  {Gasparrini}, {Gehrels}, {Germani}, {Giavitto}, {Giebels}, {Giglietto},
  {Giordano}, {Glanzman}, {Godfrey}, {Grenier}, {Grondin}, {Grove},
  {Guillemot}, {Guiriec}, {Hanabata}, {Harding}, {Hays}, {Hughes}, {Jackson},
  {J{\'o}hannesson}, {Johnson}, {Johnson}, {Johnson}, {Johnston}, {Kamae},
  {Katagiri}, {Kataoka}, {Kawai}, {Kerr}, {Kn{\"o}dlseder}, {Kocian}, {Kuss},
  {Lande}, {Latronico}, {Lemoine-Goumard}, {Longo}, {Loparco}, {Lott},
  {Lovellette}, {Lubrano}, {Makeev}, {Marelli}, {Mazziotta}, {McEnery},
  {Meurer}, {Michelson}, {Mitthumsiri}, {Mizuno}, {Moiseev}, {Monte},
  {Monzani}, {Morselli}, {Moskalenko}, {Murgia}, {Nolan}, {Norris}, {Nuss},
  {Ohsugi}, {Omodei}, {Orlando}, {Ormes}, {Paneque}, {Parent}, {Pelassa},
  {Pepe}, {Pesce-Rollins}, {Piron}, {Porter}, {Rain{\`o}}, {Rando}, {Ray},
  {Razzano}, {Reimer}, {Reimer}, {Reposeur}, {Ritz}, {Roberts}, {Rochester},
  {Rodriguez}, {Ro'mani}, {Roth}, {Ryde}, {Sadrozinski}, {Sanchez}, {Sander},
  {Saz Parkinson}, {Scargle}, {Sgr{\`o}}, {Siskind}, {Smith}, {Smith},
  {Spandre}, {Spinelli}, {Strickman}, {Suson}, {Tajima}, {Takahashi}, {Tanaka},
  {Thayer}, {Thayer}, {Theureau}, {Thompson}, {Tibaldo}, {Tibolla}, {Torres},
  {Tosti}, {Tramacere}, {Uchiyama}, {Usher}, {Van Etten}, {Vasileiou},
  {Venter}, {Vilchez}, {Vitale}, {Waite}, {Wang}, {Watters}, {Winer}, {Wolff},
  {Wood}, {Ylinen}, \& {Ziegler}}]{aaa+10}
---. 2010, \apj, 711, 64, \dodoi{10.1088/0004-637X/711/1/64}

\bibitem[{Alpar {et~al.}(1982)Alpar, Cheng, Ruderman, \& Shaham}]{acrs82}
Alpar, M.~A., Cheng, A.~F., Ruderman, M.~A., \& Shaham, J. 1982, Nature, 300,
  728

\bibitem[{{Antoniadis} {et~al.}(2013){Antoniadis}, {Freire}, {Wex}, {Tauris},
  {Lynch}, {van Kerkwijk}, {Kramer}, {Bassa}, {Dhillon}, {Driebe}, {Hessels},
  {Kaspi}, {Kondratiev}, {Langer}, {Marsh}, {McLaughlin}, {Pennucci}, {Ransom},
  {Stairs}, {van Leeuwen}, {Verbiest}, \& {Whelan}}]{afw+13}
{Antoniadis}, J., {Freire}, P. C.~C., {Wex}, N., {et~al.} 2013, Science, 340,
  448, \dodoi{10.1126/science.1233232}

\bibitem[{Backer {et~al.}(1982)Backer, Kulkarni, Heiles, Davis, \&
  Goss}]{bkh+82}
Backer, D.~C., Kulkarni, S.~R., Heiles, C., Davis, M.~M., \& Goss, W.~M. 1982,
  Nature, 300, 615

\bibitem[{{Barr} {et~al.}(2013){Barr}, {Champion}, {Kramer}, {Eatough},
  {Freire}, {Karuppusamy}, {Lee}, {Verbiest}, {Bassa}, {Lyne}, {Stappers},
  {Lorimer}, \& {Klein}}]{bck+13}
{Barr}, E.~D., {Champion}, D.~J., {Kramer}, M., {et~al.} 2013, \mnras, 435,
  2234, \dodoi{10.1093/mnras/stt1440}

\bibitem[{{Bates} {et~al.}(2011){Bates}, {Bailes}, {Bhat}, {Burgay},
  {Burke-Spolaor}, {D'Amico}, {Jameson}, {Johnston}, {Keith}, {Kramer},
  {Levin}, {Lyne}, {Milia}, {Possenti}, {Stappers}, \& {van Straten}}]{bbb+11}
{Bates}, S.~D., {Bailes}, M., {Bhat}, N.~D.~R., {et~al.} 2011, \mnras, 416,
  2455, \dodoi{10.1111/j.1365-2966.2011.18416.x}

\bibitem[{{Bhat} {et~al.}(2023){Bhat}, {Swainston}, {McSweeney}, {Xue},
  {Meyers}, {Kudale}, {Dai}, {Tremblay}, {van Straten}, {Shannon}, {Smith},
  {Sokolowski}, {Ord}, {Sleap}, {Williams}, {Hancock}, {Lange}, {Tocknell},
  {Johnston-Hollitt}, {Kaplan}, {Tingay}, \& {Walker}}]{bsm+23}
{Bhat}, N.~D.~R., {Swainston}, N.~A., {McSweeney}, S.~J., {et~al.} 2023, \pasa,
  40, e021, \dodoi{10.1017/pasa.2023.17}

\bibitem[{{Bhattacharyya} {et~al.}(2016){Bhattacharyya}, {Cooper}, {Malenta},
  {Roy}, {Chengalur}, {Keith}, {Kudale}, {McLaughlin}, {Ransom}, {Ray}, \&
  {Stappers}}]{bcm+16}
{Bhattacharyya}, B., {Cooper}, S., {Malenta}, M., {et~al.} 2016, \apj, 817,
  130, \dodoi{10.3847/0004-637X/817/2/130}

\bibitem[{{Boyles} {et~al.}(2013){Boyles}, {Lynch}, {Ransom}, {Stairs},
  {Lorimer}, {McLaughlin}, {Hessels}, {Kaspi}, {Kondratiev}, {Archibald},
  {Berndsen}, {Cardoso}, {Cherry}, {Epstein}, {Karako-Argaman}, {McPhee},
  {Pennucci}, {Roberts}, {Stovall}, \& {van Leeuwen}}]{blr+13}
{Boyles}, J., {Lynch}, R.~S., {Ransom}, S.~M., {et~al.} 2013, \apj, 763, 80,
  \dodoi{10.1088/0004-637X/763/2/80}

\bibitem[{{Burgay} {et~al.}(2006){Burgay}, {Rea}, {Israel}, {Possenti},
  {Burderi}, {di Salvo}, {D'Amico}, \& {Stella}}]{bri+06}
{Burgay}, M., {Rea}, N., {Israel}, G.~L., {et~al.} 2006, \mnras, 372, 410,
  \dodoi{10.1111/j.1365-2966.2006.10872.x}

\bibitem[{{Burgay} {et~al.}(2019){Burgay}, {Stappers}, {Bailes}, {Barr},
  {Bates}, {Bhat}, {Burke-Spolaor}, {Cameron}, {Champion}, {Eatough}, {Flynn},
  {Jameson}, {Johnston}, {Keith}, {Keane}, {Kramer}, {Levin}, {Ng}, {Petroff},
  {Possenti}, {van Straten}, {Tiburzi}, {Bondonneau}, \& {Lyne}}]{bsb+19}
{Burgay}, M., {Stappers}, B., {Bailes}, M., {et~al.} 2019, \mnras, 484, 5791,
  \dodoi{10.1093/mnras/stz401}

\bibitem[{{Burke-Spolaor} {et~al.}(2011){Burke-Spolaor}, {Bailes}, {Johnston},
  {Bates}, {Bhat}, {Burgay}, {D'Amico}, {Jameson}, {Keith}, {Kramer}, {Levin},
  {Milia}, {Possenti}, {Stappers}, \& {van Straten}}]{bbj+11}
{Burke-Spolaor}, S., {Bailes}, M., {Johnston}, S., {et~al.} 2011, \mnras, 416,
  2465, \dodoi{10.1111/j.1365-2966.2011.18521.x}

\bibitem[{{Caleb} {et~al.}(2022){Caleb}, {Heywood}, {Rajwade}, {Malenta},
  {Stappers}, {Barr}, {Chen}, {Morello}, {Sanidas}, {van den Eijnden},
  {Kramer}, {Buckley}, {Brink}, {Motta}, {Woudt}, {Weltevrede}, {Jankowski},
  {Surnis}, {Buchner}, {Bezuidenhout}, {Driessen}, \& {Fender}}]{chr+22}
{Caleb}, M., {Heywood}, I., {Rajwade}, K., {et~al.} 2022, Nature Astronomy, 6,
  828, \dodoi{10.1038/s41550-022-01688-x}

\bibitem[{{Camilo} {et~al.}(1996){Camilo}, {Nice}, {Shrauner}, \&
  {Taylor}}]{cns+96}
{Camilo}, F., {Nice}, D.~J., {Shrauner}, J.~A., \& {Taylor}, J.~H. 1996, \apj,
  469, 819, \dodoi{10.1086/177829}

\bibitem[{{Caraveo} {et~al.}(1998){Caraveo}, {Lattanzi}, {Massone}, {Mignani},
  {Makarov}, {Perryman}, \& {Bignami}}]{clm+98}
{Caraveo}, P.~A., {Lattanzi}, M.~G., {Massone}, G., {et~al.} 1998, \aap, 329,
  L1, \dodoi{10.48550/arXiv.astro-ph/9711029}

\bibitem[{{Champion} {et~al.}(2008){Champion}, {Ransom}, {Lazarus}, {Camilo},
  {Bassa}, {Kaspi}, {Nice}, {Freire}, {Stairs}, {van Leeuwen}, {Stappers},
  {Cordes}, {Hessels}, {Lorimer}, {Arzoumanian}, {Backer}, {Bhat},
  {Chatterjee}, {Cognard}, {Deneva}, {Faucher-Gigu{\`e}re}, {Gaensler}, {Han},
  {Jenet}, {Kasian}, {Kondratiev}, {Kramer}, {Lazio}, {McLaughlin},
  {Venkataraman}, \& {Vlemmings}}]{crl+08}
{Champion}, D.~J., {Ransom}, S.~M., {Lazarus}, P., {et~al.} 2008, Science, 320,
  1309, \dodoi{10.1126/science.1157580}

\bibitem[{{Clark} {et~al.}(2015){Clark}, {Pletsch}, {Wu}, {Guillemot},
  {Ackermann}, {Allen}, {de Angelis}, {Aulbert}, {Baldini}, {Ballet},
  {Barbiellini}, {Bastieri}, {Bellazzini}, {Bissaldi}, {Bock}, {Bonino},
  {Bottacini}, {Brandt}, {Bregeon}, {Bruel}, {Buson}, {Caliandro}, {Cameron},
  {Caragiulo}, {Caraveo}, {Cecchi}, {Champion}, {Charles}, {Chekhtman},
  {Chiang}, {Chiaro}, {Ciprini}, {Claus}, {Cohen-Tanugi}, {Cu{\'e}llar},
  {Cutini}, {D'Ammando}, {Desiante}, {Drell}, {Eggenstein}, {Favuzzi},
  {Fehrmann}, {Ferrara}, {Focke}, {Franckowiak}, {Fusco}, {Gargano},
  {Gasparrini}, {Giglietto}, {Giordano}, {Glanzman}, {Godfrey}, {Grenier},
  {Grove}, {Guiriec}, {Harding}, {Hays}, {Hewitt}, {Hill}, {Horan}, {Hou},
  {Jogler}, {Johnson}, {J{\'o}hannesson}, {Kramer}, {Krauss}, {Kuss}, {Laffon},
  {Larsson}, {Latronico}, {Li}, {Li}, {Longo}, {Loparco}, {Lovellette},
  {Lubrano}, {Machenschalk}, {Manfreda}, {Marelli}, {Mayer}, {Mazziotta},
  {Michelson}, {Mizuno}, {Monzani}, {Morselli}, {Moskalenko}, {Murgia}, {Nuss},
  {Ohsugi}, {Orienti}, {Orlando}, {de Palma}, {Paneque}, {Pesce-Rollins},
  {Piron}, {Pivato}, {Rain{\`o}}, {Rando}, {Razzano}, {Reimer}, {Saz
  Parkinson}, {Schaal}, {Schulz}, {Sgr{\`o}}, {Siskind}, {Spada}, {Spandre},
  {Spinelli}, {Suson}, {Takahashi}, {Thayer}, {Tibaldo}, {Torne}, {Torres},
  {Tosti}, {Troja}, {Vianello}, {Wood}, {Wood}, \& {Yassine}}]{cpw+15}
{Clark}, C.~J., {Pletsch}, H.~J., {Wu}, J., {et~al.} 2015, \apjl, 809, L2,
  \dodoi{10.1088/2041-8205/809/1/L2}

\bibitem[{{Clark} {et~al.}(2017){Clark}, {Wu}, {Pletsch}, {Guillemot}, {Allen},
  {Aulbert}, {Beer}, {Bock}, {Cu{\'e}llar}, {Eggenstein}, {Fehrmann}, {Kramer},
  {Machenschalk}, \& {Nieder}}]{cwp+17}
{Clark}, C.~J., {Wu}, J., {Pletsch}, H.~J., {et~al.} 2017, \apj, 834, 106,
  \dodoi{10.3847/1538-4357/834/2/106}

\bibitem[{Clifton {et~al.}(1992)Clifton, Lyne, Jones, McKenna, \&
  Ashworth}]{clj+92}
Clifton, T.~R., Lyne, A.~G., Jones, A.~W., McKenna, J., \& Ashworth, M. 1992,
  MNRAS, 254, 177

\bibitem[{{Cordes}(2004)}]{ne2001}
{Cordes}, J.~M. 2004, in Astronomical Society of the Pacific Conference Series,
  Vol. 317, Milky Way Surveys: The Structure and Evolution of our Galaxy, ed.
  D.~{Clemens}, R.~{Shah}, \& T.~{Brainerd}, 211

\bibitem[{{Cordes} \& {Chatterjee}(2019)}]{cc19}
{Cordes}, J.~M., \& {Chatterjee}, S. 2019, \araa, 57, 417,
  \dodoi{10.1146/annurev-astro-091918-104501}

\bibitem[{{Cordes} \& {Lazio}(2002)}]{cl02}
{Cordes}, J.~M., \& {Lazio}, T.~J.~W. 2002

\bibitem[{{Cordes} {et~al.}(2006){Cordes}, {Freire}, {Lorimer}, {Camilo},
  {Champion}, {Nice}, {Ramachandran}, {Hessels}, {Vlemmings}, {van Leeuwen},
  {Ransom}, {Bhat}, {Arzoumanian}, {McLaughlin}, {Kaspi}, {Kasian}, {Deneva},
  {Reid}, {Chatterjee}, {Han}, {Backer}, {Stairs}, {Deshpande}, \&
  {Faucher-Gigu{\`e}re}}]{cfl+06}
{Cordes}, J.~M., {Freire}, P.~C.~C., {Lorimer}, D.~R., {et~al.} 2006, \apj,
  637, 446, \dodoi{10.1086/498335}

\bibitem[{{Cromartie}(2020)}]{cro20}
{Cromartie}, H.~T. 2020, PhD thesis, University of Virginia

\bibitem[{{Cruces} {et~al.}(2021){Cruces}, {Champion}, {Li}, {Kramer}, {Zhu},
  {Wang}, {Cameron}, {Chen}, {Hobbs}, {Freire}, {Graikou}, {Krco}, {Liu},
  {Miao}, {Niu}, {Pan}, {Qian}, {Xue}, {Xie}, {You}, {Yu}, {Yuan}, {Yue},
  {Zhu}, {Zhu}, {Lackeos}, {Porayko}, {Wongphecauxon}, {Main}, \& {Crafts
  Collaboration}}]{ccl+21}
{Cruces}, M., {Champion}, D.~J., {Li}, D., {et~al.} 2021, \mnras, 508, 300,
  \dodoi{10.1093/mnras/stab2540}

\bibitem[{Cusumano {et~al.}(2000)Cusumano, Maccarone, Nicastro, Sacco, \&
  Kaaret}]{cmn+00}
Cusumano, G., Maccarone, M.~C., Nicastro, L., Sacco, B., \& Kaaret, P. 2000,
  ApJ, 528, L25

\bibitem[{{Deneva} {et~al.}(2024){Deneva}, {McLaughlin}, {Olszanski}, {Lewis},
  {Pang}, {Freire}, {Bagchi}, \& {Stovall}}]{dmo+24}
{Deneva}, J.~S., {McLaughlin}, M., {Olszanski}, T.~E.~E., {et~al.} 2024, \apjs,
  271, 23, \dodoi{10.3847/1538-4365/ad19da}

\bibitem[{{Deneva} {et~al.}(2013){Deneva}, {Stovall}, {McLaughlin}, {Bates},
  {Freire}, {Martinez}, {Jenet}, \& {Bagchi}}]{dsm+13}
{Deneva}, J.~S., {Stovall}, K., {McLaughlin}, M.~A., {et~al.} 2013, \apj, 775,
  51, \dodoi{10.1088/0004-637X/775/1/51}

\bibitem[{{Deneva} {et~al.}(2009){Deneva}, {Cordes}, {McLaughlin}, {Nice},
  {Lorimer}, {Crawford}, {Bhat}, {Camilo}, {Champion}, {Freire}, {Edel},
  {Kondratiev}, {Hessels}, {Jenet}, {Kasian}, {Kaspi}, {Kramer}, {Lazarus},
  {Ransom}, {Stairs}, {Stappers}, {van Leeuwen}, {Brazier}, {Venkataraman},
  {Zollweg}, \& {Bogdanov}}]{dcm+09}
{Deneva}, J.~S., {Cordes}, J.~M., {McLaughlin}, M.~A., {et~al.} 2009, \apj,
  703, 2259, \dodoi{10.1088/0004-637X/703/2/2259}

\bibitem[{{Deneva} {et~al.}(2016){Deneva}, {Stovall}, {McLaughlin}, {Bagchi},
  {Bates}, {Freire}, {Martinez}, {Jenet}, \& {Garver-Daniels}}]{dsm+16}
{Deneva}, J.~S., {Stovall}, K., {McLaughlin}, M.~A., {et~al.} 2016, \apj, 821,
  10, \dodoi{10.3847/0004-637X/821/1/10}

\bibitem[{{Edwards} {et~al.}(2001){Edwards}, {Bailes}, {van Straten}, \&
  {Britton}}]{ebv+01}
{Edwards}, R.~T., {Bailes}, M., {van Straten}, W., \& {Britton}, M.~C. 2001,
  \mnras, 326, 358, \dodoi{10.1046/j.1365-8711.2001.04637.x}

\bibitem[{{Frail} {et~al.}(1999){Frail}, {Kulkarni}, \& {Bloom}}]{fkb+99}
{Frail}, D.~A., {Kulkarni}, S.~R., \& {Bloom}, J.~S. 1999, \nat, 398, 127,
  \dodoi{10.1038/18163}

\bibitem[{{Freire} {et~al.}(2001){Freire}, {Kramer}, \& {Lyne}}]{fkl01}
{Freire}, P.~C., {Kramer}, M., \& {Lyne}, A.~G. 2001, MNRAS, 322, 885

\bibitem[{Gaensler {et~al.}(2005)Gaensler, Haverkorn, Staveley-Smith, Dickey,
  McClure-Griffiths, Dickel, \& Wolleben}]{ghs+05}
Gaensler, B.~M., Haverkorn, M., Staveley-Smith, L., {et~al.} 2005, Science,
  307, 1610

\bibitem[{{Gao} {et~al.}(2011){Gao}, {Sun}, {Han}, {Reich}, {Reich}, \&
  {Wielebinski}}]{gsh+11}
{Gao}, X.~Y., {Sun}, X.~H., {Han}, J.~L., {et~al.} 2011, \aap, 532, A144,
  \dodoi{10.1051/0004-6361/201117179}

\bibitem[{{Gogus} {et~al.}(2008){Gogus}, {Woods}, \& {Kouveliotou}}]{gwk+08}
{Gogus}, E., {Woods}, P., \& {Kouveliotou}, C. 2008, GRB Coordinates Network,
  8118, 1

\bibitem[{{Good} {et~al.}(2021){Good}, {Andersen}, {Chawla}, {Crowter}, {Dong},
  {Fonseca}, {Meyers}, {Ng}, {Pleunis}, {Ransom}, {Stairs}, {Tan}, {Bhardwaj},
  {Boyle}, {Dobbs}, {Gaensler}, {Kaspi}, {Masui}, {Naidu}, {Rafiei-Ravandi},
  {Scholz}, {Smith}, \& {Tendulkar}}]{gac+21}
{Good}, D.~C., {Andersen}, B.~C., {Chawla}, P., {et~al.} 2021, \apj, 922, 43,
  \dodoi{10.3847/1538-4357/ac1da6}

\bibitem[{{Gotthelf} {et~al.}(2011){Gotthelf}, {Halpern}, {Terrier}, \&
  {Mattana}}]{ghtm11}
{Gotthelf}, E.~V., {Halpern}, J.~P., {Terrier}, R., \& {Mattana}, F. 2011,
  \apjl, 729, L16, \dodoi{10.1088/2041-8205/729/2/L16}

\bibitem[{{Gotthelf} \& {Vasisht}(1998)}]{gv98}
{Gotthelf}, E.~V., \& {Vasisht}, G. 1998, New Astr., 3, 293

\bibitem[{{G{\"o}{\v{g}}{\"u}{\c{s}}}
  {et~al.}(2010){G{\"o}{\v{g}}{\"u}{\c{s}}}, {Woods}, {Kouveliotou}, {Kaneko},
  {Gaensler}, \& {Chatterjee}}]{gwk+10}
{G{\"o}{\v{g}}{\"u}{\c{s}}}, E., {Woods}, P.~M., {Kouveliotou}, C., {et~al.}
  2010, \apj, 722, 899, \dodoi{10.1088/0004-637X/722/1/899}

\bibitem[{Halpern \& Holt(1992)}]{hh92}
Halpern, J.~P., \& Holt, S.~S. 1992, Nature, 357, 222

\bibitem[{{Han} {et~al.}(2016){Han}, {Wang}, {Xu}, \& {Han}}]{hwxh16}
{Han}, J., {Wang}, C., {Xu}, J., \& {Han}, J.-L. 2016, Research in Astronomy
  and Astrophysics, 16, 159, \dodoi{10.1088/1674-4527/16/10/159}

\bibitem[{{Han} {et~al.}(2021){Han}, {Wang}, {Wang}, {Wang}, {Zhou}, {Sun},
  {Yan}, {Su}, {Jing}, {Chen}, {Gao}, {Hou}, {Xu}, {Lee}, {Wang}, {Jiang},
  {Xu}, {Yan}, {Gan}, {Guan}, {Huang}, {Jiang}, {Li}, {Men}, {Sun}, {Wang},
  {Wang}, {Wang}, {Xie}, {Xu}, {Yao}, {You}, {Yu}, {Yuan}, {Yuen}, {Zhang}, \&
  {Zhu}}]{hww+21}
{Han}, J.~L., {Wang}, C., {Wang}, P.~F., {et~al.} 2021, Research in Astronomy
  and Astrophysics, 21, 107, \dodoi{10.1088/1674-4527/21/5/107}

\bibitem[{{Hessels} {et~al.}(2008){Hessels}, {Ransom}, {Kaspi}, {Roberts},
  {Champion}, \& {Stappers}}]{hrk+08}
{Hessels}, J.~W.~T., {Ransom}, S.~M., {Kaspi}, V.~M., {et~al.} 2008, in
  American Institute of Physics Conference Series, Vol. 983, 40 Years of
  Pulsars: Millisecond Pulsars, Magnetars and More, ed. C.~{Bassa}, Z.~{Wang},
  A.~{Cumming}, \& V.~M. {Kaspi} (AIP), 613--615, \dodoi{10.1063/1.2900310}

\bibitem[{{Hessels} {et~al.}(2006){Hessels}, {Ransom}, {Stairs}, {Freire},
  {Kaspi}, \& {Camilo}}]{hrs+06}
{Hessels}, J. W.~T., {Ransom}, S.~M., {Stairs}, I.~H., {et~al.} 2006, Science,
  311, 1901, \dodoi{10.1126/science.1123430}

\bibitem[{{Hessels} {et~al.}(2007){Hessels}, {Ransom}, {Stairs}, {Kaspi}, \&
  {Freire}}]{hrs+07}
{Hessels}, J.~W.~T., {Ransom}, S.~M., {Stairs}, I.~H., {Kaspi}, V.~M., \&
  {Freire}, P.~C.~C. 2007, \apj, 670, 363, \dodoi{10.1086/521780}

\bibitem[{Hobbs {et~al.}(2005)Hobbs, Lorimer, Lyne, \& Kramer}]{hllk05}
Hobbs, G., Lorimer, D.~R., Lyne, A.~G., \& Kramer, M. 2005, MNRAS, 360, 974

\bibitem[{{Hong} {et~al.}(2022){Hong}, {Han}, {Hou}, {Gao}, {Wang}, \&
  {Wang}}]{hhh+22}
{Hong}, T., {Han}, J., {Hou}, L., {et~al.} 2022, Science China Physics,
  Mechanics, and Astronomy, 65, 129702, \dodoi{10.1007/s11433-022-2040-8}

\bibitem[{{Hotan} {et~al.}(2004){Hotan}, {van Straten}, \&
  {Manchester}}]{hvm04}
{Hotan}, A.~W., {van Straten}, W., \& {Manchester}, R.~N. 2004, PASA, 21, 302

\bibitem[{{Hou} {et~al.}(2022){Hou}, {Han}, {Hong}, {Gao}, \& {Wang}}]{hhh+22a}
{Hou}, L., {Han}, J., {Hong}, T., {Gao}, X., \& {Wang}, C. 2022, Science China
  Physics, Mechanics, and Astronomy, 65, 129703,
  \dodoi{10.1007/s11433-022-2039-8}

\bibitem[{{Hou} \& {Han}(2014)}]{hh14}
{Hou}, L.~G., \& {Han}, J.~L. 2014, \aap, 569, A125,
  \dodoi{10.1051/0004-6361/201424039}

\bibitem[{{Jacoby} {et~al.}(2009){Jacoby}, {Bailes}, {Ord}, {Edwards}, \&
  {Kulkarni}}]{jbo+09}
{Jacoby}, B.~A., {Bailes}, M., {Ord}, S.~M., {Edwards}, R.~T., \& {Kulkarni},
  S.~R. 2009, \apj, 699, 2009, \dodoi{10.1088/0004-637X/699/2/2009}

\bibitem[{{Jiang} {et~al.}(2020){Jiang}, {Tang}, {Hou}, {Liu}, {Kr{\v{c}}o},
  {Qian}, {Sun}, {Ching}, {Liu}, {Duan}, {Yue}, {Gan}, {Yao}, {Li}, {Pan},
  {Yu}, {Liu}, {Li}, {Peng}, {Yan}, \& {FAST Collaboration}}]{jth+20}
{Jiang}, P., {Tang}, N.-Y., {Hou}, L.-G., {et~al.} 2020, Research in Astronomy
  and Astrophysics, 20, 064, \dodoi{10.1088/1674-4527/20/5/64}

\bibitem[{Johnston {et~al.}(1992)Johnston, Lyne, Manchester, Kniffen, D'Amico,
  Lim, \& Ashworth}]{jlm+92}
Johnston, S., Lyne, A.~G., Manchester, R.~N., {et~al.} 1992, MNRAS, 255, 401

\bibitem[{Kaspi {et~al.}(1996)Kaspi, Manchester, Johnston, Lyne, \&
  D'Amico}]{kmj+96}
Kaspi, V.~M., Manchester, R.~N., Johnston, S., Lyne, A.~G., \& D'Amico, N.
  1996, AJ, 111, 2028

\bibitem[{{Keane} {et~al.}(2011){Keane}, {Kramer}, {Lyne}, {Stappers}, \&
  {McLaughlin}}]{kkl+11}
{Keane}, E.~F., {Kramer}, M., {Lyne}, A.~G., {Stappers}, B.~W., \&
  {McLaughlin}, M.~A. 2011, \mnras, 415, 3065,
  \dodoi{10.1111/j.1365-2966.2011.18917.x}

\bibitem[{{Keane} {et~al.}(2018){Keane}, {Barr}, {Jameson}, {Morello}, {Caleb},
  {Bhandari}, {Petroff}, {Possenti}, {Burgay}, {Tiburzi}, {Bailes}, {Bhat},
  {Burke-Spolaor}, {Eatough}, {Flynn}, {Jankowski}, {Johnston}, {Kramer},
  {Levin}, {Ng}, {van Straten}, \& {Krishnan}}]{kbj+18}
{Keane}, E.~F., {Barr}, E.~D., {Jameson}, A., {et~al.} 2018, \mnras, 473, 116,
  \dodoi{10.1093/mnras/stx2126}

\bibitem[{{Keith} {et~al.}(2010){Keith}, {Jameson}, {van Straten}, {Bailes},
  {Johnston}, {Kramer}, {Possenti}, {Bates}, {Bhat}, {Burgay}, {Burke-Spolaor},
  {D'Amico}, {Levin}, {McMahon}, {Milia}, \& {Stappers}}]{kjv+10}
{Keith}, M.~J., {Jameson}, A., {van Straten}, W., {et~al.} 2010, \mnras, 409,
  619, \dodoi{10.1111/j.1365-2966.2010.17325.x}

\bibitem[{{Knispel} {et~al.}(2013){Knispel}, {Eatough}, {Kim}, {Keane},
  {Allen}, {Anderson}, {Aulbert}, {Bock}, {Crawford}, {Eggenstein}, {Fehrmann},
  {Hammer}, {Kramer}, {Lyne}, {Machenschalk}, {Miller}, {Papa}, {Rastawicki},
  {Sarkissian}, {Siemens}, \& {Stappers}}]{kek+13}
{Knispel}, B., {Eatough}, R.~P., {Kim}, H., {et~al.} 2013, \apj, 774, 93,
  \dodoi{10.1088/0004-637X/774/2/93}

\bibitem[{{Kouveliotou} {et~al.}(1999){Kouveliotou}, {Strohmayer}, {Hurley},
  {Van Paradijs}, {Finger}, {Dieters}, {Woods}, {Thompson}, \&
  {Duncan}}]{ksh+99}
{Kouveliotou}, C., {Strohmayer}, T., {Hurley}, K., {et~al.} 1999, ApJ, 510,
  L115

\bibitem[{Kramer {et~al.}(1999)Kramer, Lange, Lorimer, Backer, Xilouris,
  Jessner, \& Wielebinski}]{kll+99}
Kramer, M., Lange, C., Lorimer, D.~R., {et~al.} 1999, ApJ, 526, 957

\bibitem[{{Kramer} {et~al.}(2006){Kramer}, {Lyne}, {O'Brien}, {Jordan}, \&
  {Lorimer}}]{klo+06}
{Kramer}, M., {Lyne}, A.~G., {O'Brien}, J.~T., {Jordan}, C.~A., \& {Lorimer},
  D.~R. 2006, Science, 312, 549, \dodoi{10.1126/science.1124060}

\bibitem[{{Li} \& {Han}(2003)}]{lh03}
{Li}, X.~H., \& {Han}, J.~L. 2003, \aap, 410, 253,
  \dodoi{10.1051/0004-6361:20031190}

\bibitem[{{Liu} {et~al.}(2024){Liu}, {Zhong}, {Chen}, {Wang}, {Zhou}, {Yue}, \&
  {Li}}]{lzc+24}
{Liu}, Q.-C., {Zhong}, W.-J., {Chen}, Y., {et~al.} 2024, \mnras, 528, 6761,
  \dodoi{10.1093/mnras/stae351}

\bibitem[{{Lorimer} {et~al.}(2007){Lorimer}, {Bailes}, {McLaughlin},
  {Narkevic}, \& {Crawford}}]{lbm+07}
{Lorimer}, D.~R., {Bailes}, M., {McLaughlin}, M.~A., {Narkevic}, D.~J., \&
  {Crawford}, F. 2007, Science, 318, 777, \dodoi{10.1126/science.1147532}

\bibitem[{{Lorimer} {et~al.}(2013){Lorimer}, {Camilo}, \&
  {McLaughlin}}]{lcm+13}
{Lorimer}, D.~R., {Camilo}, F., \& {McLaughlin}, M.~A. 2013, \mnras, 434, 347,
  \dodoi{10.1093/mnras/stt1023}

\bibitem[{{Lorimer} {et~al.}(2006){Lorimer}, {Faulkner}, {Lyne}, {Manchester},
  {Kramer}, {McLaughlin}, {Hobbs}, {Possenti}, {Stairs}, {Camilo}, {Burgay},
  {D'Amico}, {Corongiu}, \& {Crawford}}]{lfl+06}
{Lorimer}, D.~R., {Faulkner}, A.~J., {Lyne}, A.~G., {et~al.} 2006, \mnras, 372,
  777, \dodoi{10.1111/j.1365-2966.2006.10887.x}

\bibitem[{Lyne \& Lorimer(1994)}]{ll94}
Lyne, A.~G., \& Lorimer, D.~R. 1994, Nature, 369, 127

\bibitem[{Lyne \& Manchester(1988)}]{lm88}
Lyne, A.~G., \& Manchester, R.~N. 1988, MNRAS, 234, 477

\bibitem[{{Lyne} {et~al.}(2017){Lyne}, {Stappers}, {Freire}, {Hessels},
  {Kaspi}, {Allen}, {Bogdanov}, {Brazier}, {Camilo}, {Cardoso}, {Chatterjee},
  {Cordes}, {Crawford}, {Deneva}, {Ferdman}, {Jenet}, {Knispel}, {Lazarus},
  {van Leeuwen}, {Lynch}, {Madsen}, {McLaughlin}, {Parent}, {Patel}, {Ransom},
  {Scholz}, {Seymour}, {Siemens}, {Spitler}, {Stairs}, {Stovall}, {Swiggum},
  {Wharton}, \& {Zhu}}]{lsf+17}
{Lyne}, A.~G., {Stappers}, B.~W., {Freire}, P.~C.~C., {et~al.} 2017, \apj, 834,
  72, \dodoi{10.3847/1538-4357/834/1/72}

\bibitem[{Manchester {et~al.}(1996)Manchester, Lyne, D'Amico, Bailes, Johnston,
  Lorimer, Harrison, Nicastro, \& Bell}]{mld+96}
Manchester, R.~N., Lyne, A.~G., D'Amico, N., {et~al.} 1996, MNRAS, 279, 1235

\bibitem[{Manchester {et~al.}(2001)Manchester, Lyne, Camilo, Bell, Kaspi,
  D'Amico, McKay, Crawford, Stairs, Possenti, Morris, \& Sheppard}]{mlc+01}
Manchester, R.~N., Lyne, A.~G., Camilo, F., {et~al.} 2001, MNRAS, 328, 17

\bibitem[{{McEwen} {et~al.}(2024){McEwen}, {Lynch}, {Kaplan}, {Bolda},
  {Sengar}, {Fonseca}, {Agoudemos}, {Boyles}, {Chatterjee}, {Cohen},
  {Crawford}, {DeCesar}, {Ehlke}, {Fernandez}, {Ferrara}, {Fiore}, {Gilhaus},
  {Gleiter}, {Hessels}, {Holman}, {Joy}, {Kaspi}, {Kondratiev}, {Leon},
  {Levin}, {Lorenz}, {Lorimer}, {Madison}, {McLaughlin}, {Meyers}, {Parent},
  {Patron}, {Ransom}, {Ray}, {Roberts}, {Roch}, {Siemens}, {Stearns},
  {Swiggum}, {Stairs}, {Stovall}, {Tan}, {Valentine}, \& {van
  Leeuwen}}]{mlk+24}
{McEwen}, A.~E., {Lynch}, R.~S., {Kaplan}, D.~L., {et~al.} 2024, \apj, 969,
  118, \dodoi{10.3847/1538-4357/ad47f0}

\bibitem[{{McLaughlin} {et~al.}(2006){McLaughlin}, {Lyne}, {Lorimer}, {Kramer},
  {Faulkner}, {Manchester}, {Cordes}, {Camilo}, {Possenti}, {Stairs}, {Hobbs},
  {D'Amico}, {Burgay}, \& {O'Brien}}]{mll+06}
{McLaughlin}, M.~A., {Lyne}, A.~G., {Lorimer}, D.~R., {et~al.} 2006, \nat, 439,
  817, \dodoi{10.1038/nature04440}

\bibitem[{{Mereghetti} {et~al.}(2006){Mereghetti}, {Esposito}, {Tiengo},
  {Zane}, {Turolla}, {Stella}, {Israel}, {G{\"o}tz}, \& {Feroci}}]{met+06}
{Mereghetti}, S., {Esposito}, P., {Tiengo}, A., {et~al.} 2006, \apj, 653, 1423,
  \dodoi{10.1086/508682}

\bibitem[{{Michilli} {et~al.}(2018){Michilli}, {Hessels}, {Lyon}, {Tan},
  {Bassa}, {Cooper}, {Kondratiev}, {Sanidas}, {Stappers}, \& {van
  Leeuwen}}]{mhl+18}
{Michilli}, D., {Hessels}, J.~W.~T., {Lyon}, R.~J., {et~al.} 2018, \mnras, 480,
  3457, \dodoi{10.1093/mnras/sty2072}

\bibitem[{{Morello} {et~al.}(2022){Morello}, {Rajwade}, \& {Stappers}}]{mrs22}
{Morello}, V., {Rajwade}, K.~M., \& {Stappers}, B.~W. 2022, \mnras, 510, 1393,
  \dodoi{10.1093/mnras/stab3493}

\bibitem[{{Nan}(2006)}]{nan06}
{Nan}, R. 2006, Science in China: Physics, Mechanics and Astronomy, 49, 129,
  \dodoi{10.1007/s11433-006-0129-9}

\bibitem[{{Ng} {et~al.}(2015){Ng}, {Champion}, {Bailes}, {Barr}, {Bates},
  {Bhat}, {Burgay}, {Burke-Spolaor}, {Flynn}, {Jameson}, {Johnston}, {Keith},
  {Kramer}, {Levin}, {Petroff}, {Possenti}, {Stappers}, {van Straten},
  {Tiburzi}, {Eatough}, \& {Lyne}}]{ncb+15}
{Ng}, C., {Champion}, D.~J., {Bailes}, M., {et~al.} 2015, \mnras, 450, 2922,
  \dodoi{10.1093/mnras/stv753}

\bibitem[{{Nice} {et~al.}(2013){Nice}, {Altiere}, {Bogdanov}, {Cordes},
  {Farrington}, {Hessels}, {Kaspi}, {Lyne}, {Popa}, {Ransom}, {Sanpa-arsa},
  {Stappers}, {Wang}, {Allen}, {Bhat}, {Brazier}, {Camilo}, {Champion},
  {Chatterjee}, {Crawford}, {Deneva}, {Desvignes}, {Freire}, {Jenet},
  {Knispel}, {Lazarus}, {Lee}, {van Leeuwen}, {Lorimer}, {Lynch}, {McLaughlin},
  {Scholz}, {Siemens}, {Stairs}, {Stovall}, {Venkataraman}, \& {Zhu}}]{nab+13}
{Nice}, D.~J., {Altiere}, E., {Bogdanov}, S., {et~al.} 2013, \apj, 772, 50,
  \dodoi{10.1088/0004-637X/772/1/50}

\bibitem[{{Niu} {et~al.}(2021){Niu}, {Li}, {Luo}, {Wang}, {Yao}, {Zhang},
  {Zhu}, {Wang}, {Ye}, {Zhang}, {Niu}, {Tang}, {Duan}, {Krco}, {Dai}, {Feng},
  {Miao}, {Pan}, {Qian}, {Xue}, {Yuan}, {Yue}, {Zhang}, \& {Zhang}}]{nll+21}
{Niu}, C.-H., {Li}, D., {Luo}, R., {et~al.} 2021, \apjl, 909, L8,
  \dodoi{10.3847/2041-8213/abe7f0}

\bibitem[{{Niu} {et~al.}(2022){Niu}, {Aggarwal}, {Li}, {Zhang}, {Chatterjee},
  {Tsai}, {Yu}, {Law}, {Burke-Spolaor}, {Cordes}, {Zhang}, {Ocker}, {Yao},
  {Wang}, {Feng}, {Niino}, {Bochenek}, {Cruces}, {Connor}, {Jiang}, {Dai},
  {Luo}, {Li}, {Miao}, {Niu}, {Anna-Thomas}, {Sydnor}, {Stern}, {Wang}, {Yuan},
  {Yue}, {Zhou}, {Yan}, {Zhu}, \& {Zhang}}]{nal+22}
{Niu}, C.~H., {Aggarwal}, K., {Li}, D., {et~al.} 2022, \nat, 606, 873,
  \dodoi{10.1038/s41586-022-04755-5}

\bibitem[{{Padmanabh} {et~al.}(2023){Padmanabh}, {Barr}, {Sridhar}, {Rugel},
  {Damas-Segovia}, {Jacob}, {Balakrishnan}, {Berezina}, {Bernadich},
  {Brunthaler}, {Champion}, {Freire}, {Khan}, {Kl{\"o}ckner}, {Kramer}, {Ma},
  {Mao}, {Men}, {Menten}, {Sengupta}, {Venkatraman Krishnan}, {Wucknitz},
  {Wyrowski}, {Bezuidenhout}, {Buchner}, {Burgay}, {Chen}, {Clark},
  {K{\"u}nkel}, {Nieder}, {Stappers}, {Legodi}, \& {Nyamai}}]{pbs+23}
{Padmanabh}, P.~V., {Barr}, E.~D., {Sridhar}, S.~S., {et~al.} 2023, \mnras,
  524, 1291, \dodoi{10.1093/mnras/stad1900}

\bibitem[{{Pan} {et~al.}(2021){Pan}, {Qian}, {Ma}, {Liu}, {Wang}, {Luo}, {Yan},
  {Ransom}, {Lorimer}, {Li}, \& {Jiang}}]{pqm+21}
{Pan}, Z., {Qian}, L., {Ma}, X., {et~al.} 2021, \apjl, 915, L28,
  \dodoi{10.3847/2041-8213/ac0bbd}

\bibitem[{{Pan} {et~al.}(2023){Pan}, {Lu}, {Jiang}, {Han}, {Chen}, {Han},
  {Liu}, {Qian}, {Xu}, {Zhang}, {Luo}, {Yan}, {Yang}, {Zhou}, {Wang}, {Wang},
  {Li}, \& {Zhu}}]{plj+23}
{Pan}, Z., {Lu}, J.~G., {Jiang}, P., {et~al.} 2023, \nat, 620, 961,
  \dodoi{10.1038/s41586-023-06308-w}

\bibitem[{{Parent} {et~al.}(2022){Parent}, {Sewalls}, {Freire}, {Matheny},
  {Lyne}, {Perera}, {Cardoso}, {McLaughlin}, {Allen}, {Brazier}, {Camilo},
  {Chatterjee}, {Cordes}, {Crawford}, {Deneva}, {Dong}, {Ferdman}, {Fonseca},
  {Hessels}, {Kaspi}, {Knispel}, {van Leeuwen}, {Lynch}, {Meyers}, {McKee},
  {Mickaliger}, {Patel}, {Ransom}, {Rochon}, {Scholz}, {Stairs}, {Stappers},
  {Tan}, \& {Zhu}}]{psf+22}
{Parent}, E., {Sewalls}, H., {Freire}, P.~C.~C., {et~al.} 2022, \apj, 924, 135,
  \dodoi{10.3847/1538-4357/ac375d}

\bibitem[{{Patel} {et~al.}(2018){Patel}, {Agarwal}, {Bhardwaj}, {Boyce},
  {Brazier}, {Chatterjee}, {Chawla}, {Kaspi}, {Lorimer}, {McLaughlin},
  {Parent}, {Pleunis}, {Ransom}, {Scholz}, {Wharton}, {Zhu}, {Alam}, {Caballero
  Valdez}, {Camilo}, {Cordes}, {Crawford}, {Deneva}, {Ferdman}, {Freire},
  {Hessels}, {Nguyen}, {Stairs}, {Stovall}, \& {van Leeuwen}}]{pab+18}
{Patel}, C., {Agarwal}, D., {Bhardwaj}, M., {et~al.} 2018, \apj, 869, 181,
  \dodoi{10.3847/1538-4357/aaee65}

\bibitem[{{Pletsch} {et~al.}(2012{\natexlab{a}}){Pletsch}, {Guillemot},
  {Allen}, {Kramer}, {Aulbert}, {Fehrmann}, {Ray}, {Barr}, {Belfiore},
  {Camilo}, {Caraveo}, {{\c{C}}elik}, {Champion}, {Dormody}, {Eatough},
  {Ferrara}, {Freire}, {Hessels}, {Keith}, {Kerr}, {de Luca}, {Lyne},
  {Marelli}, {McLaughlin}, {Parent}, {Ransom}, {Razzano}, {Reich}, {Saz
  Parkinson}, {Stappers}, \& {Wolff}}]{pga+12}
{Pletsch}, H.~J., {Guillemot}, L., {Allen}, B., {et~al.} 2012{\natexlab{a}},
  \apj, 744, 105, \dodoi{10.1088/0004-637X/744/2/105}

\bibitem[{{Pletsch} {et~al.}(2012{\natexlab{b}}){Pletsch}, {Guillemot},
  {Allen}, {Kramer}, {Aulbert}, {Fehrmann}, {Ray}, {Barr}, {Belfiore},
  {Camilo}, {Caraveo}, {{\c{C}}elik}, {Champion}, {Dormody}, {Eatough},
  {Ferrara}, {Freire}, {Hessels}, {Keith}, {Kerr}, {de Luca}, {Lyne},
  {Marelli}, {McLaughlin}, {Parent}, {Ransom}, {Razzano}, {Reich}, {Saz
  Parkinson}, {Stappers}, \& {Wolff}}]{pha+12}
---. 2012{\natexlab{b}}, \apj, 744, 105, \dodoi{10.1088/0004-637X/744/2/105}

\bibitem[{{Pletsch} {et~al.}(2013){Pletsch}, {Guillemot}, {Allen}, {Anderson},
  {Aulbert}, {Bock}, {Champion}, {Eggenstein}, {Fehrmann}, {Hammer},
  {Karuppusamy}, {Keith}, {Kramer}, {Machenschalk}, {Ng}, {Papa}, {Ray}, \&
  {Siemens}}]{pga+13}
---. 2013, \apjl, 779, L11, \dodoi{10.1088/2041-8205/779/1/L11}

\bibitem[{{Pleunis} {et~al.}(2021){Pleunis}, {Good}, {Kaspi}, {Mckinven},
  {Ransom}, {Scholz}, {Bandura}, {Bhardwaj}, {Boyle}, {Brar}, {Cassanelli},
  {Chawla}, {(Adam) Dong}, {Fonseca}, {Gaensler}, {Josephy}, {Kaczmarek},
  {Leung}, {Lin}, {Masui}, {Mena-Parra}, {Michilli}, {Ng}, {Patel},
  {Rafiei-Ravandi}, {Rahman}, {Sanghavi}, {Shin}, {Smith}, {Stairs}, \&
  {Tendulkar}}]{pgk+21}
{Pleunis}, Z., {Good}, D.~C., {Kaspi}, V.~M., {et~al.} 2021, \apj, 923, 1,
  \dodoi{10.3847/1538-4357/ac33ac}

\bibitem[{{Ransom}(2001{\natexlab{a}})}]{Ransom01}
{Ransom}, S.~M. 2001{\natexlab{a}}, PhD thesis, Harvard University,
  Massachusetts

\bibitem[{{Ransom}(2001{\natexlab{b}})}]{ran01}
---. 2001{\natexlab{b}}, PhD thesis, Harvard University

\bibitem[{{Ray} {et~al.}(2011){Ray}, {Kerr}, {Parent}, {Abdo}, {Guillemot},
  {Ransom}, {Rea}, {Wolff}, {Makeev}, {Roberts}, {Camilo}, {Dormody}, {Freire},
  {Grove}, {Gwon}, {Harding}, {Johnston}, {Keith}, {Kramer}, {Michelson},
  {Romani}, {Saz Parkinson}, {Thompson}, {Weltevrede}, {Wood}, \&
  {Ziegler}}]{rkp+11}
{Ray}, P.~S., {Kerr}, M., {Parent}, D., {et~al.} 2011, \apjs, 194, 17,
  \dodoi{10.1088/0067-0049/194/2/17}

\bibitem[{{Reid} {et~al.}(2019){Reid}, {Menten}, {Brunthaler}, {Zheng}, {Dame},
  {Xu}, {Li}, {Sakai}, {Wu}, {Immer}, {Zhang}, {Sanna}, {Moscadelli}, {Rygl},
  {Bartkiewicz}, {Hu}, {Quiroga-Nu{\~n}ez}, \& {van Langevelde}}]{rmb+19}
{Reid}, M.~J., {Menten}, K.~M., {Brunthaler}, A., {et~al.} 2019, \apj, 885,
  131, \dodoi{10.3847/1538-4357/ab4a11}

\bibitem[{{Sakamoto} {et~al.}(2011){Sakamoto}, {Barbier}, {Barthelmy},
  {Cummings}, {Fenimore}, {Gehrels}, {Krimm}, {Markwardt}, {Palmer}, {Parsons},
  {Sato}, {Stamatikos}, \& {Tueller}}]{sbb+11}
{Sakamoto}, T., {Barbier}, L., {Barthelmy}, S.~D., {et~al.} 2011, Advances in
  Space Research, 47, 1346, \dodoi{10.1016/j.asr.2010.08.004}

\bibitem[{{Sanidas} {et~al.}(2019){Sanidas}, {Cooper}, {Bassa}, {Hessels},
  {Kondratiev}, {Michilli}, {Stappers}, {Tan}, {van Leeuwen}, {Cerrigone},
  {Fallows}, {Iacobelli}, {Orr{\'u}}, {Pizzo}, {Shulevski}, {Toribio}, {ter
  Veen}, {Zucca}, {Bondonneau}, {Grie{\ss}meier}, {Karastergiou}, {Kramer}, \&
  {Sobey}}]{scb+19}
{Sanidas}, S., {Cooper}, S., {Bassa}, C.~G., {et~al.} 2019, \aap, 626, A104,
  \dodoi{10.1051/0004-6361/201935609}

\bibitem[{{Saz Parkinson} {et~al.}(2010){Saz Parkinson}, {Dormody}, {Ziegler},
  {Ray}, {Abdo}, {Ballet}, {Baring}, {Belfiore}, {Burnett}, {Caliandro},
  {Camilo}, {Caraveo}, {de Luca}, {Ferrara}, {Freire}, {Grove}, {Gwon},
  {Harding}, {Johnson}, {Johnson}, {Johnston}, {Keith}, {Kerr},
  {Kn{\"o}dlseder}, {Makeev}, {Marelli}, {Michelson}, {Parent}, {Ransom},
  {Reimer}, {Romani}, {Smith}, {Thompson}, {Watters}, {Weltevrede}, {Wolff}, \&
  {Wood}}]{sdz+10}
{Saz Parkinson}, P.~M., {Dormody}, M., {Ziegler}, M., {et~al.} 2010, \apj, 725,
  571, \dodoi{10.1088/0004-637X/725/1/571}

\bibitem[{{Sengar} {et~al.}(2023){Sengar}, {Bailes}, {Balakrishnan},
  {Bernadich}, {Burgay}, {Barr}, {Flynn}, {Stevenson}, \&
  {Wongphechauxsorn}}]{sbb+23}
{Sengar}, R., {Bailes}, M., {Balakrishnan}, V., {et~al.} 2023, \mnras, 522,
  1071, \dodoi{10.1093/mnras/stad508}

\bibitem[{{Smith} {et~al.}(2023){Smith}, {Abdollahi}, {Ajello}, {Bailes},
  {Baldini}, {Ballet}, {Baring}, {Bassa}, {Gonzalez}, {Bellazzini}, {Berretta},
  {Bhattacharyya}, {Bissaldi}, {Bonino}, {Bottacini}, {Bregeon}, {Bruel},
  {Burgay}, {Burnett}, {Cameron}, {Camilo}, {Caputo}, {Caraveo}, {Cavazzuti},
  {Chiaro}, {Ciprini}, {Clark}, {Cognard}, {Corongiu}, {Orestano},
  {Crnogorcevic}, {Cuoco}, {Cutini}, {D'Ammando}, {de Angelis}, {DeCesar}, {De
  Gaetano}, {de Menezes}, {Deneva}, {de Palma}, {Di Lalla}, {Dirirsa}, {Di
  Venere}, {Dom{\'\i}nguez}, {Dumora}, {Fegan}, {Ferrara}, {Fiori},
  {Fleischhack}, {Flynn}, {Franckowiak}, {Freire}, {Fukazawa}, {Fusco},
  {Galanti}, {Gammaldi}, {Gargano}, {Gasparrini}, {Giacchino}, {Giglietto},
  {Giordano}, {Giroletti}, {Green}, {Grenier}, {Guillemot}, {Guiriec},
  {Gustafsson}, {Harding}, {Hays}, {Hewitt}, {Horan}, {Hou}, {Jankowski},
  {Johnson}, {Johnson}, {Johnston}, {Kataoka}, {Keith}, {Kerr}, {Kramer},
  {Kuss}, {Latronico}, {Lee}, {Li}, {Li}, {Limyansky}, {Longo}, {Loparco},
  {Lorusso}, {Lovellette}, {Lower}, {Lubrano}, {Lyne}, {Maan}, {Maldera},
  {Manchester}, {Manfreda}, {Marelli}, {Mart{\'\i}-Devesa}, {Mazziotta},
  {McEnery}, {Mereu}, {Michelson}, {Mickaliger}, {Mitthumsiri}, {Mizuno},
  {Moiseev}, {Monzani}, {Morselli}, {Negro}, {Nemmen}, {Nieder}, {Nuss},
  {Omodei}, {Orienti}, {Orlando}, {Ormes}, {Palatiello}, {Paneque},
  {Panzarini}, {Parthasarathy}, {Persic}, {Pesce-Rollins}, {Pillera}, {Poon},
  {Porter}, {Possenti}, {Principe}, {Rain{\`o}}, {Rando}, {Ransom}, {Ray},
  {Razzano}, {Razzaque}, {Reimer}, {Reimer}, {Renault-Tinacci}, {Romani},
  {S{\'a}nchez-Conde}, {Parkinson}, {Scotton}, {Serini}, {Sgr{\`o}}, {Shannon},
  {Sharma}, {Shen}, {Siskind}, {Spandre}, {Spinelli}, {Stappers}, {Stephens},
  {Suson}, {Tabassum}, {Tajima}, {Tak}, {Theureau}, {Thompson}, {Tibolla},
  {Torres}, {Valverde}, {Venter}, {Wadiasingh}, {Wang}, {Wang}, {Wang},
  {Weltevrede}, {Wood}, {Yan}, {Zaharijas}, {Zhang}, \& {Zhu}}]{saa+23}
{Smith}, D.~A., {Abdollahi}, S., {Ajello}, M., {et~al.} 2023, \apj, 958, 191,
  \dodoi{10.3847/1538-4357/acee67}

\bibitem[{{Stokes} {et~al.}(1985){Stokes}, {Taylor}, {Welsberg}, \&
  {Dewey}}]{stw+85}
{Stokes}, G.~H., {Taylor}, J.~H., {Welsberg}, J.~M., \& {Dewey}, R.~J. 1985,
  \nat, 317, 787, \dodoi{10.1038/317787a0}

\bibitem[{{Su} {et~al.}(2023){Su}, {Han}, {Wang}, {Yuan}, {Wang}, {Zhou},
  {Wang}, {Yan}, {Jing}, {Yang}, {Cai}, {Chen}, {Xu}, {Xie}, {Wang}, {Xu}, \&
  {You}}]{shw+23}
{Su}, W.~Q., {Han}, J.~L., {Wang}, P.~F., {et~al.} 2023, \mnras, 526, 2645,
  \dodoi{10.1093/mnras/stad2159}

\bibitem[{{Tam} {et~al.}(2006){Tam}, {Kaspi}, {Gaensler}, \&
  {Gotthelf}}]{tkg+06}
{Tam}, C.~R., {Kaspi}, V.~M., {Gaensler}, B.~M., \& {Gotthelf}, E.~V. 2006,
  \apj, 652, 548, \dodoi{10.1086/507262}

\bibitem[{{Tan} {et~al.}(2020){Tan}, {Bassa}, {Cooper}, {Hessels},
  {Kondratiev}, {Michilli}, {Sanidas}, {Stappers}, {van Leeuwen}, {Donner},
  {Grie{\ss}meier}, {Kramer}, {Tiburzi}, {Weltevrede}, {Ciardi}, {Hoeft},
  {Mann}, {Miskolczi}, {Schwarz}, {Vocks}, \& {Wucknitz}}]{tbc+20}
{Tan}, C.~M., {Bassa}, C.~G., {Cooper}, S., {et~al.} 2020, \mnras, 492, 5878,
  \dodoi{10.1093/mnras/staa113}

\bibitem[{Torii {et~al.}(1998)Torii, Kinugasa, Katayama, Tsunemi, \&
  Yamauchi}]{tkk+98}
Torii, K., Kinugasa, K., Katayama, K., Tsunemi, H., \& Yamauchi, S. 1998, ApJ,
  503, 843

\bibitem[{{Tyul'bashev} {et~al.}(2020){Tyul'bashev}, {Kitaeva}, {Tyul'bashev},
  {Malofeev}, \& {Tyul'basheva}}]{tkt+20}
{Tyul'bashev}, S.~A., {Kitaeva}, M.~A., {Tyul'bashev}, V.~S., {Malofeev},
  V.~M., \& {Tyul'basheva}, G.~E. 2020, Astronomy Reports, 64, 526,
  \dodoi{10.1134/S1063772920060074}

\bibitem[{{Tyul'bashev} {et~al.}(2022){Tyul'bashev}, {Kitaeva}, \&
  {Tyul'basheva}}]{tkt22}
{Tyul'bashev}, S.~A., {Kitaeva}, M.~A., \& {Tyul'basheva}, G.~E. 2022, \mnras,
  517, 1112, \dodoi{10.1093/mnras/stac2404}

\bibitem[{{Tyul'bashev} \& {Tyul'basheva}(2023)}]{tt23}
{Tyul'bashev}, S.~A., \& {Tyul'basheva}, G.~E. 2023, Astronomy Reports, 67,
  172, \dodoi{10.1134/S1063772923020099}

\bibitem[{{Tyul'bashev} {et~al.}(2024){Tyul'bashev}, {Tyul'basheva}, {Kitaeva},
  {Ovchinnikov}, {Oreshko}, \& {Logvinenko}}]{ttk+24}
{Tyul'bashev}, S.~A., {Tyul'basheva}, G.~E., {Kitaeva}, M.~A., {et~al.} 2024,
  \mnras, 528, 2220, \dodoi{10.1093/mnras/stae070}

\bibitem[{{Wang} {et~al.}(2023){Wang}, {Han}, {Xu}, {Wang}, {Yan}, {Jing},
  {Su}, {Zhou}, \& {Wang}}]{whx+23}
{Wang}, P.~F., {Han}, J.~L., {Xu}, J., {et~al.} 2023, Research in Astronomy and
  Astrophysics, 23, 104002, \dodoi{10.1088/1674-4527/acea1f}

\bibitem[{Wang {et~al.}(2025)Wang, Han, Yang, Wang, Wang, Su, Xu, Zhou, Yan,
  Jing, Cai, Yuan, Xu, Wang, \& You}]{why+24}
Wang, P.~F., Han, J.~L., Yang, Z.~L., {et~al.} 2025, Research in Astronomy and
  Astrophysics, 25, 014003, \dodoi{10.1088/1674-4527/ada3b8}

\bibitem[{{Wang} {et~al.}(2021){Wang}, {Zhu}, {Li}, {Pan}, {Wang}, {Cordes},
  {Chatterjee}, {Yao}, {Qian}, {Yue}, {Zhang}, {Zhao}, {Wang}, {Niu}, {Yuan},
  {Miao}, {Xie}, {Liu}, {Yu}, {You}, {Meng}, \& {Collaboration}}]{wzl+21}
{Wang}, S., {Zhu}, W.-W., {Li}, D., {et~al.} 2021, Research in Astronomy and
  Astrophysics, 21, 251, \dodoi{10.1088/1674-4527/21/10/251}

\bibitem[{{Wu} {et~al.}(2023){Wu}, {Yuan}, {Wang}, {Li}, {Wang}, {Xue}, {Zhu},
  {Miao}, {Yan}, {Wang}, {Yao}, {Wang}, {Sun}, {Kou}, {Tu}, {Xie}, {Pan},
  {Zhao}, {Chen}, {Dang}, {Feng}, {Liu}, {Miao}, {Meng}, {Yuan}, {Niu}, {Niu},
  {Qian}, {Wang}, {Xie}, {Xiao}, {Yue}, {You}, {Yu}, {Zhao}, {Zhang}, {Yuen},
  {Wen}, \& {Tedila}}]{wyw+23}
{Wu}, Q.~D., {Yuan}, J.~P., {Wang}, N., {et~al.} 2023, \mnras, 522, 5152,
  \dodoi{10.1093/mnras/stad1323}

\bibitem[{{Xu} {et~al.}(2022){Xu}, {Han}, {Wang}, \& {Yan}}]{xhw+22}
{Xu}, J., {Han}, J., {Wang}, P., \& {Yan}, Y. 2022, Science China Physics,
  Mechanics, and Astronomy, 65, 129704, \dodoi{10.1007/s11433-022-2033-2}

\bibitem[{{Xu} {et~al.}(2011){Xu}, {Wang}, {Han}, \& {Hu}}]{xwhh11}
{Xu}, X., {Wang}, C., {Han}, J., \& {Hu}, L. 2011, Science China Physics,
  Mechanics, and Astronomy, 54, 552, \dodoi{10.1007/s11433-011-4260-x}

\bibitem[{{Yang} {et~al.}(2023){Yang}, {Han}, {Jing}, \& {Su}}]{yhj+23}
{Yang}, Z.~L., {Han}, J.~L., {Jing}, W.~C., \& {Su}, W.~Q. 2023, \apjl, 956,
  L39, \dodoi{10.3847/2041-8213/acfe6e}

\bibitem[{{Yang} {et~al.}(2024){Yang}, {Han}, {Zhou}, {Jing}, {Chen}, \& {et
  al.}}]{yhz+24}
{Yang}, Z.~L., {Han}, J.~L., {Zhou}, D.~J., {et~al.} 2024, Science, in press

\bibitem[{Yang {et~al.}(2025)Yang, Han, Wang, Wang, Su, Chen, Wang, Zhou, Yan,
  Jing, Cai, Xie, Xu, Wang, \& Xu}]{yhw+24}
Yang, Z.~L., Han, J.~L., Wang, T., {et~al.} 2025, Research in Astronomy and
  Astrophysics, 25, 014002, \dodoi{10.1088/1674-4527/ada3b5}

\bibitem[{{Yao} {et~al.}(2017){Yao}, {Manchester}, \& {Wang}}]{ymw17}
{Yao}, J.~M., {Manchester}, R.~N., \& {Wang}, N. 2017, \apj, 835, 29,
  \dodoi{10.3847/1538-4357/835/1/29}

\bibitem[{{Zhi} {et~al.}(2024){Zhi}, {Bai}, {Dai}, {Xu}, {Dang}, {Shang},
  {Zhao}, {Li}, {Zhu}, {Wang}, {Yuan}, {Wang}, {Zhang}, {Feng}, {Wang}, {Wang},
  {Wu}, {Dong}, {Yang}, {Tian}, {Zhong}, {Luo}, {Filipovi{\'c}}, \&
  {Qiao}}]{zbd+24}
{Zhi}, Q.~J., {Bai}, J.~T., {Dai}, S., {et~al.} 2024, \apj, 960, 79,
  \dodoi{10.3847/1538-4357/ad0eca}

\bibitem[{{Zhou} {et~al.}(2022){Zhou}, {Han}, {Zhang}, {Lee}, {Zhu}, {Li},
  {Jing}, {Wang}, {Zhang}, {Jiang}, {Niu}, {Luo}, {Xu}, {Zhang}, {Wang}, {Xu},
  {Wang}, {Yang}, \& {Feng}}]{zhz+22}
{Zhou}, D.~J., {Han}, J.~L., {Zhang}, B., {et~al.} 2022, Research in Astronomy
  and Astrophysics, 22, 124001, \dodoi{10.1088/1674-4527/ac98f8}

\bibitem[{{Zhou} {et~al.}(2023{\natexlab{a}}){Zhou}, {Han}, {Xu}, {Wang},
  {Wang}, {Wang}, {Jing}, {Chen}, {Yan}, {Su}, {Gan}, {Jiang}, {Sun}, {Wang},
  {Wang}, {Wang}, {Xu}, \& {You}}]{zhx+23}
{Zhou}, D.~J., {Han}, J.~L., {Xu}, J., {et~al.} 2023{\natexlab{a}}, Research in
  Astronomy and Astrophysics, 23, 104001, \dodoi{10.1088/1674-4527/accc76}

\bibitem[{{Zhou} {et~al.}(2023{\natexlab{b}}){Zhou}, {Han}, {Jing}, {Wang},
  {Wang}, {Wang}, {Wang}, {Luo}, {Xu}, {Xu}, \& {Wang}}]{zhj+23}
{Zhou}, D.~J., {Han}, J.~L., {Jing}, W.~C., {et~al.} 2023{\natexlab{b}},
  \mnras, 526, 2657, \dodoi{10.1093/mnras/stad2769}

\bibitem[{{Zhou} {et~al.}(2014){Zhou}, {Chen}, {Li}, {Safi-Harb}, {Mendez},
  {Terada}, {Sun}, \& {Ge}}]{zcl+14}
{Zhou}, P., {Chen}, Y., {Li}, X.-D., {et~al.} 2014, \apjl, 781, L16,
  \dodoi{10.1088/2041-8205/781/1/L16}

\bibitem[{{Zhu} {et~al.}(2020){Zhu}, {Li}, {Luo}, {Miao}, {Zhang}, {Spitler},
  {Lorimer}, {Kramer}, {Champion}, {Yue}, {Cameron}, {Cruces}, {Duan}, {Feng},
  {Han}, {Hobbs}, {Niu}, {Niu}, {Pan}, {Qian}, {Shi}, {Tang}, {Wang}, {Wang},
  {Yuan}, {Zhang}, {Zhang}, {Cao}, {Feng}, {Gan}, {Gao}, {Gu}, {Guo}, {Hao},
  {Huang}, {Huang}, {Jiang}, {Jin}, {Li}, {Li}, {Li}, {Liu}, {Pan}, {Peng},
  {Qian}, {Shi}, {Song}, {Song}, {Sun}, {Sun}, {Wang}, {Wang}, {Wang}, {Xie},
  {Yan}, {Yang}, {Yang}, {Yao}, {Yu}, {Yu}, {Zhang}, {Zhang}, {Zhang}, {Zheng},
  {Zhou}, {Zhu}, {Zhu}, {Zhu}, {Zhu}, \& {Zhu}}]{zll+20}
{Zhu}, W., {Li}, D., {Luo}, R., {et~al.} 2020, \apjl, 895, L6,
  \dodoi{10.3847/2041-8213/ab8e46}

\end{thebibliography}


\setcounter{figure}{0}
\setcounter{table}{0}
\captionsetup[table]{name={\bf Table A}}
\captionsetup[figure]{name={\bf Figure A}}


\section*{Appendix: Data and profiles of newly discovered pulsars}

We discovered 473 pulsars, in addition to 201 pulsars and 1 RRAT in \citet{hww+21} and 76 RRATs in \citet{zhx+23}. For completeness, we here list all objects from gpps0202 in Table~A\ref{gppsPSRtab1}, with the references for objects studied in other papers. 

We present in Figure~A\ref{gppsPSRprof} the mean profiles of pulsars, except for the RRATs. The stacking pulses of 20 RRATs are presented in Figure~A\ref{19RRATs}.


\begin{table*}[h]
\centering
\caption{A list of pulsars discovered by the GPPS survey. See http://zmtt.bao.ac.cn/GPPS/ for updates.}
\label{gppsPSRtab1}
\setlength{\tabcolsep}{4pt} 
\footnotesize
\renewcommand{\arraystretch}{0.85}  
\begin{tabular}{lccrccccrrrrr}
\hline\noalign{\smallskip}
 Name$^*$    &	gpps No.  &  Period    & DM      &  RA(2000)   &   DEC(2000)  &   GL  &    GB  &  $S_{\rm 1.25GHz}$  & $D_{\rm NE2001}$ & $D_{\rm YMW16}$ & FWHM & Ref.\\
             &            &   (s)   &  &  (hh:mm:ss) & ($\pm$dd:mm) &  ($^{\circ}$) & ($^{\circ}$) & ($\mu$Jy) & (kpc)  & (kpc)   &  ($\degr$)   &  \\ 
\hline
J1853$-$0009g  & gpps0202 & 0.72110 &  443.1 & 18:53:36.2 & $-$00:09 &  33.0354 & $-$0.5556 &  12.7 & 7.2 & 5.5 &    7.0\\
J1906$+$0352g  & gpps0203 & 0.28560 &  372.0 & 19:06:27.2 & $+$03:52 &  38.0986 & $-$1.5628 &  33.7 & 7.9 & 11.2&    8.3\\
J1837$+$0528g  & gpps0204 & 0.00626 &  120.8 & 18:37:40.5 & $+$05:28 &  36.2535 & $+$5.5477 &  18.9 & 3.8 & 5.9 &   17.9\\
J1840$+$0151g  & gpps0205 & 1.75041 &   73.6 & 18:40:58.6 & $+$01:51 &  33.3997 & $+$3.1769 &  20.6 & 2.5 & 2.7 &    5.2\\
J1854$+$0358g  & gpps0206 & 0.59084 &  486.5 & 18:54:04.0 & $+$03:58 &  36.7715 & $+$1.2287 &  18.7 & 9.2 & 11.9&    5.9\\
J1858$-$0055g  & gpps0207 & 2.84662 &  270.7 & 18:58:05.8 & $-$00:55 &  32.8614 & $-$1.9064 &  16.5 & 6.3 & 5.6 &    7.0\\
J1939$+$2452g  & gpps0208 & 2.40253 &  240.5 & 19:39:09.8 & $+$24:52 &  60.3194 & $+$1.4247 &   7.7 & 8.1 & 8.6 &    3.2\\
J1933$+$2037g  & gpps0209 & 0.79886 &  100.0 & 19:33:01.7 & $+$20:37 &  55.9258 & $+$0.5929 &  23.6 & 4.2 & 3.3 &   11.6\\
J1951$+$2528g  & gpps0210 & 0.00231 &  205.1 & 19:51:27.0 & $+$25:28 &  62.2360 & $-$0.6780 & 286.8 & 7.3 & 7.9 &   63.8\\
J1927$+$1457g  & gpps0211 & 0.91963 &  211.8 & 19:27:06.8 & $+$14:57 &  50.2759 & $-$0.8925 &  11.0 & 6.7 & 5.5 &    6.2\\
J1953$+$3303g  & gpps0212 & 0.71289 &  290.7 & 19:53:09.3 & $+$33:03 &  68.9375 & $+$2.8809 &  21.6 & 11.2& 11.3&    6.9\\
J1954$+$2833g  & gpps0213 & 0.02721 &   37.0 & 19:54:11.3 & $+$28:33 &  65.1956 & $+$0.3812 & 123.5 & 2.9 & 2.4 &   12.3\\
J1921$+$1329g  & gpps0214 & 0.40842 &  573.5 & 19:21:26.3 & $+$13:29 &  48.3261 & $-$0.3790 &  11.8 & 14.4& 11.9&   19.9\\
J1919$+$1341g  & gpps0215 & 0.01166 &  394.5 & 19:19:23.0 & $+$13:41 &  48.2513 & $+$0.1680 & 153.9 & 9.9 & 8.7 &   45.1 & [J] \\
J1933$+$1454g  & gpps0216 & 1.34613 &  204.8 & 19:33:35.3 & $+$14:54 &  50.9850 & $-$2.2875 &  17.4 & 6.8 & 5.5 &    3.4\\
J1839$+$0221g  & gpps0217 & 0.27277 &  117.3 & 18:39:18.3 & $+$02:21 &  33.6438 & $+$3.7704 &  46.1 & 3.5 & 4.3 &    6.4\\
J1828$+$0020g  & gpps0218 & 0.00294 &   40.4 & 18:28:21.6 & $+$00:20 &  30.6114 & $+$5.2954 &   9.5 & 1.6 & 1.3 &  195.0\\
J1917$+$1710g  & gpps0219 & 0.41934 &   41.6 & 19:17:37.8 & $+$17:10 &  51.1564 & $+$2.1606 &  32.3 & 2.6 & 1.7 &   14.7\\
J1908$+$0457g  & gpps0220 & 0.97647 &  356.7 & 19:08:30.4 & $+$04:57 &  39.2861 & $-$1.5248 &  30.0 & 7.8 & 11.4&    1.9\\
J1835$-$0011g  & gpps0221 & 0.00323 &   36.2 & 18:35:46.0 & $-$00:11 &  30.9675 & $+$3.3961 & 205.5 & 1.5 & 1.2 &  197.2\\
J1913$+$0458g  & gpps0222 & 0.44479 &  103.8 & 19:13:37.9 & $+$04:58 &  39.8936 & $-$2.6478 &  17.8 & 3.7 & 4.1 &    7.3\\
J1848$+$1245g  & gpps0223 & 0.24823 &  100.0 & 18:48:07.3 & $+$12:45 &  43.9652 & $+$6.5043 & 335.4 & 3.8 & 5.3 &    7.8\\
J1845$+$0104g  & gpps0224 & 0.00671 &   94.9 & 18:45:50.4 & $+$01:04 &  33.2530 & $+$1.7367 & 102.6 & 2.9 & 3.1 &  106.2\\
J1945$+$2410g  & gpps0225 & 2.37759 &  478.4 & 19:45:16.0 & $+$24:10 &  60.4016 & $-$0.1271 &  41.9 & 14.9& 16.2&    5.6\\
J1906$+$0335g  & gpps0226 & 1.29639 &  213.0 & 19:06:48.8 & $+$03:35 &  37.8755 & $-$1.7792 &  8.0  & 5.3 & 6.0 &    2.8 & [2]\\
J1943$+$2210g  & gpps0227 & 0.01287 &  110.7 & 19:43:53.7 & $+$22:10 &  58.5182 & $-$0.8487 & 513.6 & 4.6 & 4.0 &    6.6\\
J1946$+$2757g  & gpps0228 & 0.74303 &  458.0 & 19:46:44.0 & $+$27:57 &  63.8469 & $+$1.4908 &  92.9 & 50.0& 15.3&   44.7 & [J] \\
J1905$+$0649g  & gpps0229 & 0.02747 &  187.7 & 19:05:05.2 & $+$06:49 &  40.5808 & $+$0.0796 & 163.5 & 4.9 & 4.4 &   21.1\\
J1920$+$0129g  & gpps0230 & 0.00358 &  104.1 & 19:20:39.9 & $+$01:29 &  37.6166 & $-$5.8097 & 449.9 & 3.5 & 4.9 &   16.9\\
J1847$+$0148g  & gpps0231 & 0.60246 &  314.7 & 18:47:09.5 & $+$01:48 &  34.0601 & $+$1.7794 &  21.9 & 7.0 & 6.5 &   22.9\\
J2036$+$3506g  & gpps0232 & 1.37092 &  108.4 & 20:36:11.6 & $+$35:06 &  75.5900 & $-$3.3891 &  13.2 & 4.9 & 5.4 &   12.5\\
J1931$+$1841g  & gpps0233 & 2.59411 &  268.8 & 19:31:12.7 & $+$18:41 &  54.0179 & $+$0.0307 &  48.6 & 8.2 & 5.4 &    4.9\\
J1921$+$0851g  & gpps0234 & 0.95671 &  101.0 & 19:21:11.3 & $+$08:51 &  44.2102 & $-$2.4994 &  72.9 & 4.2 & 4.8 &    3.2 & [2]\\
J1911$+$1206g  & gpps0235 & 0.00344 &  181.7 & 19:11:24.5 & $+$12:06 &  45.9711 & $+$1.1440 & 172.1 & 5.5 & 6.3 &   40.3\\
J1857$+$0642g  & gpps0236 & 0.00353 &   21.6 & 18:57:58.6 & $+$06:42 &  39.6395 & $+$1.6144 & 1009.4 & 1.6 & 1.0 &   44.6\\
J1921$+$1216g  & gpps0237 & 0.00300 &  256.1 & 19:21:03.3 & $+$12:16 &  47.2142 & $-$0.8655 & 600.1 & 7.2 & 6.8 &   44.3 & [J] \\
J1846$+$0153g  & gpps0238 & 2.05252 &  322.7 & 18:46:34.3 & $+$01:53 &  34.0687 & $+$1.9485 &  11.2 & 7.3 & 6.9 &    3.1\\
J1929$+$2355g  & gpps0239 & 0.00479 &  206.8 & 19:29:44.8 & $+$23:55 &  58.4448 & $+$2.8383 & 310.8 & 7.6 & 8.9 &   30.9\\
J1937$+$1358g  & gpps0240 & 2.64537 &  174.8 & 19:37:26.4 & $+$13:58 &  50.6178 & $-$3.5567 & 132.1 & 6.2 & 5.7 &   27.9\\
J1904$+$0553Bg & gpps0241 & 0.57454 &  140.9 & 19:04:17.6 & $+$05:53 &  39.6422 & $-$0.1600 &   4.9 & 4.2 & 3.7 &    6.2\\
J1946$+$0904g  & gpps0242 & 0.02577 &   37.2 & 19:46:56.8 & $+$09:04 &  47.4641 & $-$7.9780 & 647.2 & 2.2 & 1.6 &   29.6\\
J1953$+$1006g  & gpps0243 & 0.00259 &   57.2 & 19:53:33.9 & $+$10:06 &  49.1723 & $-$8.8962 &  91.5 & 2.8 & 2.5 &   18.0\\
J1948$+$1801g  & gpps0244 & 0.53624 &  265.1 & 19:48:39.9 & $+$18:01 &  55.4851 & $-$3.8973 &  12.5 & 9.9 & 12.9 &   14.3\\
J1838$+$1507g  & gpps0245 & 0.00382 &   54.8 & 18:38:36.6 & $+$15:07 &  45.0882 & $+$9.6166 & 362.1 & 2.6 & 2.4 &   58.9\\
J1851$+$0347g  & gpps0246 & 2.14145 &  439.6 & 18:51:57.2 & $+$03:47 &  36.3591 & $+$1.6099 &  10.0 & 9.1 & 12.0 &   4.9\\
J1912$+$0735g  & gpps0247 & 3.68507 &  160.4 & 19:12:41.2 & $+$07:35 &  42.1068 & $-$1.2286 &  31.9 & 4.8 & 4.9 &   10.4\\
J1937$+$1941g  & gpps0248 & 0.16676 &  333.8 & 19:37:16.2 & $+$19:41 &  55.5835 & $-$0.7367 &  49.5 & 9.7 & 8.8 &   26.6\\
J1935$+$1836g  & gpps0249 & 0.39850 &  230.3 & 19:35:02.9 & $+$18:36 &  54.3849 & $-$0.8048 &  13.3 & 7.3 & 5.0 &   13.3\\
J1809$-$0116g  & gpps0250 & 1.88738 &  128.5 & 18:09:43.1 & $-$01:16 &  26.9905 & $+$8.6810 &  59.7 & 3.9 & 9.1 &    3.2\\
J2015$+$3423g  & gpps0251 & 0.78465 &  219.1 & 20:15:46.9 & $+$34:23 &  72.5819 & $-$0.3813 &  17.0 & 7.1 & 6.6 &   14.7\\
J1843$-$0157g  & gpps0252 & 0.50886 &  751.0 & 18:43:36.3 & $-$01:57 &  30.2935 & $+$0.8474 & 209.0 & 10.3 & 10.7 &   28.1  & [J] \\
J1835$-$0113g  & gpps0253 & 0.59963 &  316.7 & 18:35:08.4 & $-$01:13 &  29.9865 & $+$3.0685 &  54.8 & 7.1 & 11.9 &    8.1\\
J1952$+$2818g  & gpps0254 & 0.00309 &  250.2 & 19:52:36.7 & $+$28:18 &  64.8041 & $+$0.5506 &  22.4 & 8.0 & 8.1 &   72.8\\
\multicolumn{13}{r}{ ... to be continued. } \\
\hline 
\end{tabular}
\tablecomments{
``g" indicates the temporary nature, due to position uncertainty of about 1.5'; 
$^{s}$ discovered via single pulse search module; 
References [1]: \citet{hww+21}; [2]: \citet{zhx+23}; [3]: parameters updated by \citet{zbd+24}; [J]: Jing WC. et al. (2024).} 
\end{table*}
\addtocounter{table}{-1}
\begin{table*}
\centering
\caption{A list of pulsars discovered by the GPPS survey -- {\it continued.}}
\setlength{\tabcolsep}{4pt} 
\footnotesize
\renewcommand{\arraystretch}{0.85}  
\begin{tabular}{lccrccccrrrrr}
\hline\noalign{\smallskip}
 Name$^*$    &	gpps No.  &  Period    & DM      &  RA(2000)   &   DEC(2000)  &   GL  &    GB  &  $S_{\rm 1.25GHz}$  & $D_{\rm NE2001}$ & $D_{\rm YMW16}$ & FWHM & Ref.\\
             &            &   (s)   &  & (hh:mm:ss) & ($\pm$dd:mm) &  ($^{\circ}$) & ($^{\circ}$) & ($\mu$Jy) & (kpc)  & (kpc)      &  ($\degr$)  & \\ 
\hline
J1929$+$1504g  & gpps0255 & 0.23278 &   91.9 & 19:29:14.4 & $+$15:04 &  50.6170 & $-$1.2917 &  11.2 & 4.0 & 2.8 &    8.7\\
J2024$+$3751g  & gpps0256 & 0.21164 &  446.6 & 20:24:32.9 & $+$37:51 &  76.4462 & $+$0.1272 &  11.6 & 50.0 & 15.4 &   13.1\\
J1850$-$0050g  & gpps0257 & 0.22152 & 1060.0 & 18:50:27.3 & $-$00:50 &  32.0765 & $-$0.1622 &  93.8 & 12.9 & 7.0 &   90.9 & [J] \\
J1851$+$1021g  & gpps0258 & 0.86786 &  191.2 & 18:51:34.6 & $+$10:21 &  42.1870 & $+$4.6724 &  52.5 & 6.5 & 13.3 &    6.0\\
J1904$+$0945g  & gpps0259 & 1.56779 &  326.6 & 19:04:20.6 & $+$09:45 &  43.0757 & $+$1.5973 &  21.4 & 8.1 & 11.3 &    4.7\\
J1903$+$0949g  & gpps0260 & 0.32544 &  107.9 & 19:03:41.7 & $+$09:49 &  43.0697 & $+$1.7739 &  22.8 & 3.9 & 3.9 &    3.9\\
\hline 
\end{tabular}
\end{table*}
\addtocounter{table}{-1}
\begin{table*}
\centering
\caption{A list of pulsars discovered by the GPPS survey -- {\it continued.}}
\setlength{\tabcolsep}{4pt} 
\footnotesize
\renewcommand{\arraystretch}{0.85}  
\begin{tabular}{lccrccccrrrrr}
\hline\noalign{\smallskip}
 Name$^*$    &	gpps No.  &  Period    & DM      &  RA(2000)   &   DEC(2000)  &   GL  &    GB  &  $S_{\rm 1.25GHz}$  & $D_{\rm NE2001}$ & $D_{\rm YMW16}$ & FWHM & Ref.\\
             &            &   (s)   &  & (hh:mm:ss) & ($\pm$dd:mm) &  ($^{\circ}$) & ($^{\circ}$) & ($\mu$Jy) & (kpc)  & (kpc)      &  ($\degr$)  & \\ 
\hline 
J1912$+$1000g  & gpps0321 & 3.05280 &  147.0 & 19:12:46.3 & $+$10:00 &  44.2592 & $-$0.1271 &  3.7  & 3.2 & 4.1 & 2.1    & [2]\\
J1919$+$1113g  & gpps0322 & 0.76593 &  288.0 & 19:19:27.1 & $+$11:13 &  46.0935 & $-$1.0167 &  4.8  & 7.6 & 7.5 & 4.2    & [2]\\
J1921$+$1227g  & gpps0323 & 1.59764 &  259.0 & 19:21:22.4 & $+$12:27 &  47.4143 & $-$0.8472 &  0.27 & 7.3 & 6.8 & 2.8    & [2]\\
J1940$+$2231g  & gpps0324 & 5.68230 &  198.0 & 19:40:55.1 & $+$22:31 &  58.4695 & $-$0.0847 &  2.9  & 6.8 & 7.6 & 3.3    & [2]\\
J1926$+$1803g  & gpps0325 & 1.03465 &  315.5 & 19:26:38.8 & $+$18:03 &  52.9418 & $+$0.6779 &  15.1 & 9.2 & 6.1 &    1.6\\
J1837$-$0048g  & gpps0326 & 0.00273 &  195.5 & 18:37:35.9 & $-$00:48 &  30.6358 & $+$2.7110 & 331.0 & 4.7 & 5.0 &   40.0\\
\hline 
\end{tabular}
\end{table*}
\addtocounter{table}{-1}
\begin{table*}
\centering
\caption{A list of pulsars discovered by the GPPS survey -- {\it continued.}}
\setlength{\tabcolsep}{4pt} 
\footnotesize
\renewcommand{\arraystretch}{0.85}  
\begin{tabular}{lccrccccrrrrr}
\hline\noalign{\smallskip}
 Name$^*$    &	gpps No.  &  Period    & DM      &  RA(2000)   &   DEC(2000)  &   GL  &    GB  &  $S_{\rm 1.25GHz}$  & $D_{\rm NE2001}$ & $D_{\rm YMW16}$ & FWHM & Ref.\\
             &            &   (s)   &  & (hh:mm:ss) & ($\pm$dd:mm) &  ($^{\circ}$) & ($^{\circ}$) & ($\mu$Jy) & (kpc)  & (kpc)      &  ($\degr$)   & \\ 
\hline
J1927$+$1849g  & gpps0389 & 0.31203 &  200.0 & 19:27:46.2 & $+$18:49 &  53.7437 & $+$0.8092 &  20.8 & 6.5 & 4.6 &   63.3 & [2]\\
J1938$+$2248g  & gpps0390 & 0.79392 &  128.4 & 19:38:38.7 & $+$22:48 &  58.4633 & $+$0.5143 &  11.7 & 5.0 & 4.5 &    4.3\\
J1938$+$2301g  & gpps0391 & 0.65507 &  398.9 & 19:38:12.5 & $+$23:01 &  58.6030 & $+$0.7085 & 137.1 &12.0 &12.0 &    2.1\\
J1938$+$2302g  & gpps0392 & 0.05276 &  303.4 & 19:38:11.6 & $+$23:02 &  58.6030 & $+$0.7085 & 402.1 & 9.1 & 8.9 &   13.0\\
J2018$+$3518g  & gpps0393 & 0.03131 &  267.0 & 20:18:48.1 & $+$35:18 &  73.6865 & $-$0.3455 & 152.5 & 8.4 & 7.1 &   18.3\\
J1836$-$0150g  & gpps0394 & 0.00546 &  199.4 & 18:36:05.8 & $-$01:50 &  29.5372 & $+$2.5683 & 661.6 & 4.8 & 4.9 &   76.2\\
\hline 
\end{tabular}
\end{table*}
\addtocounter{table}{-1}
\begin{table*}
\centering
\caption{A list of pulsars discovered by the GPPS survey -- {\it continued.}}
\setlength{\tabcolsep}{4pt} 
\footnotesize
\renewcommand{\arraystretch}{0.85}  
%
\caption{Integrated profiles of newly discovered pulsars, scaled to
the peak and plotted in rotation phase of 360$^{\circ}$ for a full
period. The pulsar name, gpps number, period and DM are noted in
each panel. RRATs are not here. -- {\it to be continued}. }
\label{gppsPSRprof}
\addtocounter{figure}{-1}
\end{figure*}
\begin{figure*}
\centering
\caption{Integrated profiles of newly discovered pulsars -- {\it to be continued}.}
\addtocounter{figure}{-1}
\end{figure*}
\begin{figure*}
\centering
\caption{Integrated profiles of newly discovered pulsars -- {\it to be continued}.}
\addtocounter{figure}{-1}
\end{figure*}
\begin{figure*}
\centering
\caption{Integrated profiles of newly discovered pulsars -- {\it to be continued}.}
\addtocounter{figure}{-1}
\end{figure*}
\begin{figure*}
\centering
\caption{Integrated profiles of newly discovered pulsars --  {\it continued and ended}.}
\end{figure*}

\begin{figure*}
\centering
    \includegraphics[width=0.29\textwidth]{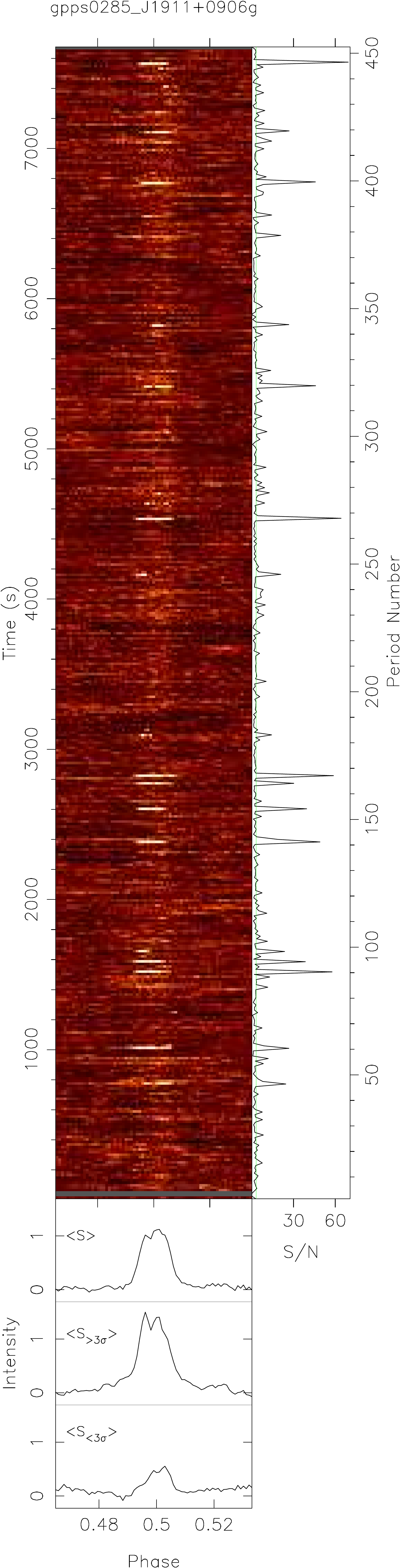}
    \includegraphics[width=0.29\textwidth]{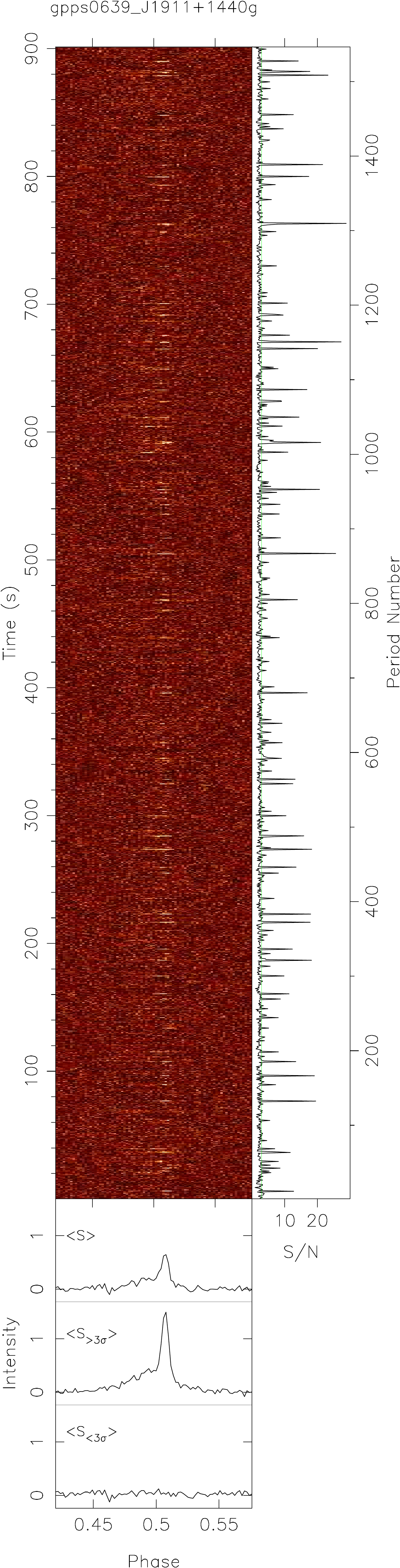}
    \includegraphics[width=0.29\textwidth]{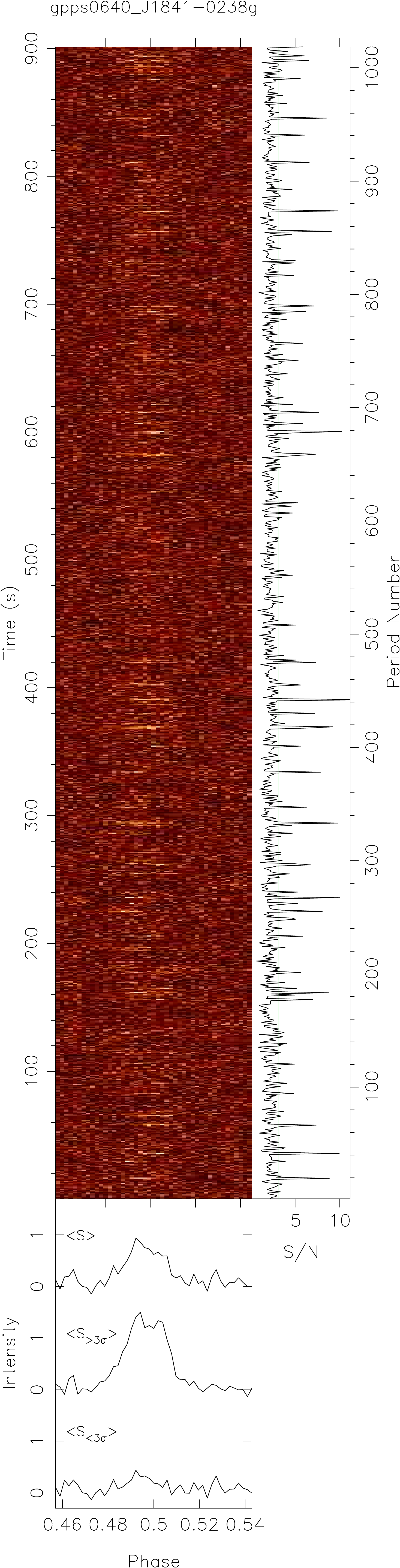}
\caption{The Pulse-stacks of 20 RRATs newly discovered in the GPPS survey. Pulse-stack is shown in the main left panel, where only a few pulses are occasionally emitted. The right panel shows the curve of S/N over pulse number, with an indication of $3\sigma$, here the sigma is calculated from a given width of the off-pulse phase range. Three sub-panels below the main panel are the averaged profiles of all periods and of single pulses with the signal-to-noise ratio $>3\sigma$ and $<3\sigma$  -- to be continued.}
\label{19RRATs}
\end{figure*}
\addtocounter{figure}{-1}
\begin{figure*}
\centering
    \includegraphics[width=0.31\textwidth]{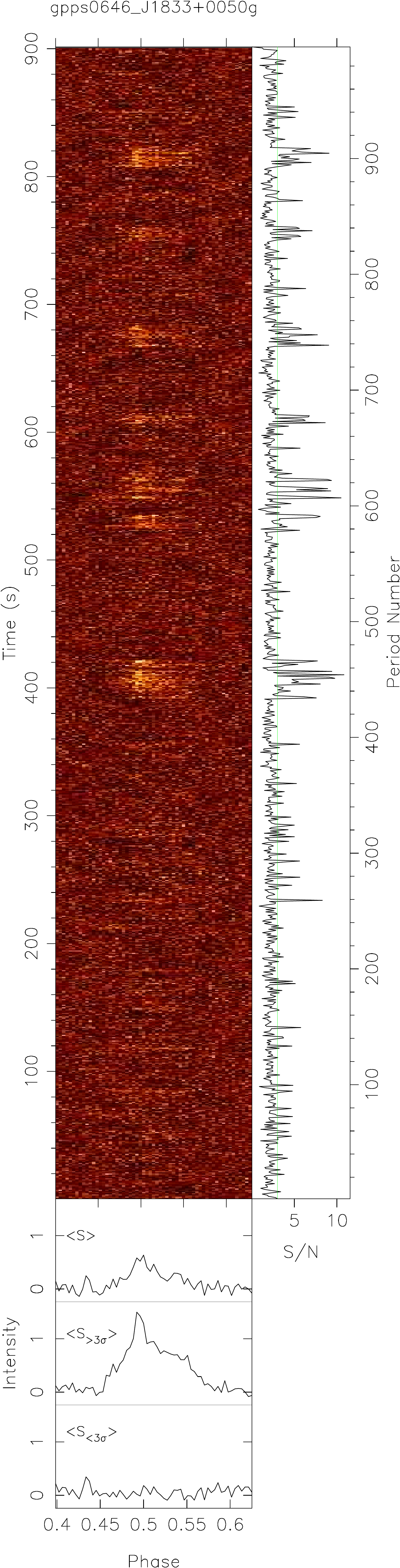}
    \includegraphics[width=0.31\textwidth]{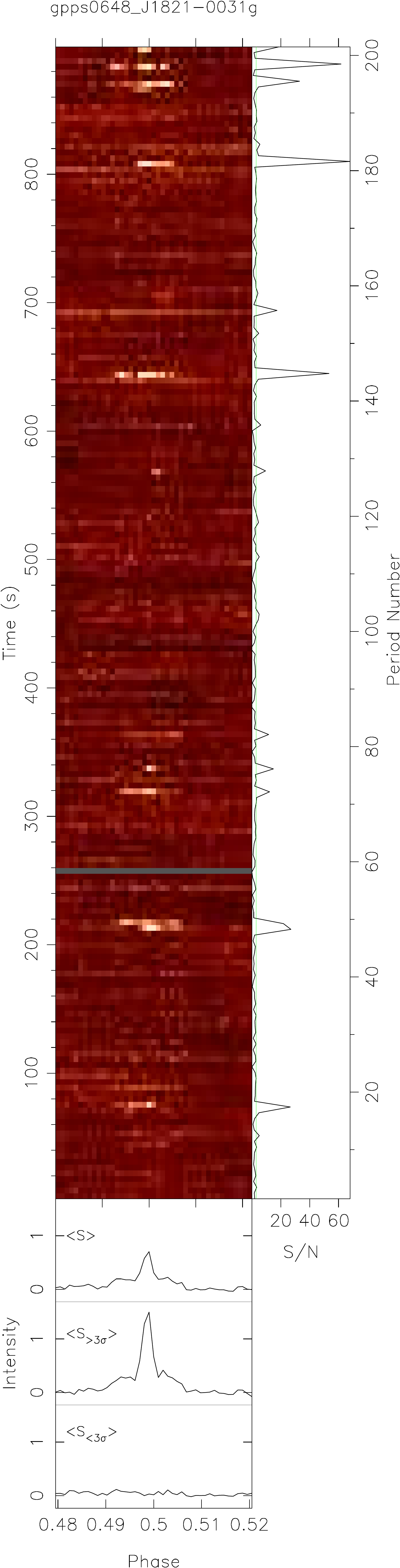}
    \includegraphics[width=0.31\textwidth]{RRAT/RAA-2024-0315.R2fig10-1.pdf}
\caption{The Pulse-stacks of 20 RRATs newly discovered in the GPPS survey  -- to be continued.}
\end{figure*}
\addtocounter{figure}{-1}
\begin{figure*}
\centering
    \includegraphics[width=0.31\textwidth]{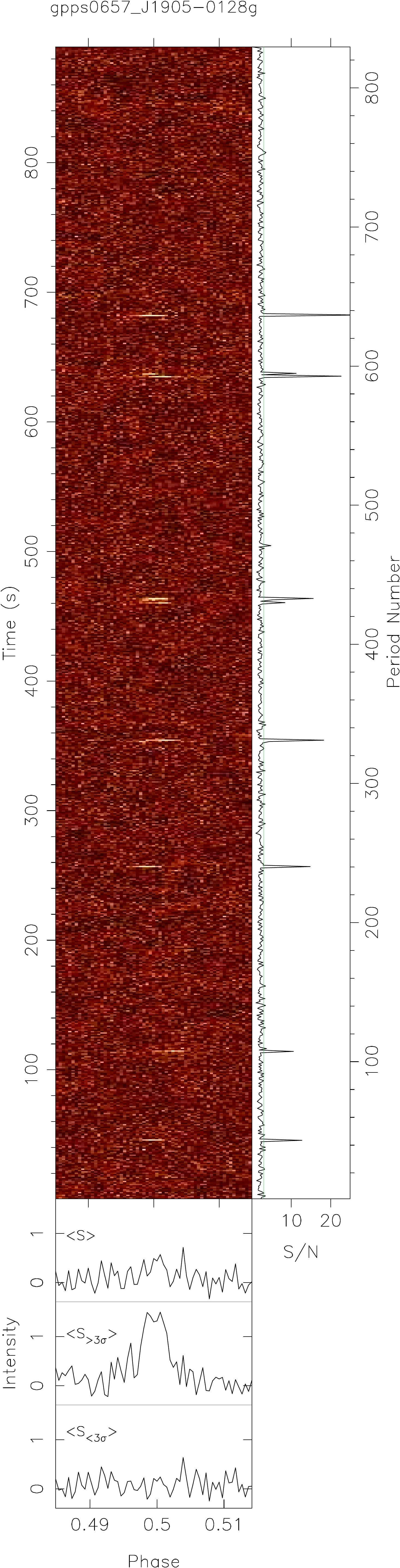}
    \includegraphics[width=0.31\textwidth]{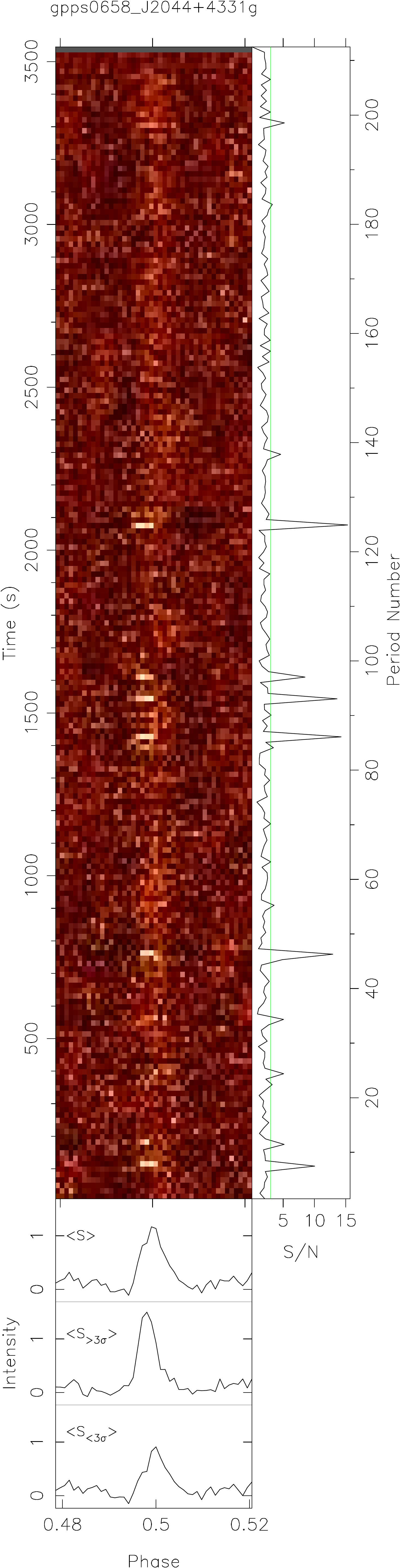}
    \includegraphics[width=0.31\textwidth]{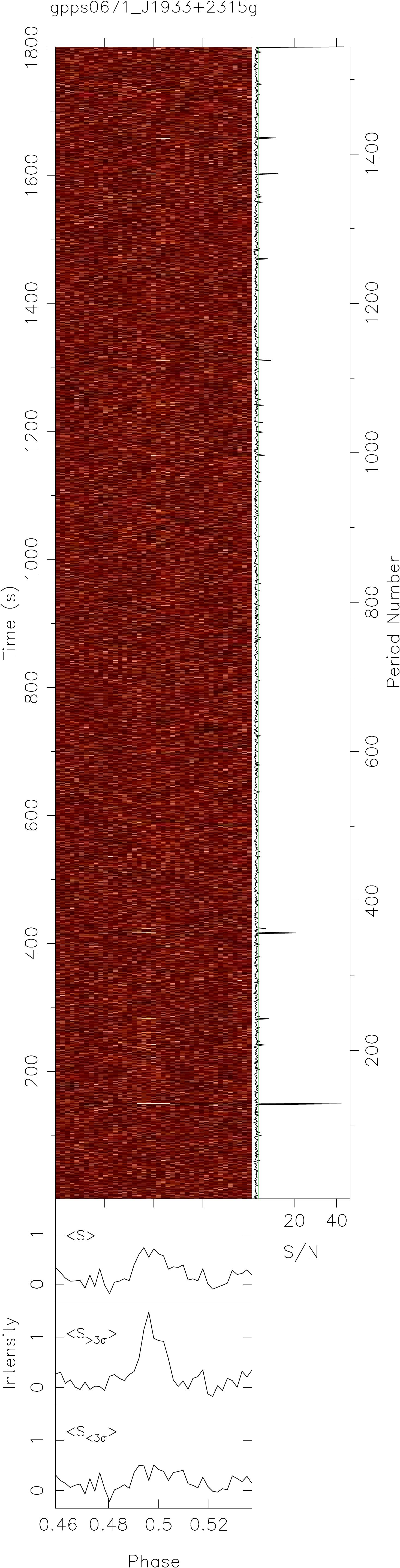}
\caption{The Pulse-stacks of 20 RRATs newly discovered in the GPPS survey  -- to be continued.}
\end{figure*}
\addtocounter{figure}{-1}
\begin{figure*}
\centering
    \includegraphics[width=0.31\textwidth]{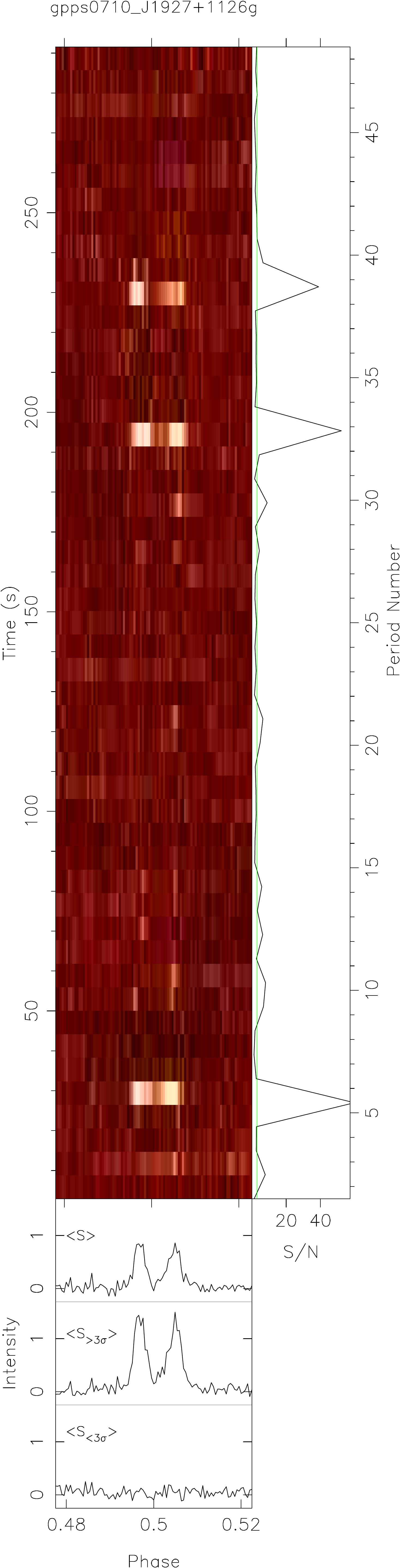}
    \includegraphics[width=0.31\textwidth]{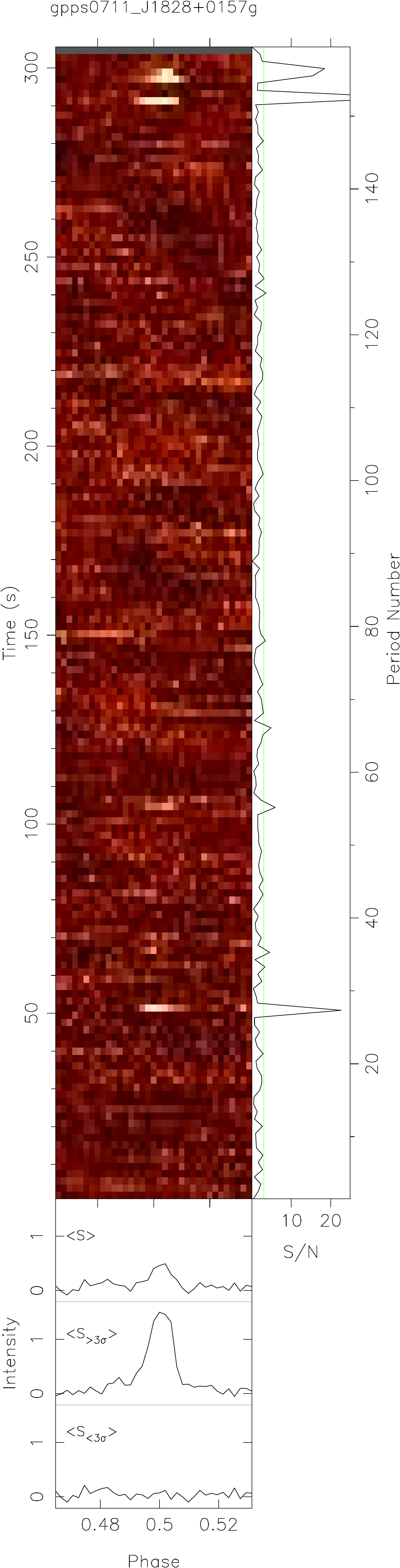}
    \includegraphics[width=0.31\textwidth]{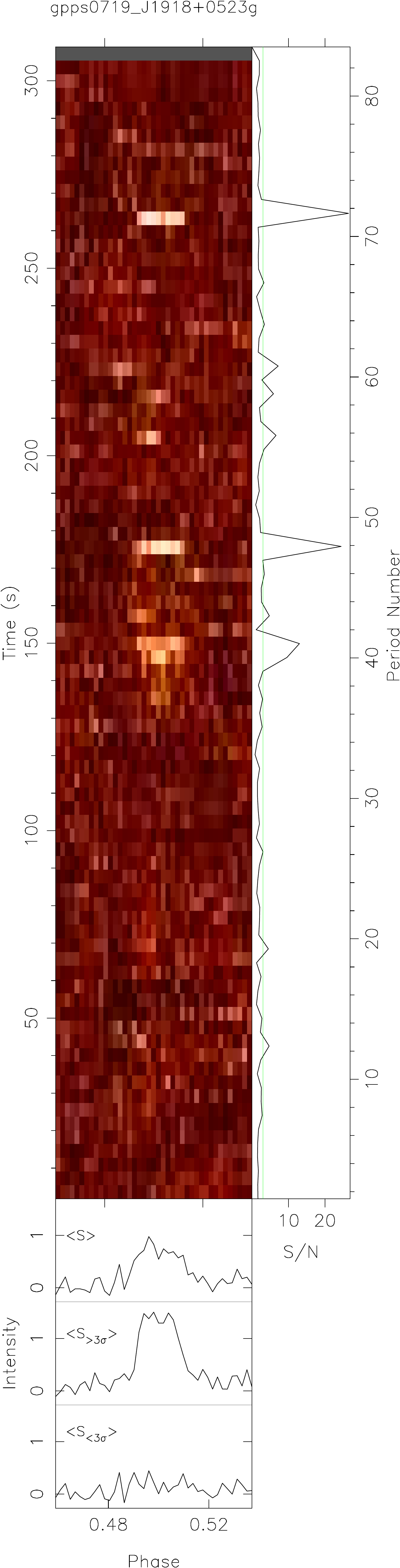}
\caption{The Pulse-stacks of 20 RRATs newly discovered in the GPPS survey  -- to be continued.}
\end{figure*}
\addtocounter{figure}{-1}
\begin{figure*}
\centering
    \includegraphics[width=0.31\textwidth]{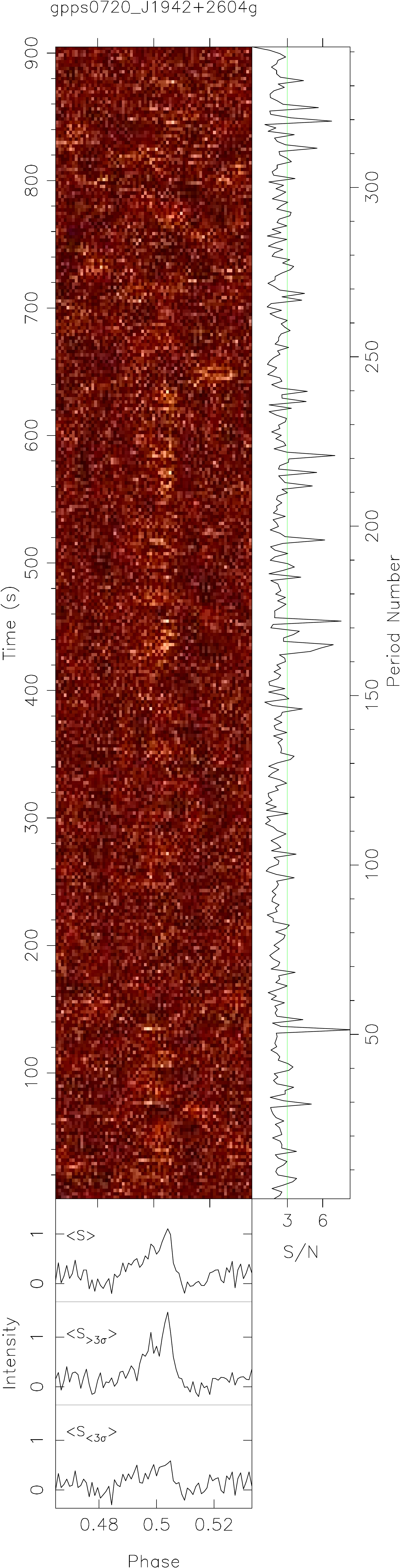}
    \includegraphics[width=0.31\textwidth]{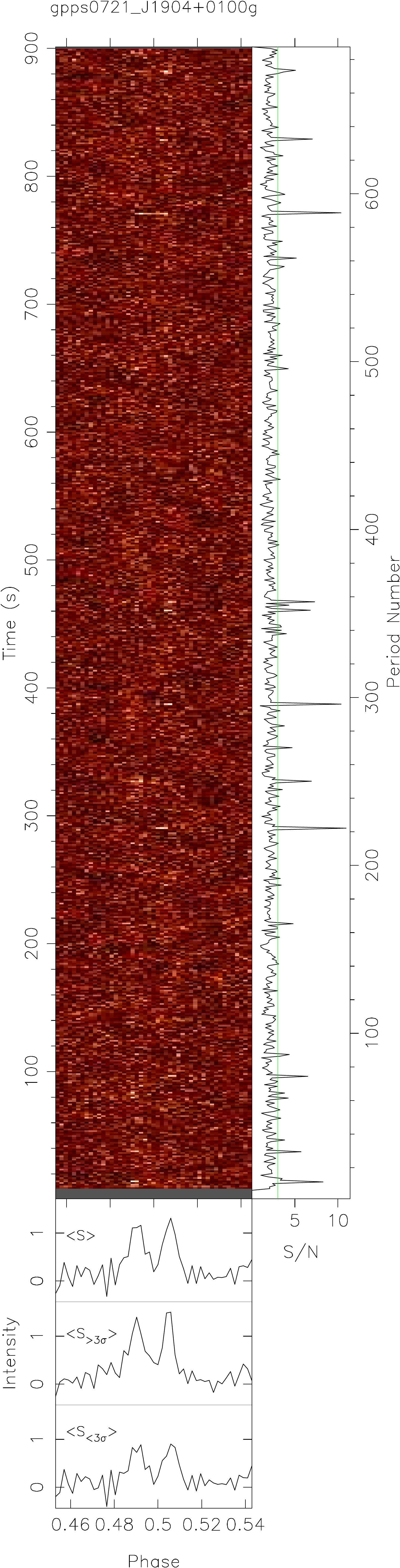}
    \includegraphics[width=0.31\textwidth]{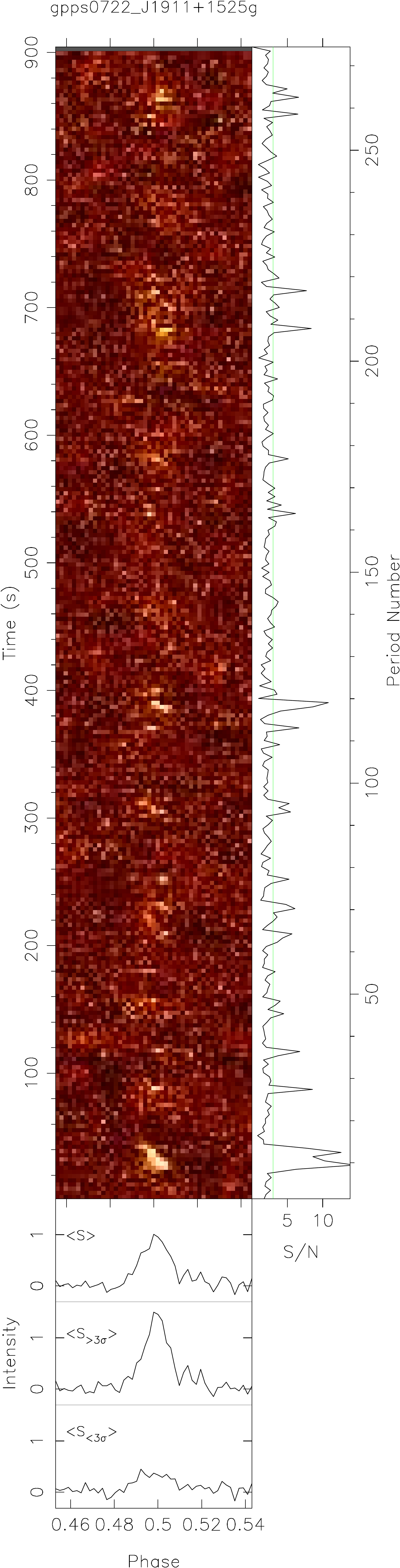}
\caption{The Pulse-stacks of 20 RRATs newly discovered in the GPPS survey  -- to be continued.}
\end{figure*}
\addtocounter{figure}{-1}
\begin{figure*}
\centering
    \includegraphics[width=0.31\textwidth]{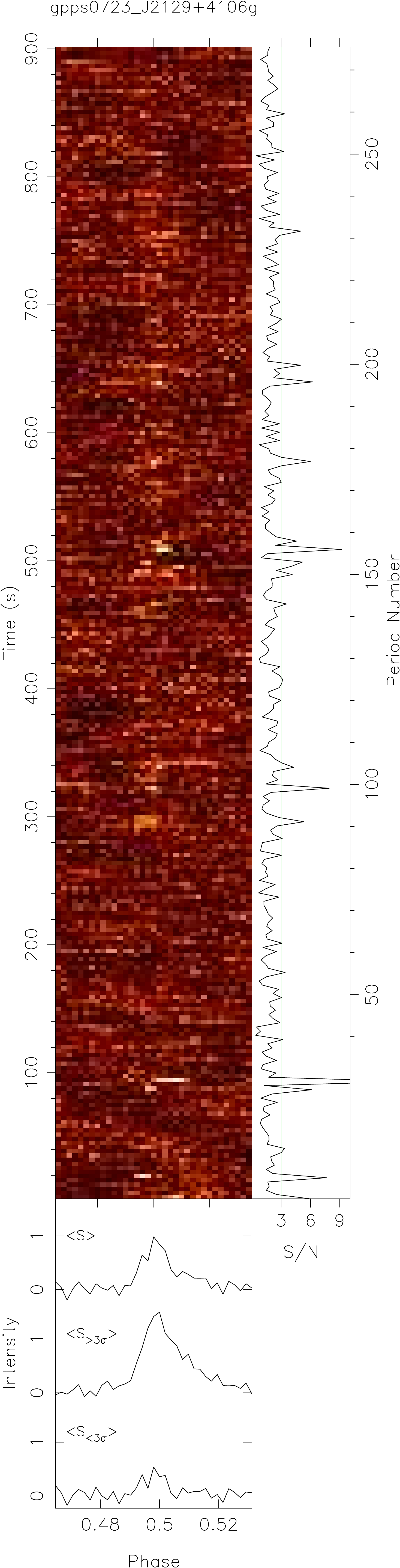}
    \includegraphics[width=0.31\textwidth]{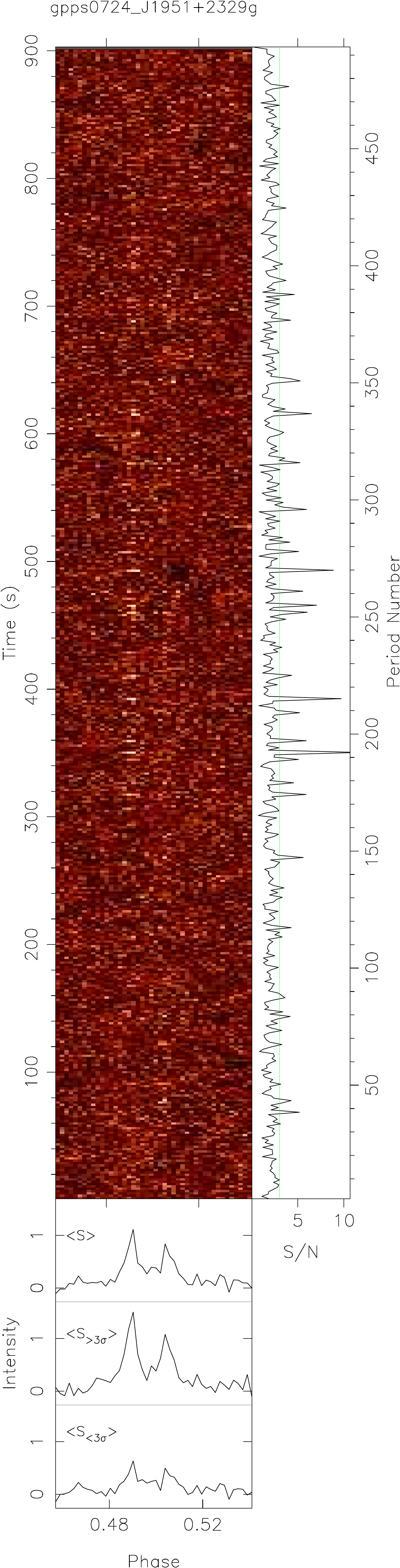}
    \includegraphics[width=0.31\textwidth]{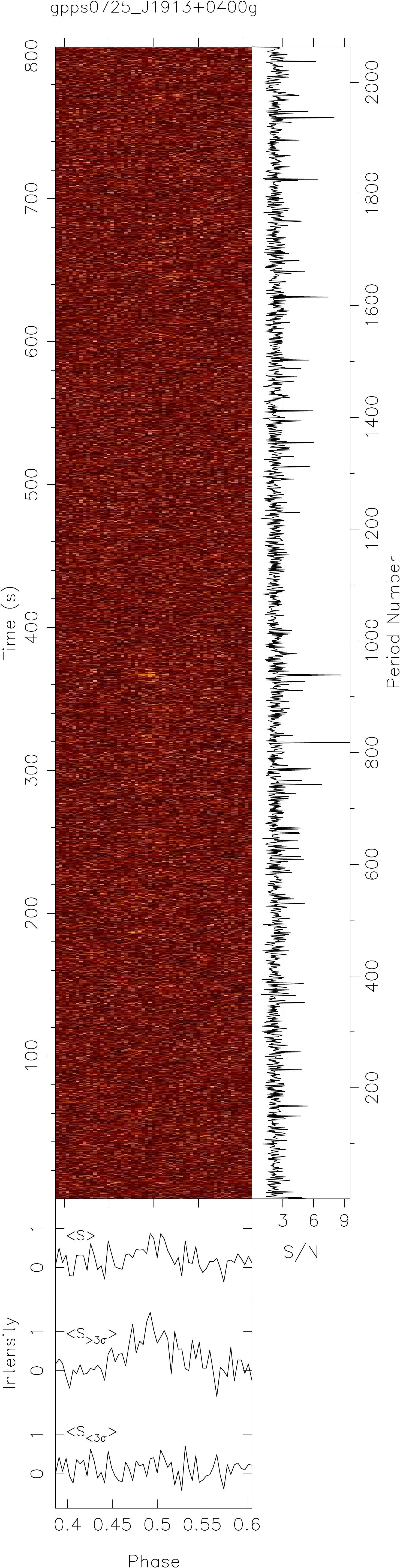}
\caption{The Pulse-stacks of 20 RRATs newly discovered in the GPPS survey  -- to be continued.}
\end{figure*}
\addtocounter{figure}{-1}
\begin{figure*}
\centering
    \includegraphics[width=0.30\textwidth]{RRAT/RAA-2024-0315.R2fig10-2.pdf}
    \includegraphics[width=0.30\textwidth]{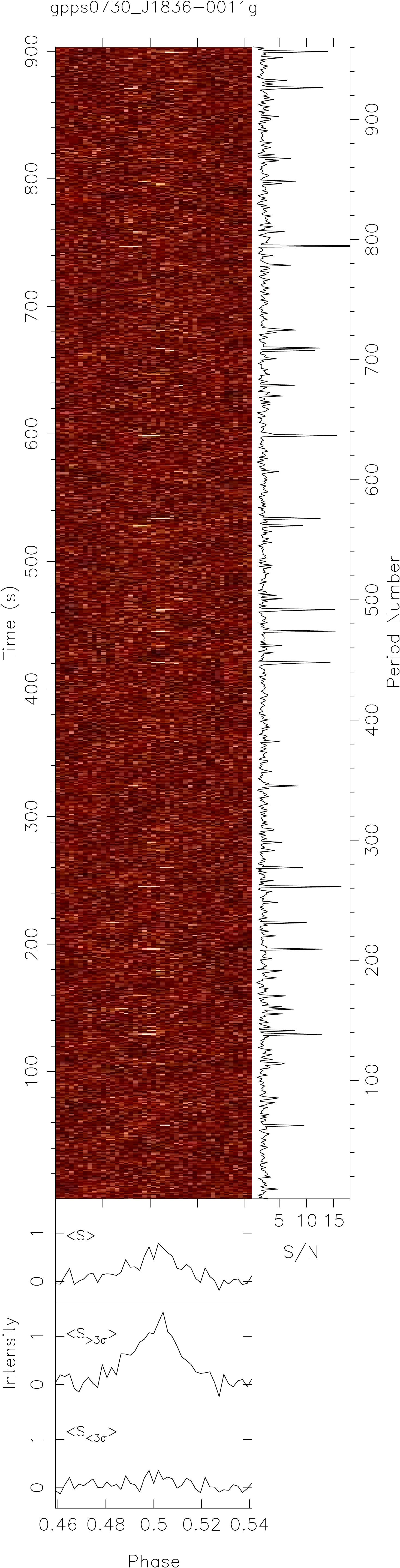}
\caption{The Pulse-stacks of 20 RRATs newly discovered in the GPPS survey  -- continued and ended.}
\end{figure*}

\end{document}